\documentclass[aps,
longbibliography,notitlepage,reprint,superscriptaddress]{revtex4-2}
\usepackage[colorlinks,linkcolor=blue,citecolor=blue,urlcolor=red]{hyperref}
\usepackage{soul}
\usepackage{changes}
\usepackage{float}
\usepackage{CJK}
\usepackage{graphicx}
\usepackage{amsmath,amssymb,amsfonts}
\usepackage{siunitx}
\usepackage{soul}
\usepackage{chngcntr}
\usepackage{chemformula}
\usepackage{MnSymbol}
\usepackage{dsfont}	% proper mathbb format
\usepackage{appendix}

\newcommand{\SiN}{$\mathrm{Si_3N_4}$}
\newcommand{\EPFL}{Swiss Federal Institute of Technology Lausanne (EPFL), CH-1015 Lausanne, Switzerland}
\newcommand{\EPFLQ}{Center for Quantum Science and Engineering, EPFL, CH-1015 Lausanne, Switzerland}
\newcommand{\CTH}{Department of Microtechnology and Nanoscience (MC2),
	Chalmers University of Technology, SE-412 96 G\"oteborg, Sweden}

\begin{document}

\author{Guanhao Huang}
\thanks{These authors contributed equally to this work.}
\affiliation{\EPFL}
\affiliation{\EPFLQ}
\author{Alberto Beccari}
\thanks{These authors contributed equally to this work.}
\affiliation{\EPFL}
\affiliation{\EPFLQ}
\author{Nils J. Engelsen}
\email{nils.engelsen@chalmers.se}
\affiliation{\EPFL}
\affiliation{\EPFLQ}
\affiliation{\CTH}
\author{Tobias J. Kippenberg}
\email{tobias.kippenberg@epfl.ch}
\affiliation{\EPFL}
\affiliation{\EPFLQ}

\title{Room-temperature quantum optomechanics using an ultra-low noise cavity}

\maketitle

	\begin{bfseries}
		Ponderomotive squeezing of light, where a mechanical oscillator creates quantum correlations between the phase and amplitude of the interacting light field, is a canonical signature of the quantum regime of optomechanics \cite{purdy2013observation}. At room temperature, this has only been reached in pioneering experiments where an optical restoring force controls the oscillator stiffness, akin to the vibrational motion of atoms in an optical lattice. These include both levitated nanoparticles \cite{magrini2022squeezed} and optically-trapped cantilevers \cite{aggarwal2020roomtemperature}. Recent advances in engineered mechanical resonators, where the restoring force is provided by material rigidity rather than an external optical potential, have realized ultra-high quality factors ($Q$) by exploiting `soft clamping' \cite{tsaturyan2017ultracoherent,ghadimi2018elastic,hoj2021ultracoherent,bereyhi2022hierarchical,bereyhi2022perimeter,shin2022spiderweb}. However entering the quantum regime with such resonators, has so far been prevented by optical cavity frequency fluctuations \cite{saarinen2023laser} and thermal intermodulation noise \cite{fedorov2020thermal,pluchar2023thermal}. Here, we overcome this challenge and demonstrate optomechanical squeezing at room temperature in a phononic-engineered ``membrane-in-the-middle" system. By using a high finesse cavity whose mirrors are patterned with phononic crystal structures, we reduce cavity frequency noise by more than 700-fold. In this ultra-low noise cavity, we introduce a silicon nitride membrane oscillator whose density is modulated by silicon nano-pillars~\cite{hoj2022ultracoherent}, yielding both high thermal conductance and a localized mechanical mode with $Q$ of $1.8\times10^8$. These advances enable operation of the optomechanical displacement sensor within a factor of 2.5 of the Heisenberg limit \cite{aspelmeyer2014cavity}, leading to squeezing of the probing laser field by \SI{1.09}{dB} below the vacuum fluctuations. Moreover, the long thermal decoherence time of the membrane oscillator (more than 30 vibrational periods) allows us to obtain conditional displaced thermal states of motion with an occupation of 0.97 phonons, using a multimode Kalman filter. Our work extends quantum control of engineered macroscopic oscillators to room temperature.
	\end{bfseries}	
	
	\section{Introduction}
%%%%%%%%%%
The fragile nature of quantum systems \cite{haroche2006exploring} renders them susceptible to the influence of the thermal environment. This presents a significant challenge in quantum science and technology, especially for solid-state systems that are strongly coupled to their environment. Nevertheless, over the past decade, quantum control has been extended to solid-state mechanical resonators, both with radiation pressure optomechanical coupling \cite{aspelmeyer2014cavity} and piezoelectric coupling with superconducting qubits \cite{chu2017quantum,satzinger2018quantum}. 
Cavity optomechanics, where the mechanical oscillator is dispersively coupled to an optical cavity, has enabled numerous advances including ground state cooling \cite{chan2011laser,rossi2018measurementbased}, optomechanical squeezing of light \cite{safavi-naeini2013squeezed,nielsen2017multimode,purdy2013strong,aggarwal2020roomtemperature,mason2019continuous} and entanglement of separate mechanical oscillators \cite{riedinger2018remote,mercierdelepinay2021quantum,kotler2021direct}. 
In addition, optomechanical systems have explored the quantum limits of displacement sensing---a state-of-the-art system demonstrated a backaction-imprecision product exceeding the Heisenberg limit by only \SI{22}{\%} \cite{rossi2018measurementbased}, and showed sensitivities beyond the standard quantum limit (SQL) \cite{mason2019continuous}.
This architecture has also been extended to quantum transducers~\cite{delaney2022superconductingqubit} and may even allow macroscopic tests of quantum mechanics~\cite{schrinski2023macroscopic}. 
Yet, all these advances necessitate cryogenic cooling to reduce thermal fluctuations. Room temperature operation, on the other hand, is beneficial to the accessibility and widespread adoption of technology, as witnessed in other branches of physical science~\cite{alferov2001nobel,bloch2008manybody,bongs2019taking}. 
Developing room temperature quantum optomechanical systems would imply a drastic reduction of experimental complexity by removing the limitations imposed by cryocoolers, such as poor thermalization, excess acoustic noise and limited optical access. Room temperature operation would stimulate applications such as coupling to atomic systems \cite{moller2017quantum}, force microscopy \cite{halg2021membranebased} and variational displacement measurements \cite{mason2019continuous}.

To enter the quantum regime of optomechanics, the product between the total force noise $\bar{S}_{FF}^\mathrm{tot}$ (including environment thermal force $\bar{S}_{FF}^\mathrm{th}$ as well as measurement-induced quantum backaction $\bar{S}_{FF}^\mathrm{qba}$) and the displacement measurement imprecision $\bar{S}_{xx}^\mathrm{imp}$ must approach the limit $\sqrt{\bar{S}_{xx}^\mathrm{imp}\bar{S}_{FF}^\mathrm{tot}} \geq \hbar/2$ set by the Heisenberg uncertainty principle \cite{aspelmeyer2014cavity}. A necessary condition imposed by this limit is that the quantum backaction (QBA) rate of the light field $\Gamma_\mathrm{qba}=x_\mathrm{zpf}^2\bar{S}_\mathrm{FF}^\mathrm{qba}/\hbar^2$($x_\mathrm{zpf}$ being the oscillator's zero-point displacement fluctuation amplitude) must exceed the thermal decoherence rate $\Gamma_{\text{th}}$ of the mechanical oscillator, which is determined by the bath temperature $T$ and by the quality factor $Q$ as $\Gamma_{\text{th}}=k_B T/(\hbar Q)$. 
In addition, the required measurement imprecision is far below the SQL, imposing strict limitations on the amount of tolerable cavity frequency noise.

Over the past decade, multiple pioneering approaches were pursued to reach the ultralow mechanical dissipation required to enter the quantum backaction-dominated regime at room temperature, including levitated nanoparticles \cite{chang2010cavity} and micromechanical objects whose rigidity is controlled by an optical field \cite{corbitt2006squeezedstate,cripe2019measurement}. These methods enhance the mechanical $Q$ by optical trapping, and resulted in recent observations of quantum backaction~\cite{cripe2019measurement}, optomechanical squeezing of light~\cite{aggarwal2020roomtemperature,magrini2022squeezed} and ground state cooling~\cite{magrini2021realtime}---all at room temperature. 
Yet, room temperature quantum optomechanical phenomena have not been accessible with engineered ultracoherent solid-state mechanical resonators, due to thermal intermodulation noise \cite{fedorov2020thermal}, vibrations of the cavity mirror substrates \cite{saarinen2023laser} and optical heating-induced instability \cite{metzger2008selfinduced}. These thermal effects result in excess imprecision and backaction noise, preventing their product from reaching the Heisenberg limit.

Here we demonstrate optomechanical squeezing of light at room temperature using a membrane-in-the-middle system with an ultracoherent membrane \cite{tsaturyan2017ultracoherent}. Our system also allows a measurement of mechanical motion strong enough to project the mechanical mode to a displaced thermal state with occupation of a single phonon. This implies that the strength of the measurement is sufficient to implement measurement-based quantum state preparation protocols, e.g. feedback cooling to the ground state \cite{bowen2015quantum}. We achieve this by eliminating cavity frequency noise with a phononic-engineered Fabry-P\'erot cavity, and mitigating intermodulation noise through single port homodyne detection.

\begin{figure}[t]
\includegraphics[width = 0.495\textwidth, page = 1]{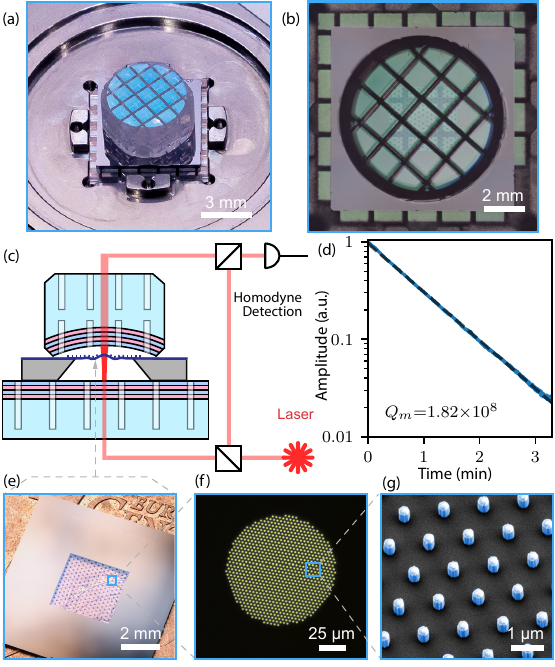}	\caption{\textbf{Ultra-low noise phononic-engineered membrane cavity. (a)} Photograph of the membrane-in-the-middle assembly. \textbf{(b)} Optical microscope image of the MIM assembly from the top, showing the overlapping square unit cells of the top and bottom phononic crystal mirrors and the density-modulated membrane.  \textbf{(c)} Setup schematic. \textbf{(d)} Mechanical ringdown measurement of the pillar membrane's soft clamped mode quality factor. The ringdown was acquired with the membrane installed in the MIM cavity. \textbf{(e, f, g)} Overview image and details of a pillar membrane sample at different length scales.}
\label{fig:fig1}
\end{figure}

\section{Ultra-low noise optomechanical cavity}
\begin{figure*}[t]
\includegraphics[width = 1\textwidth, page = 1]{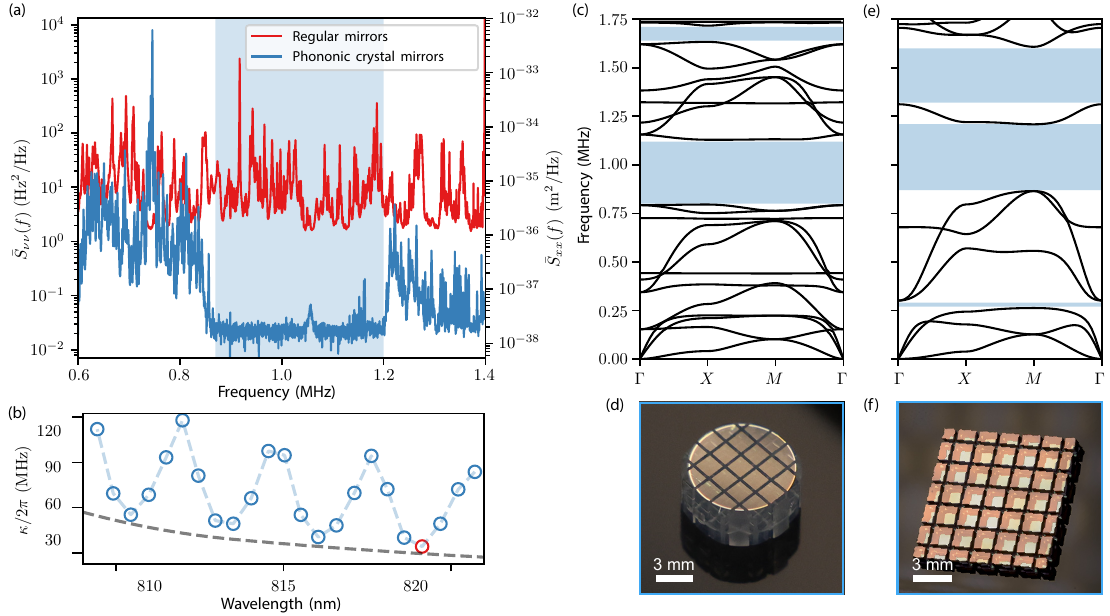} \caption{
\textbf{Suppression of cavity frequency noise in the phononic bandgap. (a)} Cavity frequency/displacement noise comparison between regular mirror assembly and phononic crystal mirror assembly, showing a 700-fold total noise reduction within the blue-shaded region. The vertical axis is calibrated in cavity frequency noise units (left) and in equivalent mirror mechanical displacement units (right). \textbf{(b)} MIM cavity optical linewidth as a function of wavelength. Blue circles: measured optical cavity linewidths. Red circle: the optical mode used for experiments. The modulation of the cavity linewidth is due to the presence of the membrane in the cavity. Dashed gray line: the ideal empty cavity linewidth based on the measured mirror transmission after deposition of the high-reflectivity coating but before definition of the PnC. \textbf{(c,d)} Band diagram and photograph of the top phononic crystal mirror.  \textbf{(e,f)} Band diagram and photograph of the bottom phononic crystal mirror. }
\label{fig:fig2}
\end{figure*}
The quantum cooperativity,
$\mathcal{C}_q=4\overline{n}_{\mathrm{cav}}g_0^2/(\kappa\Gamma_{\mathrm{th}})=\Gamma_{\mathrm{qba}}/\Gamma_{\mathrm{th}}$, is the ratio between quantum backaction and thermal decoherence rate ($\overline{n}_{\mathrm{cav}}$ is the mean intracavity photon number, $g_0$ is the vacuum optomechanical coupling rate and $\kappa$ is the cavity linewidth). To achieve Heisenberg-limited operation, the quantum cooperativity must exceed unity.
To reach this regime, we adopt a modular approach and employ the membrane-in-the-middle (MIM) architecture \cite{thompson2008strong} using an optical Fabry-Perot cavity [see Fig.~\ref{fig:fig1}(a,b,c)]. The high-finesse cavity ($\mathcal{F}\sim\SI{e4}{}$) allows operation at high quantum cooperativity while keeping the optical probe power below $\SI{1}{mW}$, where quantum-noise-limited laser operation can be achieved. 
Heisenberg-limited operation further requires low displacement measurement imprecision, i.e. $\bar{S}_{xx}^\mathrm{imp}<x_\mathrm{zpf}^2/\Gamma_\mathrm{th}$. This is particularly challenging at room temperature, as the required imprecision scales inversely with the number of quanta in the mechanical oscillator's thermal bath. For our device, we estimate the bound to be $\SI{e-35}{m^2/Hz}$.
The cavity frequency noise is thus required to be extremely small:
the frequency noise $\bar{S}_{\nu\nu}(f)=\bar{S}_{xx}^\mathrm{imp,cav}(f)\times g_0^2/(2\pi x_\mathrm{zpf})^2$ should satisfy $\bar{S}_{\nu\nu}(f)<(g_0/2\pi)^2/\Gamma_\mathrm{th}$ to allow ground state cooling \cite{rabl2009phasenoise} and significant optomechanical squeezing. This bound is well below the typical thermal fluctuations of the cavity mirrors, even with state-of-the-art mechanical resonators.  In Ref.~\cite{saarinen2023laser}, a phononic shield was used to address this problem by producing a phononic bandgap which suppressed the driven response of the mirror. A significant reduction of thermomechanical cavity mirror noise, however, has remained elusive.

We overcome this challenge by engineering both cavity mirrors' vibrational spectra with phononic crystal structures (PnC) [see Fig.~\ref{fig:fig2}(d,f)]. A precision circular saw is used to pattern the phononic structure on glass mirror substrates endowed with high-reflectivity dielectric coatings. The phononic unit cell dimensions (see Supplementary Information) are chosen such that mechanical motion in the frequency band of \SI{0.87}{MHz} to \SI{1.2}{MHz} is prohibited [see Fig.~\ref{fig:fig2}(c,e)]. The thermomechanical noise density $\bar{S}_{\nu\nu}(f)$ in this frequency band is reduced by a factor of more than 700 as shown in Fig.~\ref{fig:fig2}(a), where the estimation of the mirror noise suppression is limited by laser noise. 
This noise reduction greatly relaxes the requirements to observe quantum optomechanical effects at room temperature.
Furthermore, we show linewidth measurements of 23 TE\textsubscript{00} cavity resonances with the membrane chip loaded in Fig.~\ref{fig:fig2}(b). As can be seen from comparison with the ideal empty cavity linewidth (dashed gray line), the phononic crystal patterning did not result in significant excess optical losses, thereby maintaining high cavity out-coupling efficiency as required for the observation of optomechanical squeezing and measurement-based ground state cooling. We use the optical mode at \SI{819}{nm} for the experiment, which has an out-coupling efficiency of $>\SI{80}{\%}$ with an optical linewidth of $\kappa/2\pi = \SI{34.2}{MHz}$. 

A suitable ultracoherent mechanical membrane resonator is vital for operation of the MIM system at room temperature. To this end, recently-demonstrated phononic density-modulated membranes are promising \cite{hoj2022ultracoherent}. Compared to earlier stress-modulated, perforated, soft-clamped membranes \cite{tsaturyan2017ultracoherent}, this design maintains higher material stress which enhances dissipation dilution. Furthermore, the unperforated membrane reduces optical losses due to scattering from \SiN-vacuum interfaces, and increases heat dissipation, diminishing the thermal effects due to optical absorption. 
After reproducing this design, we found however that the membrane had high optical absorption, which led to strong cavity bi-stability \cite{an1997optical} and mechanical instability \cite{metzger2008selfinduced}. We therefore developed a fabrication process that minimizes optical absorption, such that photothermal effects are absent for the optical mode and optical powers used in this experiment.

The mechanical resonator consists of a \SiN\ nanomechanical membrane patterned with \ch{aSi}-\ch{HfO2} nanopillars (700-nm diameter) much smaller than the acoustic wavelength, implementing phononic density modulation \cite{hoj2022ultracoherent}. Microscope images of the device at different length scales are shown in Fig.~\ref{fig:fig1}(e,f,g). The periodic density modulation creates a mechanical bandgap, that spectrally isolates a 7-ng high-$Q$ soft-clamped defect mode with a mechanical frequency of $\Omega_m/2\pi = \SI{1.16}{MHz}$ and a damping rate of $\Gamma_m/2\pi=\SI{6.41}{mHz}$, corresponding to a room temperature thermal occupancy of $\overline{n}_\mathrm{th}=\SI{5.3e6}{}$ and zero-point fluctuations of $x_\mathrm{zpf}=\SI{1.0}{fm}$. A ringdown measurement of mechanical $Q$ is shown in Fig.~\ref{fig:fig1}(d), giving $Q = \SI{1.8e8}{}$. 
More details on the device fabrication are given in the Supplementary Information. By clamping the density-modulated membrane chip in-between the phononic crystal mirrors, we construct a MIM system with $g_0/2\pi=\SI{160}{Hz}$ and cavity frequency noise satisfying the $\bar{S}_{\nu\nu}(f)<\SI{0.11}{Hz^2/Hz}$ requirement, which allows high quantum cooperativity operation with quantum-noise-limited measurement imprecision and backaction.

\section{Optomechanical squeezing of light}
\begin{figure*}[!t]
 	\includegraphics[width = 1\textwidth]{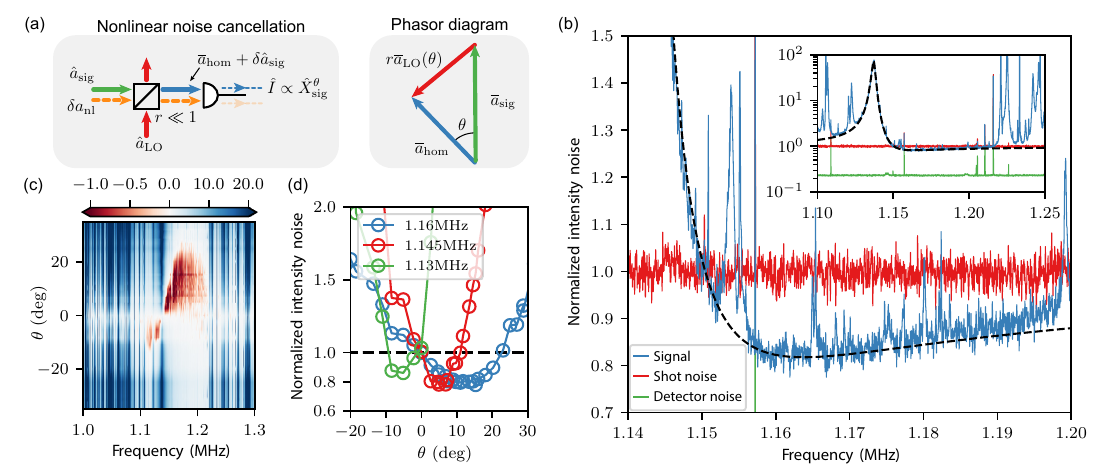}	\caption{\textbf{Optomechanical generation of squeezed light. (a)} Homodyne detection scheme that cancels nonlinear mixing noise $\delta a_\mathrm{nl}$ from the records of cavity transmission. For each quadrature angle $\theta$, a specific LO amplitude is needed in order to cancel the nonlinear mixing noise. 
  \textbf{(b)} Detected power spectral density (PSD) at the quadrature angle $\theta = \SI{15.3}{\degree}$ at frequencies above the mechanical resonance, compared to the measured shot noise (SN) and detector noise. Inset: the same PSD is displayed on broader frequency and power ranges. \textbf{(c)} Collection of homodyne photocurrent spectra of the cavity output field, normalized to the shot noise level, measured at different quadrature angles. The magnitude and bandwidth of squeezing depends on the homodyne quadrature angle. \textbf{(d)} Averaged homodyne PSD within three different frequency bands (\SI{5}{kHz} bandwidth) at different quadrature angles, showing a maximum squeezing depth of \SI{22.2}{\%} (\SI{1.09}{dB}).}
	\label{fig:fig3}
\end{figure*}

To demonstrate that the system operates in the quantum regime at room temperature, we generate optomechanical squeezing---the quantum signature most robust against calibration errors.  
In the textbook description of cavity optomechanics, when the cavity is driven on resonance, the vacuum fluctuations of the laser amplitude drive the mechanical oscillator and are imprinted onto the mechanical motion. The linear response of the cavity also transduces the mechanical motion into phase fluctuations of the light field. This results in correlations between the amplitude and phase quadratures of the cavity output field, and manifests as a noise reduction below the shot noise level (squeezing). However, the nonlinear transduction response of the cavity produces mixing products at sum and difference frequencies of the mechanical modes, giving rise to excess nonlinear noise beyond vacuum fluctuations that does not naturally fit in the linear framework of optical quadratures. Due to the high number of modes of the membrane and the large Brownian motion at room temperature, the mixing products manifest as a broadband noise, termed thermal intermodulation noise (TIN) \cite{fedorov2020thermal}. TIN results in intracavity photon number fluctuations that have significant power even at frequencies within the mechanical bandgap, thereby degrading the measurement signal-to-noise ratio (SNR), and inducing additional mechanical decoherence~\cite{pluchar2023thermal}. 

To eliminate TIN intracavity photon number fluctuations, we pump the cavity with the laser detuned by $2\Delta/\kappa = -1/\sqrt{3}$ (magic detuning; $\Delta$ is the frequency detuning of the input field from cavity resonance), where the quadratic term of the cavity response vanishes. Pumping the cavity at this detuning has the additional effect of inducing a redshift and cooling the defect mode to an occupancy $\bar{n}_{\mathrm{eff}}\approx 20$ phonons, via dynamical backaction cooling \cite{aspelmeyer2014cavity}
(lower effective phonon occupancy of $\bar{n}_{\mathrm{eff}}\approx 5.7$ can be achieved with a narrower-linewidth cavity mode at \SI{862}{nm}, cf. Supplementary Information). Furthermore, to eliminate TIN in the optical detection, we deploy a specialized homodyne detection scheme using only one detector, shown in Fig.~\ref{fig:fig3}(a). Instead of balanced homodyne detection, a single detector offers the required photodetection nonlinearity\cite{fedorov2020thermal} to eliminate TIN at arbitrary optical quadrature angles. We achieve this by carefully selecting the local oscillator power for each quadrature angle we detect. A detailed description of this cancellation strategy is provided in the Supplementary Information. 

We measure the noise of the cavity output field at optical quadrature angles ranging from \SI{-33}{\degree} to \SI{33}{\degree} [see Fig.~\ref{fig:fig3}(c)], where the \SI{0}{\degree} quadrature is defined as the one with no mechanical displacement information. Depending on the quadrature angle, we observed optical squeezing (up to \SI{50}{kHz} bandwidth) on either side of the defect mode, whose extent is limited by the membrane modes at the edge of the bandgap [see Fig.~\ref{fig:fig3}(b) for a representative spectrum]. For the three frequency bands that are devoid of parasitic modes and nonlinear mixing noise peaks, we compute the average intensity noise over a bandwidth of \SI{5}{kHz}, as a function of quadrature angle [see Fig.~\ref{fig:fig3}(d)]. We observe a maximum squeezing of \SI{22.2}{\%} (\SI{1.09}{dB}) below the shot noise level, which matches well with the theory prediction. To calibrate the shot noise level, we perform a separate measurement where we direct the same amount of optical power onto the same photodetector and measure the laser intensity noise. More details on the shot noise calibration, including an analysis of classical noise and the detector linearity, are described in the Supplementary Information. 

\section{Conditional quantum state preparation}
\begin{figure}[!t]
 	\includegraphics[width = 0.495\textwidth, page = 1]{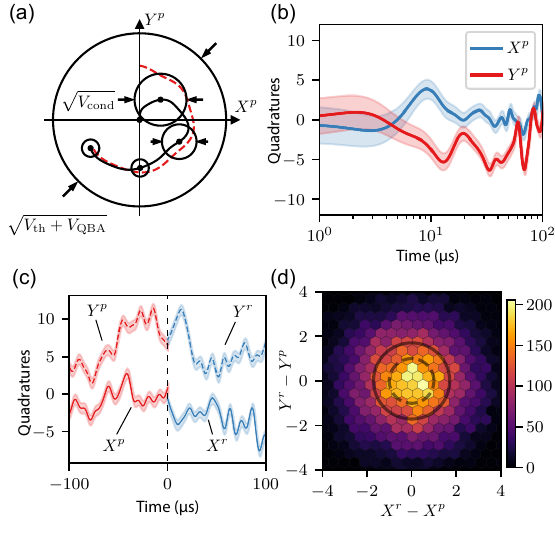}	\caption{\textbf{Conditional state preparation and verification. (a)} Schematic of state preparation by continuous displacement measurement. The mechanical mode is initially in a thermal state. The state is continuously estimated and purified over time given the measurement record. \textbf{(b)} Exemplary quantum trajectory of the mechanical quadratures $\mathbf{r}^p=(X^p, Y^p)$. The shaded width corresponds to one standard deviation $\sigma$ of uncertainty in the estimated quadratures. \textbf{(c)} Using state retrodiction $\mathbf{r}^r=(X^r, Y^r)$ (blue), the differences with the prediction result $\mathbf{r}^p$ (red) at $t=0$ is employed to reconstruct the covariance matrix of the multimode system. \textbf{(d)} Phase space density heat map of the collective mode of interest, with the solid circle marking the statistical standard deviation, and the dashed circle indicating the standard deviation of a pure coherent state as a reference.}
	\label{fig:fig4}
\end{figure}

The observation of optomechanical squeezing demonstrates that we can conduct quantum measurements with high efficiency. In fact, with quantum-limited detection, the maximum squeezing equals the measurement efficiency $\eta_\mathrm{meas}=\Gamma_{\mathrm{meas}}/(\Gamma_\mathrm{th}+\Gamma_\mathrm{qba})$ of the system, with $\Gamma_\mathrm{meas}$ being the measurement rate \cite{clerk2010introduction}, and quantifies how far the measurement is from the Heisenberg uncertainty limit: $\sqrt{\bar{S}_{xx}^\mathrm{imp}\bar{S}_{FF}^\mathrm{tot}} = \hbar/\left(2\sqrt{\eta_\mathrm{meas}}\right)$.
Measurement efficiency is likewise crucial for measurement-based quantum control of mechanical motion \cite{rossi2018measurementbased}: the measurement rate $\Gamma_\mathrm{meas}=x_\mathrm{zpf}^2/\left(4\bar{S}_{xx}^\mathrm{imp}\right)$ represents the rate at which information is gained and the decoherence rate $\Gamma_\mathrm{th}+\Gamma_\mathrm{qba}$ the rate at which information is lost. We prepare conditional mechanical states by measuring the mechanical resonator at a rate close to its decoherence rate, demonstrating that our system is in a parameter regime where quantum control of mechanical motion is possible at room temperature.

We proceed by locking the laser to the cavity at the magic detuning and adjust the single-detector homodyne to measure the mechanical motion at the quadrature angle $\theta \approx \SI{-90}{\degree}$, maximizing the readout efficiency of mechanical motion. We digitize the measurement signal at a $\SI{14}{MHz}$ rate over \SI{2}{\second} for state preparation in post-processing, and measure a long-time average of the spectrum on a real-time spectrum analyzer for system parameter calibration. By fitting the measured noise spectrum with our model, we extract a total detection efficiency of $\eta_\mathrm{d} = \SI{31}{\%}$, and $\mathcal{C}_q = 0.93$ (see Supplementary Information). These parameters correspond to a measurement rate of $\Gamma_{\mathrm{meas}}=\eta_\mathrm{d} \Gamma_{\mathrm{qba}}= 2\pi\times\SI{11}{kHz}$ (equivalent imprecision noise quanta $n_\mathrm{imp}=\Gamma_m/16\Gamma_\mathrm{meas}=\SI{3.6e-8}{}$), approaching the thermal decoherence rate of $\Gamma_{\mathrm{th}}=2\pi\times\SI{34}{kHz}$ and resulting in a measurement efficiency of $\eta_\mathrm{meas}=\SI{16}{\%}$. This efficiency corresponds to an imprecision-force noise product 2.5 times the Heisenberg uncertainty limit. Compared to the maximum squeezing, the degradation of the measurement efficiency mainly comes from the lower homodyne efficiency at quadrature angle $\theta\approx\SI{-90}{\degree}$.

We conduct state preparation on the digitized signal using Kalman filtering \cite{wieczorek2015optimal,magrini2021realtime} based on the quantum master equations of the system. As is shown in Fig.~\ref{fig:fig4}(a), the mechanical mode is initially in a thermal state with phase-space variance determined by both thermal decoherence and QBA decoherence. Based on the continuous measurement result, the state estimation procedure predicts both the most probable values of the mechanical quadratures $\mathbf{r}^p=( X^p,Y^p)$, and the corresponding uncertainties in time in a theoretically-optimal fashion. 
However, as there are still parasitic modes near the mode of interest, single-mode state estimation underestimates the conditional occupancy. We therefore conduct multimode Kalman filtering and include the 9 nearest modes in the estimation procedure (details of calibration provided in the Supplementary Information). Using this method, we are able to isolate the mechanical motion of the defect mode and mitigate spectral contamination between different modes. We can then compute the quadrature trajectory and the uncertainty of the optimally-estimated defect mode [see Fig.~\ref{fig:fig4}(b)]. 

To estimate the quadrature variances $V_{X,Y}$, we follow a retrodiction procedure \cite{zhang2017prediction,rossi2019observing} (details provided in the Supplementary Information), using the displacement records in the ``future'' relative to the time of state conditioning as a separate measurement result $\mathbf{r}^r$, shown in Fig.~\ref{fig:fig4}(c). The variance of the quadrature differences between the prediction and retrodiction results should be exactly $\llangle \|\mathbf{r}^r - \mathbf{r}^p\|^2 \rrangle = 4V_{X,Y} $, where $\llangle \cdots \rrangle$ is the statistical average over the data set. 
We retrieved a thermal occupation $\overline{n}_\mathrm{cond} = V_{X,Y} -1/2 = 1.43$ of the prepared displaced thermal state, with only $3\%$ deviation from the theoretical value. We found that the multimode estimation result shows \SI{63}{\%} increased thermal occupancy compared to the idealized single-mode estimation. The degradation stems from the fact that mechanical modes, although separated in frequency, always have finite spectral overlap and cannot be fully distinguished from each other. This results in strong cross-correlations between mechanical modes, whereby collective modes with enhanced measurement rates can be defined~\cite{meng2022measurementbased}. From the reconstructed multimode covariance matrix $\mathbf{C}$ (definition of the matrix elements $C_{M,N}$ given in the Supplementary Information), we can quantify the magnitude of correlations by evaluating $C_{M_i,N_j}/\sqrt{V_{M_i}V_{N_j}}$, where $M_i$ and $N_j$ are arbitrary quadratures of the corresponding modes $i$ and $j$. The correlation can be as strong as $-0.44$ depending on the frequency separation and measurement rates. We can define a set of uncorrelated collective modes that diagonalize $\mathbf{C}$ through a symplectic transformation (for details see the Supplementary Information). In the collective mode basis, all the modes are uncorrelated, and are only weakly modified from the original basis. We conduct a similar prediction-retrodiction procedure using multimode filtering, and plot the quadrature differences $\mathbf{r}^r - \mathbf{r}^p$ in Fig.~\ref{fig:fig4}(d). We find a modified defect mode thermal occupancy of $\overline{n}_\mathrm{cond, col}=0.97$ in the new collective mode bases, with a corresponding state purity of \SI{34}{\%}.

\section{Discussion and outlook}
Using an ultra-low noise cavity in conjunction with a PnC density modulated membrane, we have been able to operate in the quantum regime of the optomechanical interaction at room temperature. With a single-detector homodyne scheme countering thermal intermodulation noise, we demonstrate optomechanical squeezing and conditional state preparation of a displaced thermal state with single-phonon occupation, which is a prerequisite for real-time optimal quantum control of this macroscopic mechanical resonator. 

With a reasonable improvement of the mechanical quality factor and a wider mechanical bandgap \cite{hoj2022ultracoherent}, together with a real-time digital feedback~\cite{rossi2018measurementbased,magrini2021realtime} using the studied optimal Kalman filters, we expect measurement-based feedback to the ground state and mechanical squeezed state preparation~\cite{szorkovszky2011mechanical} to be feasible under similar experiment conditions. Non-Gaussian states at room temperature can also be prepared using nonlinear measurements, e.g. photon counting~\cite{Galinskiy2020Phonon} which is inherently compatible with our nonlinear noise cancellation scheme. 

In the long term, the ability to observe optomechanical squeezing at room temperature is advantageous for new hybrid quantum systems~\cite{treutlein2014hybrid}, with e.g. atomic ensembles~\cite{moller2017quantum,karg2020lightmediated} and solid-state spins~\cite{fischer2019spin,kosata2020spin}, by obviating the need to operate experiments inside cryostats that have limited optical access and low thermal budget. The experimental advantage of achieving quantum control of macroscopic systems at room temperature may establish new paths for practical applications in real-world scenarios~\cite{simonsen2019sensitive,zhou2021broadband}.

\section*{Acknowledgments}
We thank Sergey A. Fedorov for setting up the initial experiment, Adrien Toros for assistance with the mirror substrate dicing, and Junxin Chen for helpful discussions. This work was supported by funding from the Swiss National Science Foundation under grant agreement No. 185870 (Ambizione) and grant agreement No. 204927 (Cavity Quantum Electro-optomechanics). We further acknowledge funding from the European Research Council (ERC) under the EU H2020 research and innovation programme, grant agreement No. 835329 (ExCOM-cCEO).

\section*{Author contributions}

G.H. conceived and simulated the PnC cavity mirrors, that were fabricated by A.B. A.B performed finite elements simulations of the density-modulated membranes with assistance from G.H. A.B. developed the membrane fabrication process and fabricated the sample used for the experiment. G.H. developed the theoretical framework, conducted the experiments and analyzed the data with assistance from N.J.E. and A.B. The manuscript was written by G.H, A.B. and N.J.E. with input from T.J.K. N.J.E. and T.J.K. supervised the project. All samples were fabricated in the Center of MicroNanoTechnology (CMi) at EPFL.

%%%%%%%%%%%%%%%%%%%%%%%%%%%%%%%%%%%%%%%%%%%%%%%%%%%%%%%%%%%%%%%%%%%%%%%%%%%%
%%%%%%%%%%%%%%%%%%%%%%%%%%%%%%%% Appendices %%%%%%%%%%%%%%%%%%%%%%%%%%%%%%%% 
%%%%%%%%%%%%%%%%%%%%%%%%%%%%%%%%%%%%%%%%%%%%%%%%%%%%%%%%%%%%%%%%%%%%%%%%%%%%

\appendixpage
\appendix

\counterwithin{figure}{section}

\section{System parameters \& calibration}

We characterized several experimental parameters of the optomechanical cavity prior to the measurements shown in the manuscript and use them as pre-determined parameters for the fits of the measured mechanical spectra. We obtained cavity resonance linewidths $\kappa$ at different optical wavelengths, shown in Fig.~\ref{fig:fig2}(b), by scanning our Titanium:Sapphire (TiSa) laser frequency at a rate of \SI{300}{MHz/s} across the optical resonances, with optical power smaller than \SI{1}{\micro\watt} to prevent significant mechanical excitation when the laser frequency crosses to the blue side of the cavity resonance \cite{aspelmeyer2014cavity}. The laser is phase modulated at \SI{211}{MHz} with modulation depth $\sim 0.1$ to generate two small sidebands, which serve as the reference points for calibrating the frequency axis. A Lorentzian function is fitted to the cavity transmission to extract the optical linewidth.

The mechanical frequency of the membrane defect mode, $\Omega_m$, is measured from the cavity transmission spectrum at low input optical power, to limit optical spring effects due to dynamical backaction. The mechanical energy decay rate $\Gamma_m$ is measured with a mechanical ringdown measurement (see Fig. \ref{fig:fig1}(d)). In the ringdown experiment, a pump laser is first tuned to the blue side of an optical cavity mode at \SI{845}{nm} (high finesse) to strongly excite the membrane defect mode (phonon lasing), then the pump laser is switched off and a weak probe laser is switched on near a cavity mode at \SI{780}{nm} (low finesse) to measure the mechanical oscillation amplitude decay through the cavity transmission signal. The same ringdown measurement is repeated at multiple probe laser detuning frequencies, to ensure no dynamical backaction damping or amplification is present during the ringdown measurement. The extracted damping rate is consistent with a previous measurement of the membrane in an optical interferometer, before cavity assembly, indicating that no significant mechanical dissipation is added to the defect mode during the MIM assembly procedure.

To obtain vacuum optomechanical coupling rates, $g_0$, we follow the procedure of \cite{gorodetksy2010determination}. We phase-modulate a weak laser probe with known depth, and compare the mechanical signal to the phase modulation tone transduced on the transmitted light intensity, directly detected at the cavity output. The probe laser is stabilized at the magic detuning to avoid heating from intracavity TIN. We assume the mechanical mode is initially thermalized to the room temperature environment, and undergoes dynamical backaction cooling by the red-detuned probe laser. The probe power is kept relatively low ($\mathcal{C}_q\ll 1$) to avoid quantum backaction heating, such that the thermal occupation of the defect mode can be reliably inferred from the dynamical backaction cooling rate found by fitting the mechanical spectrum. The probe laser is phase modulated at \SI{1.1457}{MHz} with a depth of 0.1275, and at \SI{1.19}{MHz} with a depth of 0.152. The two modulation frequencies are selected such that they are at the edges of the mechanical bandgap, and are symmetric with respect to the frequency of the defect mode. The geometric average of their transduced signals is calculated in order to cancel the OMIT frequency response \cite{weis2010optomechanically} from the defect mode and obtain an estimate of $g_0$. In the actual experiment, however, the field enhanced optomechanical coupling rate $g = g_0\overline{a}$ is the relevant parameter, where $\overline{a}$ is the cavity field amplitude. $g$ is left as a free parameter in all the fits of the measured mechanical spectra. In the experiments shown in the main text, where $\mathcal{C}_q\sim1$, we have $g/2\pi\sim$\SI{500}{kHz}. 

The aforementioned system parameters are summarized in the table below.

\begin{center}
\begin{tabular}{|c | c|} 
 \hline
 System parameter & Value \\ [0.5ex] 
 \hline\hline
 $\Omega_m/2\pi$ & \SI{1.167}{MHz}  \\ 
 \hline
 $g_0/2\pi$ & \SI{159.0}{Hz} \\ 
 \hline
 $\kappa/2\pi$ & \SI{34.2}{MHz} \\
 \hline
 $\Gamma_m/2\pi$ & \SI{6.41}{mHz} \\
 \hline
 $C_0$ & \SI{0.461}{}  \\ 
 \hline
\end{tabular}
\end{center}

We now discuss the various contributions to the detection efficiency $\eta_d$, which is an essential figure of merit for the state preparation experiment. In the experiments, we retrieve the detection efficiency by spectral fitting of the squeezing/mechanical signal, which depends sensitively on it. In the following, we individually identify and describe the calibration of some of the inefficiency contributions in the measurement chain:

\begin{itemize}

\item The detector electronics noise is measured by blocking the optical beam. The electronics noise is flat and broadband, so the inefficiency contribution from it can be retrieved by comparing it to the laser shot noise.
\item The transmission from the optical elements after the cavity output and before the detector is measured with a handheld power meter. This loss term is dominated by a series of polarizing beam splitters, a beam pick-off for generating an error signal for laser locking, several waveplates, and the highly asymmetric beam splitter which combines the local oscillator with the signal beam. 
\item The detector quantum efficiency is calculated from the detector responsivity, specified by the manufacturer. 
\item The homodyne efficiency is calculated from the mode matching efficiency of $90\%$ between the local oscillator and signal beam, given the measured interference visibility of $95\%$. The efficiency is only $75\%$ in contrast to the usual $90\%$ in conventional balanced homodyne detection, due to the fact that we are using a single detector homodyne scheme to cancel the nonlinear mixing noise. The single detector homodyne has a quadrature angle-dependent efficiency: the efficiency is lower when measuring mechanical motion, and higher for optomechanical squeezing detection. The quadrature angle-dependent homodyne efficiencies are shown in Fig.~\ref{fig:SI_hom}(d). 
\item The cavity output efficiency is hard to infer from the cavity reflection scan due to the unknown optical input mode-matching efficiency, and from the empty cavity characterization due to the redistributed intra-cavity optical power in the presence of the weakly reflective membrane. We assume a perfect cavity out-coupling efficiency of $94.8\%$, and consider the added cavity inefficiency from membrane scattering and absorption as part of the overall detection inefficiency in the measurement chain. This residual inefficiency is retrieved from the spectral fitting of the mechanical signal.
\item The defect-mode conversion to mixing sidebands at other frequencies due to the cavity transduction nonlinearity could reduce the linear transduction efficiency as well. It is in principle also included in the fitted residual inefficiency. From the measured $\langle\delta\Delta^2\rangle/\kappa^2<\SI{e-3}{}$, this contribution is estimated to be less than 2\%, and thus excluded from the analysis. 
\end{itemize}

The various inefficiency terms degrading the SNR of the mechanical displacement records are shown in the table below.

\begin{center}
\begin{tabular}{|c | c|} 
 \hline
 Inefficiency source & Value\\ [0.5ex] 
 \hline\hline
 Cavity output (ideal) & \SI{94.8}{\%} \\ 
 \hline
 Detector electronics noise & \SI{86.1}{\%} \\
 \hline
 Passive optics & \SI{66.5}{\%} \\
 \hline
 Detector quantum efficiency & \SI{90}{\%}  \\ 
 \hline
 Homodyne efficiency & \SI{75}{\%}  \\ 
 \hline
 Fitted additional loss & \SI{83.5}{\%} \\ 
 \hline \hline
 \textbf{Total efficiency} & \SI{30.6}{\%}\\
 \hline
\end{tabular}
\end{center}
\section{Signal shot noise calibration}

\begin{figure}[!t]
 	\includegraphics[width = 0.5\textwidth]{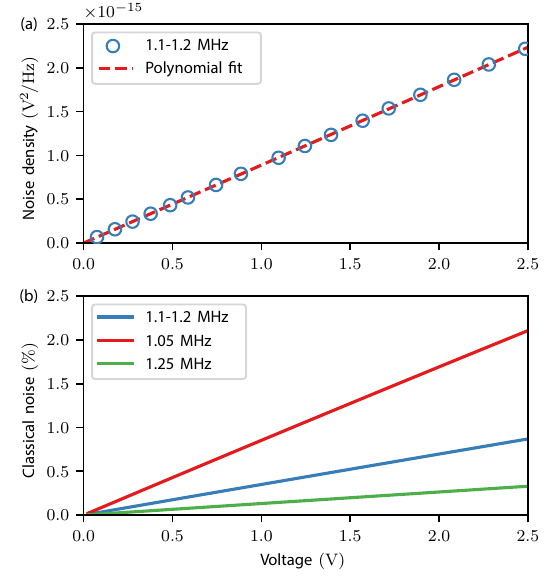}	\caption{\textbf{(a)} Laser intensity noise power (detector noise subtracted) averaged over \SI{1.1}{}-\SI{1.2}{MHz} frequency band, as a function of detector output voltage. A polynomial function consisting of linear and quadratic terms is used to fit and extract the contribution of the laser classical noise. \textbf{(b)} The laser classical noise fractional contribution to the intensity noise is plotted as a function of detector voltage, over three frequency bands. Most of the classical noise is the high frequency tail from the laser relaxation oscillation peak around \SI{300}{kHz}. At the frequency band of interest \SI{1.1}{}-\SI{1.2}{MHz} where the experiment is conducted, the classical noise contribution is less than \SI{1}{\%}. }
	\label{fig:SI_SN}
\end{figure}

To calibrate the reference shot noise power spectral density for a given signal, we proceed as follows:
\begin{enumerate}
    \item Record the average detector voltage output during data acquisition for each optical power level impinging on the detector,
    \item Block the beam from the output of the optical cavity such that only the local oscillator beam hits the detector,
    \item Adjust the local oscillator power such that the average detector voltage output is the same as the one recorded during the data acquisition. This can be done with accuracy better than \SI{1}{\%},
    \item Record the noise power spectral density of the detector voltage output through the same electrical signal chain as the data acquisition, which serves as the reference shot noise power spectral density.
\end{enumerate}
Since the detector voltage output is the same, the generated photocurrent flux and the optical noise power should be the same as well. However, it must be verified that the local oscillator beam is shot-noise-limited at the Fourier frequency of interest. The TiSa laser we use is predominantly shot-noise-limited at around \SI{1}{MHz} under our experimental conditions. We quantify this by recording the photocurrent noise scaling as the optical power is varied. The laser shot noise scales linearly with the optical power, which is proportional to the detector DC voltage output, while any classical noise contribution scales quadratically. We fit the local oscillator noise power integrated over each frequency band as a function of detector DC voltage with a polynomial function consisting only the 1st-order and the 2nd-order terms, as shown in Fig.~\ref{fig:SI_SN}(a). We confirm that within the frequency range of interest and optical powers used in the experiment, the classical noise contribution is less than \SI{1}{\%}, as shown in Fig.~\ref{fig:SI_SN}(b). Using the fit results, we correct the measured reference shot noise by subtracting the calibrated classical noise contribution to ensure that we do not overstate the amount of squeezing. We also measured the detector DC voltage output linearity with respect to optical power with a reference power meter, and found a maximum deviation of less than \SI{0.3}{\%}.

\section{Elimination of TIN in photodetection}
The thermal intermodulation noise from the nonlinear cavity transduction response can be derived from the classical Langevin equation for the evolution of the cavity field:
\begin{equation}
	\dot{a}(t)=[i(\overline{\Delta}+\delta\Delta(t)-\kappa/2]a(t)+\sum_i\sqrt{\kappa_i}a_{\mathrm{in},i}(t)
\end{equation}
where $a_{\mathrm{in},i}$ are the different coupling channels of the cavity (including the intracavity losses). $\overline{\Delta}$ is the mean value of the laser-cavity detuning, and the $\delta\Delta(t)$ term includes all cavity frequency fluctuations, such as those due to membrane motion and mirror vibrations. The frequency domain equivalent is a Fredholm integral equation of the second kind, whose solution can be written as a Liouville-Neumann series of different orders, $a(\omega) = \sum_n a_n(\omega)$. The 0\textsuperscript{th} order includes the DC response and vacuum fluctuations. We are mostly concerned with the 1\textsuperscript{st} and 2\textsuperscript{nd} orders, that provide the linear and quadratic transduction of mechanical motion:
\begin{gather}
a_1(\omega)=\overline{a}\frac{i\Delta(\omega)}{i(-\omega-\overline{\Delta})+\kappa/2},\label{eq:a1}\\
a_2(\omega)=-\overline{a}\int\frac{\Delta(\omega-\omega')}{i(-\omega-\overline{\Delta})+\kappa/2}\frac{\Delta(\omega')}{i(-\omega'-\overline{\Delta})+\kappa/2}\frac{d\omega'}{2\pi},\label{eq:a2}
\end{gather}
where $\overline{a}$ is the mean cavity field, $\overline{\Delta}$ is the mean laser detuning,  and $\Delta(\omega) = \mathrm{FT}[\delta\Delta(t)]$ is dominated by the cavity thermal mechanical noise (FT indicates a Fourier transform). 

The intracavity photon number noise is given by:
\begin{gather}
    n_\mathrm{cav}(\omega) =  \overline{a}^*\left[a_1(\omega) + a_2(\omega)\right]+h.c. \\
    + \int \frac{d\omega'}{2\pi} a_1^\dag(\omega-\omega')a_1(\omega'),
\end{gather}
and exhibits mixing noise terms both from the cavity nonlinear transduction, $a_2(\omega)$, and the intrinsic nonlinearity of the photon number operator, $\mathrm{FT}\left[a_1^\dag(t) a_1(t)\right]$. The $h.c.$ symbol indicates the hermitian conjugate of the previous term. In the bad-cavity limit ($\omega\ll\kappa$), we find that at the magic detuning, $2\overline{\Delta}/\kappa = -1/\sqrt{3}$, the nonlinear photon number noise is cancelled, and the conventional linear transduction $n_\mathrm{cav}(\omega) = \overline{a}^*a_1(\omega)+h.c.$ is kept. Operation at the magic detuning prevents excess oscillator heating due to nonlinear classical radiation pressure noise. 

\begin{figure}[!t]
 	\includegraphics[width = 0.5\textwidth]{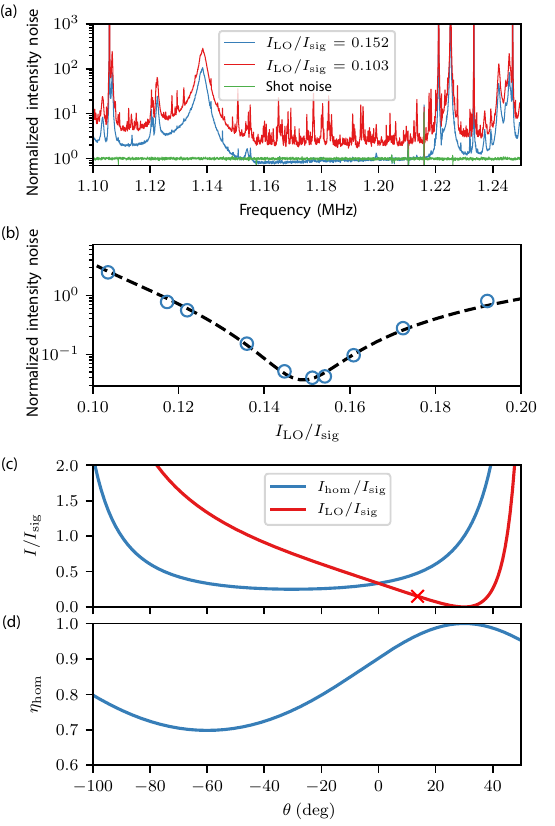}	\caption{\textbf{(a)} When the homodyne power is locked to $I_\mathrm{hom}/I_\mathrm{sig} = 0.481$, different injected local oscillator intensities $I_\mathrm{LO}$ result in different optical quadratures being probed, and drastically different nonlinear noise levels in detection. When the appropriate local oscillator intensity is injected, the nonlinear noise can be efficiently cancelled (blue). \textbf{(b)} The averaged nonlinear noise from \SI{1.16}{MHz} to \SI{1.18}{MHz}, normalized to shot noise, is plotted as a function of local oscillator intensity $I_\mathrm{LO}$ (theory fit in dashed line, including a \SI{4}{\%} noise offset from nearby mechanical modes). \textbf{(c)} Required $I_\mathrm{hom}$ and $I_\mathrm{LO}$ for nonlinear noise cancellation at different quadrature angles $\theta$. The angles are displayed after subtraction of the the cavity-induced angle rotation of \SI{-30}{\degree} at the magic detuning. The red cross marks $I_\mathrm{LO}/I_\mathrm{sig}=0.150$, the noise cancellation condition shown in \textbf{(a,b)}. \textbf{(d)} Homodyne detection efficiency at different quadrature angles, given the experimentally characterized $v=0.95$.  }
	\label{fig:SI_hom}
\end{figure}

In order to reveal the quantum correlations leading to optomechanical squeezing, we need to perform measurements at arbitrary optical quadrature angles. Balanced homodyne detection provides the possibility of tuning the optical quadrature, but does not offer enough degrees of freedom to cancel the nonlinear noise at arbitrary quadrature angles. The photocurrent signal in balanced homodyne detection is given by:
\begin{equation}
    \Delta I = I_+ - I_- \propto \overline{a}_\mathrm{LO}^*\left[a_1(\omega) + a_2(\omega)\right] + h.c.,
\end{equation}
where $\overline{a}_\mathrm{LO}$ is the local oscillator (LO) field amplitude. The field nonlinearity $a_2(\omega)$ is generally non-zero in this expression. However, if the LO is injected from a highly asymmetric beam splitter with a very small reflectivity ($r \ll 1$) and the combined field is detected on a single photodiode, the photocurrent acquires another quadratic term:
\begin{gather}
    I \propto \overline{a}_\mathrm{hom}^*\left[a_1(\omega) + a_2(\omega)\right] + h.c.\\
    - \sqrt{\kappa_\mathrm{ex}}\int \frac{d\omega'}{2\pi} a_1^\dag(\omega-\omega')a_1(\omega'),
\end{gather}
where $\kappa_\mathrm{ex} = \kappa_a$ is the cavity output coupling rate. In this case, simultaneous tuning of LO amplitude and phase enables nonlinear mixing noise cancellation at arbitrary quadrature angles. In the bad-cavity limit, the cancellation condition is:
\begin{equation}
\begin{split}
    \left|\frac{\overline{a}_\mathrm{sig}}{\overline{a}_\mathrm{hom}}\right| &= 2\mathrm{Re}\left[\frac{e^{-i
    \theta}}{(-i\overline{\Delta} + \kappa/2)^2}\right]\left[\overline{\Delta}^2 + \left(\frac{\kappa}{2}\right)^2\right] \\
    &= 2\mathrm{cos}\left[\theta - 2\mathrm{arg}\left(\chi_\mathrm{opt}(0)\right)\right],
    \label{eq:hom_cancellation}
\end{split}
\end{equation}
where $\overline{a}_\mathrm{hom} \approx \overline{a}_\mathrm{sig} + r\overline{a}_\mathrm{LO}$ is the coherent combination of the signal field $\overline{a}_\mathrm{sig} = -\sqrt{\kappa_\mathrm{ex}}\overline{a}$ and the LO field $\overline{a}_\mathrm{LO}$ (defined as the amplitude before the beam splitter), and $\theta = \theta_\mathrm{hom} - \theta_\mathrm{sig}$ is the quadrature rotation angle (cfr. phasor diagram in Fig. \ref{fig:fig3}). $\chi_\mathrm{opt}(0) = \left(\kappa/2 - i\overline{\Delta}\right)^{-1}$ is the cavity DC optical susceptibility. We show the required local oscillator intensity $I_\mathrm{LO} = |r\overline{a}_\mathrm{LO}|^2$ and the combined field intensity $I_\mathrm{hom} = |\overline{a}_\mathrm{hom}|^2$ in order to cancel the nonlinear noise as a function of quadrature angle $\theta$ in Fig.~\ref{fig:SI_hom}(c). 

In the experiment, to detect  a certain quadrature angle while cancelling nonlinear noise, we lock the homodyne power at the corresponding combined field intensity $I_\mathrm{hom}$. Then, we continuously vary the local oscillator power using a tunable neutral density filter until the noise in the mechanical bandgap is perfectly cancelled, as is shown in Fig.~\ref{fig:SI_hom}(a). The level of mixing noise is very sensitive to the local oscillator power, and therefore the cancellation point can serve as a good indicator of the measured quadrature angle $\theta$. Knowing the field amplitude ratios between $\lvert\overline{a}_\mathrm{hom}\rvert,\ \lvert\overline{a}_\mathrm{sig}\rvert$, and that $\overline{\Delta} = -\kappa/(2\sqrt{3})$, we can reconstruct the measured quadrature angles as the ones satisfying the condition in \ref{eq:hom_cancellation}. The nonlinear noise level in a frequency band inside the mechanical bandgap as a function of local oscillator power is shown in Fig.~\ref{fig:SI_hom}(b).

The detection efficiency in this setting is different from the balanced homodyne case, where it is given by the square of the interference visibility $v$ between the signal beam and the LO beam. In our case, the LO and signal beams have comparable optical powers. The added noise comes from the mode-mismatched LO intensity $\Delta I_\mathrm{LO} = |r\overline{a}_\mathrm{LO}|^2(1/v^2-1)$, which results in the reduced homodyne efficiency of $\eta_\mathrm{hom} = I_\mathrm{hom}/(I_\mathrm{hom} + \Delta I_\mathrm{LO})$. Since the local oscillator power is determined by the requirement of nonlinear noise elimination (equation \ref{eq:hom_cancellation}), the homodyne efficiency acquires a quadrature angle dependence. We plot the efficiency at different quadrature angles in Fig.~\ref{fig:SI_hom}(d) given the experimentally characterized visibility $v = 95\%$. 

Note that the nonlinear mixing noise can be perfectly cancelled only in the bad cavity limit. In reality, the level of cancellation is limited by three factors. The first is the laser detuning stability, the second is the homodyne angle stability, and the third is the partial breakdown of the bad cavity limit due to the finite cavity bandwidth ($\kappa/\Omega_m \approx 30$ in our experiment). The finite ratio results in a non-negligible noise floor for various system observables, e.g. for the intracavity photon number:
\begin{gather*}
    S_{n_\mathrm{cav}n_\mathrm{cav}}(\omega) \propto \int S_{\Delta\Delta}(\omega-\omega')S_{\Delta\Delta}(\omega')d\omega'\\
    \times\frac{(4\overline{\Delta}^2+\kappa^2)(-12\overline{\Delta}^2+\kappa^2)^2 + 8(-4\overline{\Delta}^2+3\kappa^2)\kappa^2\omega^2}{(4\overline{\Delta}^2+\kappa^2)^5}
\end{gather*}
At low optical power (output field on the order of \SI{10}{\micro\watt}) accompanied with low passive optical cooling, we find that the combination of these effects limits the cancellation factor at magic detuning to about \SI{300}{} for the intracavity photon number noise measured with direct detection. At high optical power where the experiment is operated, the passive optical cooling is strong enough that this noise floor is hardly visible in detection.

\section{Quantum Langevin equations of the optomechanical system}

Here, we derive the measurement noise spectral density expression that is used for fitting the experimental data. 
In the Heisenberg picture, the quantum Langevin equations are:
\begin{eqnarray*}
	\dot {\hat a} &=& -\left[\frac{\kappa}{2}+i(\Delta+\sqrt{2}g_0\hat Q)\right]\hat a+\sum_{i=a,b,\cdots}\sqrt{\kappa_i}\hat a_\mathrm{in,i},
\end{eqnarray*}
for the cavity mode annihilation operator $\hat a$ and input mode operators $\hat a_\mathrm{in,i}$, where the detuning is defined as $\Delta = \omega_\mathrm{cav}-\omega_\mathrm{laser}$. For the mechanical mode evolution: 
\begin{eqnarray*}
	\dot{\hat{Q}}&=&\Omega_m\hat P\\
	\dot{\hat{P}}&=&-\Omega_m \hat Q+\sqrt{2\Gamma_m}\hat P_\mathrm{in}-\Gamma_m\hat P-\sqrt{2}g_0 \hat a^\dag \hat a,
\end{eqnarray*}
where we defined the dimensionless mechanical position operator $\hat Q = (\hat b^\dag + \hat b)/\sqrt{2}$ and momentum operator $\hat P = i(\hat b^\dag - \hat b)/\sqrt{2}$; $\hat{b}$ is the mechanical mode's annihilation operator, and $\hat P_\mathrm{in}(t)=\frac{i}{\sqrt{2}}\left(\hat b^\dag_\mathrm{in}(t)-\hat b_\mathrm{in}(t)\right)$ is the input momentum fluctuation. We can also define the optical quadrature operators $\hat X=(\hat a^\dag+\hat a)/\sqrt{2}$ and $\hat Y=i(\hat a^\dag-\hat a)/\sqrt{2}$. Note that these quadrature definitions are only used within this section. It is also convenient to define the optomechanical coupling rate $g=g_0\overline{a}$. Combining the evolution equations for the quadratures, we obtain for the mechanical position operator:
\begin{eqnarray*}
	\ddot{\hat Q}+\Gamma_m\dot{\hat Q}+\Omega_m^2\hat Q&=&\sqrt{2\Gamma_m}\Omega_m\hat P_\mathrm{in}-2g\Omega_m\left(\hat X + X_\mathrm{nl}\right)\\
	\hat Q(\omega)&=&\chi_m(\omega)\left(\sqrt{2\Gamma_m}\hat P_\mathrm{in}(\omega)-2g\hat X(\omega)\right)\label{eq:Q},
\end{eqnarray*}
where we introduced the frequency-domain mechanical susceptibility $\chi_m(\omega)\equiv \Omega_m/(\Omega_m^2-\omega^2-i\omega\Gamma_m)$, the nonlinear noise $X_\mathrm{nl}$, which is cancelled at the magic detuning (and set to zero for the following derivations). The dimensionless input momentum fluctuation $\hat P_\mathrm{in}(t)$ is characterized by the following spectral correlations:
\begin{eqnarray*}
	S_{P_\mathrm{in}P_\mathrm{in}}(\omega) &=& \frac{\omega}{\Omega_m}(\overline{n}_\mathrm{th}(\omega) + 1)\\
	S_{P_\mathrm{in}P_\mathrm{in}}(-\omega) &=& \frac{\omega}{\Omega_m}\overline{n}_\mathrm{th}(\omega)
\end{eqnarray*}

Now for simplicity, the mean cavity field $\overline{a}$ is used as the phase reference $\theta=0$ for other fields, such that both $\overline{a}$ and $g$ are real valued. Considering all optical channels $\hat a_\mathrm{in, i}$, we define the following susceptibilities:
\begin{eqnarray*}
    \chi_c(\omega) &=& {2}^{-1/2}(\kappa/2 + i\Delta - i\omega)^{-1}\\
	\chi_c^X(\omega) &=& i(\chi_c^*(-\omega)-\chi_c(\omega))\\
	\chi_c^Y(\omega) &=& -(\chi_c^*(-\omega)+\chi_c(\omega))\\
	\chi_{mc}^X(\omega) &=& (1+2\sqrt{2}g^2\chi_m(\omega)\chi_c^X(\omega))^{-1}\\
	\chi_\Delta^X(\omega) &=& \overline{a}\chi_c^X(\omega)\chi_{mc}^X(\omega)\\
	\chi_{P_\mathrm{in}}^X(\omega) &=& \sqrt{2}g\chi_m(\omega)\chi_c^X(\omega)\chi_{mc}^X(\omega)\\
\chi_{a_\mathrm{in,i}}^X(\omega) &=& \sqrt{\kappa_i}\chi_c(\omega)\chi_{mc}^X(\omega)\\
	\chi_{a_\mathrm{in,i}^\dag}^X(\omega) &=& \sqrt{\kappa_i}\chi_c^*(-\omega)\chi_{mc}^X(\omega)\\
	% \chi_{mc}^Y(\omega) &=& ?\\
	\chi_\Delta^Y(\omega) &=& \chi_c^Y(\omega)[\overline{a}-2\sqrt{2}g^2\chi_m(\omega)\chi_\Delta^X(\omega)]\\
	\chi_{P_\mathrm{in}}^Y(\omega) &=& \sqrt{2}g\chi_c^Y(\omega)\chi_m(\omega)(1-2g\chi_{P_\mathrm{in}}^X(\omega))\\
	\chi_{a_\mathrm{in,i}}^Y(\omega) &=& -i\sqrt{\kappa_i}\chi_c(\omega) - 2\sqrt{2}g^2\chi_c^Y(\omega)\chi_m(\omega)\chi_{a_\mathrm{in,i}}^X(\omega)\\
	\chi_{a_\mathrm{in,i}^\dag}^Y(\omega) &=& i\sqrt{\kappa_i}\chi_c^*(-\omega) - 2\sqrt{2}g^2\chi_c^Y(\omega)\chi_m(\omega)\chi_{a_\mathrm{in,i}^\dag}^X(\omega)
\end{eqnarray*}
We obtain the following frequency domain equations for the optical and mechanical quadratures:
\begin{widetext}
\begin{eqnarray*}
	\hat X(\omega) &=& \chi_\Delta^X(\omega) \Delta(\omega) + \chi_{P_\mathrm{in}}^X(\omega) \sqrt{2\Gamma}\hat P_\mathrm{in}(\omega) + \sum_{i=a,b,...}\left(\chi_{a_\mathrm{in,i}}^X(\omega) \hat a_\mathrm{in,i}(\omega) + \chi_{a_\mathrm{in,i}^\dag}^X(\omega) \hat a_\mathrm{in,i}^\dag(\omega)\right)\\
	\hat Y(\omega) &=& \chi_\Delta^Y(\omega) \Delta(\omega) + \chi_{P_\mathrm{in}}^Y(\omega) \sqrt{2\Gamma}\hat P_\mathrm{in}(\omega) + \sum_{i=a,b,...}\left(\chi_{a_\mathrm{in,i}}^Y(\omega) \hat a_\mathrm{in,i}(\omega) + \chi_{a_\mathrm{in,i}^\dag}^Y(\omega) \hat a_\mathrm{in,i}^\dag(\omega)\right)\\
	\hat Q(\omega) &=& \chi_m(\omega)\left\{(1-2g\chi_{P_\mathrm{in}}^X)\sqrt{2\Gamma}\hat P_\mathrm{in}(\omega)-2g\left[\chi_\Delta^X \Delta(\omega) + \sum_{i=a,b,...}\left(\chi_{a_\mathrm{in,i}}^X(\omega) \hat a_\mathrm{in,i}(\omega) + \chi_{a_\mathrm{in,i}^\dag}^X(\omega) \hat a_\mathrm{in,i}^\dag(\omega)\right)\right]\right\}
\end{eqnarray*}
\end{widetext}
We also define quadratures with arbitrary angles $\hat X^\theta = {2}^{-1/2}(\hat a e^{-i\theta} + \hat a^\dag e^{i\theta}) = \hat X \cos \theta + \hat Y \sin \theta$. We pick $\hat a_\mathrm{in,a}$ to be the cavity output port, with $\hat a_\mathrm{in,b}$, $\hat a_\mathrm{in,c}$ the input and loss channels. From the input-output formalism, the output quadrature is:
\begin{eqnarray*}
	\hat X_\mathrm{out}^\theta = \hat X_\mathrm{in,a}^\theta - \sqrt{\kappa_a} \hat X^\theta.
\end{eqnarray*}
With the following susceptibilities ($k=b,c$, Fourier frequency dependence omitted):
\begin{eqnarray*}
	\chi_\Delta^\theta &=& -\sqrt{\kappa_a}(\chi_\Delta^X\cos\theta + \chi_\Delta^Y\sin\theta)\\
	\chi_{P_\mathrm{in}}^\theta &=& -\sqrt{\kappa_a}(\chi_{P_\mathrm{in}}^X\cos\theta + \chi_{P_\mathrm{in}}^Y\sin\theta)\\
\end{eqnarray*}

 \begin{eqnarray*}
	\chi_{a_\mathrm{in,a}}^\theta &=& \sqrt{2}^{-1}e^{-i\theta}-\sqrt{\kappa_a}(\chi_{a_\mathrm{in,a}}^X\cos\theta + \chi_{a_\mathrm{in,a}}^Y\sin\theta)\\
	\chi_{a_\mathrm{in,a}^\dag}^\theta &=& \sqrt{2}^{-1}e^{i\theta}-\sqrt{\kappa_a}(\chi_{a_\mathrm{in,a}^\dag}^X\cos\theta + \chi_{a_\mathrm{in,a}^\dag}^Y\sin\theta)\\
	\chi_{a_\mathrm{in,k}}^\theta &=& -\sqrt{\kappa_k}(\chi_{a_\mathrm{in,k}}^X\cos\theta + \chi_{a_\mathrm{in,k}}^Y\sin\theta)\\
	\chi_{a_\mathrm{in,k}^\dag}^\theta &=& -\sqrt{\kappa_k}(\chi_{a_\mathrm{in,k}^\dag}^X\cos\theta + \chi_{a_\mathrm{in,k}^\dag}^Y\sin\theta)
\end{eqnarray*}
and taking into account photon collection efficiency $\eta$,
the detected optical quadrature is expressed as:
\begin{widetext}
\begin{eqnarray*}
	\hat X_\mathrm{det}^\theta(\omega)  = \sqrt{\eta}\left[\chi_\Delta^\theta\Delta(\omega) + \chi_{P_\mathrm{in}}^\theta\sqrt{2\Gamma}\hat P_\mathrm{in}(\omega) + \sum_{i=a,b,c}\left(\chi_{a_\mathrm{in,i}}^\theta \hat a_\mathrm{in,i}(\omega) + \chi_{a_\mathrm{in,i}^\dag}^\theta \hat a_\mathrm{in,i}^\dag(\omega)\right)\right] + i\sqrt{1-\eta}\hat X_\mathrm{vac}(\omega).
\end{eqnarray*}
\end{widetext}
Using the following correlation relations,
\begin{eqnarray*}
	S_{AB}(\omega) &=& \int \left<A^\dag(\omega)B(\omega')\right>\frac{d\omega'}{2\pi}
\end{eqnarray*}
\newline
\newline
\begin{eqnarray*}
	\left<a_\mathrm{in}(\omega) a_\mathrm{in}^\dag(\omega')\right> &=& 2\pi \delta(\omega + \omega')\\
	\left<a_\mathrm{in}^\dag(\omega) a_\mathrm{in}(\omega')\right> &=& 0,
\end{eqnarray*}
we can calculate the two important spectral densities: 
\begin{widetext}
\begin{equation}
	S_{\hat X_\mathrm{det}^\theta\hat X_\mathrm{det}^\theta}(\omega)  = \eta\left[|\chi_\Delta^\theta(\omega)|^2S_{\Delta\Delta}(\omega) + |\chi_{P_\mathrm{in}}^\theta(\omega)|^22\Gamma S_{\hat P_\mathrm{in}\hat P_\mathrm{in}}(\omega) + \sum_{i=a,b,c}|\chi_{a_\mathrm{in,i}^\dag}^\theta(\omega)|^2\right]  + \frac{1-\eta}{2},\label{eq:S_xx}
\end{equation}

\begin{equation}
    S_{\hat Q\hat Q}(\omega) = |\chi_m(\omega)|^2\left[\vphantom{\sum_k}| 2g\chi_\Delta^X(\omega)|^2S_{\Delta\Delta} + |1-2g\chi_{P_\mathrm{in}}^X(\omega)|^22\Gamma S_{\hat P_\mathrm{in}\hat P_\mathrm{in}}(\omega) + \sum_{i=a,b,c}|2g\chi_{a_\mathrm{in,i}^\dag}^X(\omega)|^2\right]\label{eq:S_qq}
\end{equation}
\end{widetext}
The single sided spectrum of Eq.~\ref{eq:S_xx} is used in the fitting of the measured homodyne spectral data, then $\int \frac{d\omega}{2\pi} S_{\hat Q\hat Q}(\omega) = \overline{n}_m+\frac{1}{2}$ is used to extract the unconditional mechanical thermal occupancy $\overline{n}_m$.
A specific example is provided in Section~\ref{sec:SI_SC}.

\section{Multimode state estimation model}
Here, starting from the quantum master equation, we derive the equations of quadrature evolution used for implementing the Kalman filter. We work in a parameter regime where the measurement rate is significantly smaller than the frequency of the mechanical mode, such that we can perform IQ demodulation of the mechanical motion at $\Omega_m$ to obtain the slowly-varying $X,Y$ quadratures. Their evolution is described by decoupled quantum master equations~\cite{bowen2015quantum}. In the case where the measurement rate approaches the mechanical frequency, as can be in the case of resonators with a fundamental mode isolated in frequency from the higher order modes, the two equations can be coupled as was in the case in ~\cite{magrini2021realtime,cripe2019measurement}, and measurement-induced mechanical squeezing can happen. In our case, since both quadratures are measured with identical measurement rates, only thermal coherent states are prepared through the measurement process. 

Since we operate in the bad-cavity limit $\Omega_m\ll \kappa$, the cavity dynamics is simplified in our modeling. Before the IQ demodulation, the normalized photocurrent signal is described by:
\begin{equation*}
    i(t) = dW(t) + \sum_i\sqrt{8\Gamma_{\mathrm{meas}}^i}\langle \hat Q_i\rangle(t),
\end{equation*}
where multiple mechanical modes $\hat Q_i$ oscillating at $\Omega_i$ are probed with measurement rates $\Gamma_{\mathrm{meas}}^i$, and the Wiener increment $dW(t)  = \xi(t) dt$ is defined in terms of an ideal unit Gaussian white noise process $\langle\xi(t)\xi(t')\rangle=\delta (t-t')$. Since all oscillation frequencies are close to $\Omega_m$, we decompose the mechanical motion into two slowly varying quadrature observables as $\hat Q_i = \cos(\Omega_m t)\hat X_i + \sin(\Omega_m t) \hat Y_i$, and perform IQ demodulation at frequency $\Omega_m$, which leads to the signal in a vector form:
\begin{equation}
    \mathbf{i}(t) dt = d\mathbf{W}(t) + \sum_i\sqrt{4\Gamma_{\mathrm{meas}}^i}\langle \mathbf{\hat r}_i \rangle(t) dt \label{eq:iw}
\end{equation}
where $\mathbf{i} = \begin{bmatrix}
i_X\\
i_Y
\end{bmatrix}$, $\mathbf{\hat r}_i = \begin{bmatrix}
\hat X_i\\
\hat Y_i
\end{bmatrix}$ and $d\mathbf{W} = \begin{bmatrix}
dW_X\\
dW_Y
\end{bmatrix}$. 

Since the measurement is purely linear, the system stays in Gaussian state-space and the dynamics are completely captured by the expectation values of the quadratures $\langle X_i\rangle$, $\langle Y_i\rangle$, and their covariance matrix $\mathbf{C}$. To derive the time evolutions of these quantities, we start from the quantum master equation for the system density matrix $\hat \rho$ \cite{bowen2015quantum}:

\begin{equation*}
    d\hat \rho = -\frac{i}{\hbar}\left[\hat H_0,\hat \rho\right] + \mathcal{L}_{\mathrm{env}} \hat \rho dt + \sum_{j}\mathcal{D}[\hat c_j]\hat \rho dt + \sqrt{\eta_d}\sum_{j}\mathcal{H}[\hat c_j]\hat \rho dW_j,
\end{equation*}
where the index $j$ is summed over $X$ and $Y$ measurement channels, and $\eta_d$ is the total detection efficiency. The measurement observables are $\mathbf{\hat c}(t) = \sum_i\sqrt{\Gamma_{\mathrm{qba}}^i}\mathbf{\hat r}_i$. The system Hamiltonian $\hat H_0 = \sum_{i}\hbar(\Omega_i-\Omega_m)b_i^\dag b_i$ describes the oscillator unitary dynamics in the $\Omega_m$ rotating frame. The environmental coupling term $\mathcal{L}_{\mathrm{env}} \hat \rho$ describes the bath induced decoherence dynamics, and with the Markov approximation in the rotating frame we have:
\begin{equation*}
    \mathcal{L}_{\mathrm{env}} \hat \rho = \sum_i \Gamma_m^i(\overline{n}_i + 1) \mathcal{D}[\hat b_i]\hat \rho + \Gamma_m^i \overline{n}_i \mathcal{D}[\hat b_i^\dag]\hat \rho.
\end{equation*}
From here, we derive the time evolution of the quadrature expectation values as:
\begin{equation}
    d\langle \mathbf{\hat  r}_i \rangle = \mathbf{A}_i\langle \mathbf{\hat  r}_i\rangle dt + 2\mathbf{B}_id\mathbf{W}(t), \label{eq:qdt}
\end{equation}
where 
\begin{equation*}
\mathbf{A}_i = \begin{bmatrix}
-\Gamma_m^i/2 & \Omega_i - \Omega_m\\
\Omega_m - \Omega_i & -\Gamma_m^i/2
\end{bmatrix}
\end{equation*} and 

\begin{equation*}
\mathbf{B}_i= \begin{bmatrix}
\sum_{j}\sqrt{\Gamma_{\mathrm{meas}}^j}C_{\hat X_i \hat X_j} & \sum_{j}\sqrt{\Gamma_{\mathrm{meas}}^j}C_{\hat X_i \hat Y_j}\\
\sum_{j}\sqrt{\Gamma_{\mathrm{meas}}^j}C_{\hat Y_i \hat X_j} & \sum_{j}\sqrt{\Gamma_{\mathrm{meas}}^j}C_{\hat Y_i \hat Y_j}
\end{bmatrix}. 
\end{equation*}

The covariance matrix elements $C_{\hat M \hat N} = \langle \hat M  \hat N + \hat N \hat M\rangle /2 - \langle \hat M\rangle\langle \hat N \rangle$ evolve as:
\begin{gather}
    \dot{C}_{\hat M_i\hat N_j} = -\frac{\Gamma_m^i + \Gamma_m^j}{2}\dot{C}_{\hat M_i\hat N_j} + \delta_{\hat M_i, \hat N_j}\Gamma_{\mathrm{th}}^i + \delta_{M,N}\sqrt{\Gamma_{\mathrm{qba}}^i\Gamma_{\mathrm{qba}}^j}\nonumber\\
    + (-1)^{\delta_{M,Y}}(\Omega_i-\Omega_m) C_{\hat{ \mathcal{M}}_i\hat N_j} + (-1)^{\delta_{N,Y}}(\Omega_j-\Omega_m) C_{\hat{ M}_i\hat{ \mathcal{N}}_j}\nonumber\\
    -4\left(\sum_k\sqrt{\Gamma_{\mathrm{meas}}^k}C_{\hat M_i \hat X_k}\right)\left(\sum_l\sqrt{\Gamma_{\mathrm{meas}}^l}C_{\hat N_j \hat X_l}\right)\nonumber \\
    -4\left(\sum_k\sqrt{\Gamma_{\mathrm{meas}}^k}C_{\hat M_i \hat Y_k}\right)\left(\sum_l\sqrt{\Gamma_{\mathrm{meas}}^l}C_{\hat N_j \hat Y_l}\right),\label{eq:cov}
\end{gather}
where $\hat{\mathcal{M}}$  and  $\hat{\mathcal{N}}$ are the canonical conjugate observables of $\hat M$ and $\hat N$. 

Eqs.(\ref{eq:iw},\ref{eq:qdt},\ref{eq:cov}) form a closed set of update equations given the measurement record $\mathbf{i}(t)$, and allow quadrature estimations of an arbitrary number of modes and their correlations. The thermal occupancy $\overline{n}_{\mathrm{cond}}^i$ of a specific mechanical mode is determined by the quadrature phase space variances $V_{\hat X_i} = C_{\hat X_i\hat X_i} $, $V_{\hat Y_i} = C_{\hat Y_i\hat Y_i}$, which are both equal to $\overline{n}_{\mathrm{cond},i} + 1/2$ due to the symmetric measurement rates of the $X,Y$ channels.

In the case of the evolution of a single mechanical mode, the quadrature variance approaches a steady state described by an analytical solution, 
\begin{equation}
    V_{X,Y} = \frac{-\Gamma_m'+\sqrt{\Gamma_m'{}^2 + 16\Gamma_{\mathrm{meas}}(\Gamma_{\mathrm{th}}+\Gamma_{\mathrm{qba}})}}{8\Gamma_{\mathrm{meas}}}\label{eq:Vx},
\end{equation} 
where $\Gamma_m' = \Gamma_m + \Gamma_\mathrm{opt}$ takes into account the optical damping rate $\Gamma_\mathrm{opt}$. 
For a system consisting of  multiple mechanical modes, that are not sufficiently separated in frequency ($|\Omega_i - \Omega_j|$ not significantly  faster than any other rates in the system), cross correlations between different mechanical modes emerge due to common measurement imprecision noise and common quantum backaction force. This generally leads to higher quadrature variance due to the effectively reduced measurement efficiency of individual modes.

To decouple the mechanical oscillators that are interacting due to the spectral overlap and the measurement process, we define a new set of collective motional modes through a symplectic (canonical) transformation $\mathbf{U}$ of quadrature basis that diagonalizes the steady state covariance matrix $\mathbf{U}^\dag\mathbf{C}\mathbf{U} = \mathbf{V}$~\cite{andersen2010continuousvariable}. Since the covariance matrix is real and symmetric, the elements of $\mathbf{U}$ are always real, which is required for real observables. The transformation can be understood as a normal mode decomposition of the collective Gaussian state that preserves the commutation relations, as opposed to conventional diagonalization using unitary matrices. This is represented by the requirement of the symplectic transformation $\mathbf{U}\mathbf{\Omega}\mathbf{U}^\dag=\mathbf{\Omega}$ where $\mathbf{\Omega} = \begin{bmatrix}
0 & \mathbf{I}_N \\
-\mathbf{I}_N & 0
\end{bmatrix}$ is the N-mode symplectic form. 

For the multimode estimation experiment, we extract the required system parameters of the nearest 10 mechanical modes around $\Omega_m$ by fitting the measured spectral noise density, and feed the time-series signal $\mathbf{i}(t)$ to the closed update equations to retrieve all the 20 quadrature expectations and 210 independent covariance matrix elements at different times. The signal is normalized such that the Wiener increment gives the correct noise level in frequency domain, and the filtered unconditional defect mode signal has the correct laser-cooled thermal occupancy. The prediction-retrodiction method is used to reconstruct the covariance matrix $\mathbf{C}$, which is then used to compute the symplectic transformation needed to define the new collective mode bases. We derived the retrodiction update equations and found that they are identical to the prediction update equations~\cite{zhang2017prediction}, except with negative mechanical frequencies. As a result, we have the following relations between covariance matrix elements estimated by prediction and retrodiction (respectively identified by the superscripts $p$ and $r$):
\begin{gather*}
    C_{\hat X_i\hat X_j}^p = C_{\hat X_i\hat X_j}^r\\
    C_{\hat Y_i\hat Y_j}^p = C_{\hat Y_i\hat Y_j}^r\\
    C_{\hat X_i\hat Y_j}^p = -C_{\hat X_i\hat Y_j}^r.
\end{gather*}

\section{Numerical implementation of multimode state estimation}
We record the voltage output from the photodetector using an UHFLI lock-in amplifier (Zurich Instruments), for a total duration of \SI{2}{s}, digitizing the signal at \SI{14}{MHz} sampling rate, and we store the data digitally for post-processing. The noise power spectrum density of the digitized signal is compared to the one simultaneously measured on a real-time spectrum analyzer, to rule out SNR degradation from the digitization noise. Two 7th-order Butterworth bandpass filters with passband \SI{1.05}{}-\SI{1.22}{MHz} are applied in order to reduce the influence of the mechanical modes outside the bandgap, and around 50 notch filters are applied to these mechanical modes until no mechanical peak is above the shot noise level. After the filtering, only the 10 mechanical modes around the defect mode frequency $\Omega_m$ need to be kept for the multimode state estimation study.

As discussed in the previous section, we demodulate the signal at $\Omega_m$, and implement the discretized version of the update equation Eq.~\ref{eq:qdt},
\begin{equation}
    \Delta\langle \mathbf{\hat  r}_i \rangle = \mathbf{A}_i'\langle \mathbf{\hat  r}_i\rangle \Delta t + 2\mathbf{B}_i\Delta\mathbf{W}(t)\label{eq:update_disc}
\end{equation}
where $\mathbf{A}_i' = \begin{bmatrix}
-\Gamma_m'^{i}/2 & \Omega_i' - \Omega_m\\
\Omega_m - \Omega_i' & -\Gamma_m'^{i}/2
\end{bmatrix}$ contains modified mechanical parameters:
\begin{gather*}
    \Gamma_m'^{i} = \Gamma_m^{i} + 2\mathrm{Re}\left[-\frac{1-\cos\left((\Omega_i-\Omega_m)\Delta t\right)}{\Delta t}\right]\\
    \Omega_i' = \Omega_i - \mathrm{Im}\left[i(\Omega_i-\Omega_m)-\frac{e^{i(\Omega_i-\Omega_m)\Delta t}-1}{\Delta t}\right]
\end{gather*}
to compensate for the influence of discretization on the state estimation performance compared to an ideal continuous one. 
\begin{figure}[!t]
 	\includegraphics[width = 0.5\textwidth]{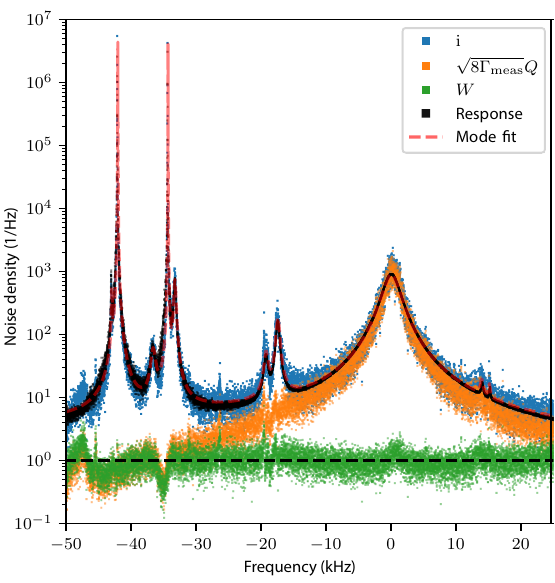}
  \caption{Multimode state estimation spectral response. The noise density of the photocurrent signal $i(t)$ is shown in blue, together with the fit (dashed red), employed to extract the system parameters for the 10 mechanical modes in this frequency band. Those parameters are used to construct the multimode state estimation filter, from which the mechanical mode signals (defect mode shown in orange) and the measurement imprecision (green) are separated. The white noise filter frequency response (gray) is numerically retrieved, and is identical to that of the spectral fit, indicating that the multimode state estimation filter is correctly implemented. }
	\label{fig:SE_resp}
\end{figure}
The matrix $\mathbf{B}_i$ evolution can be computed independently from the sampled time-domain data. Therefore, we calculate the evolution following Eq.~\ref{eq:cov}, with an update rate of \SI{140}{MHz} to mitigate the discretization effect, which is then used for the update equation Eq.~\ref{eq:update_disc} at the sampling rate of \SI{14}{MHz}. The correct implementation of the state estimation can be checked from the Kalman filter response numerically calculated from the estimated $i(t)$, $Q_i(t)$ and $W(t)$, and is shown in Fig.~\ref{fig:SE_resp}.

To experimentally reconstruct the covariance matrix from the estimated quadrature data, for each time trace slice we calculate the difference between prediction and retrodiction results $ \langle\hat{\mathbf{r}}\rangle_{r} - \langle\hat{\mathbf{r}}\rangle_{p}$, and calculate the covariance matrix via
\begin{gather*}
        \mathbf{C} =  \frac{1}{2}\llangle\left(\langle\hat{\mathbf{r}}\rangle_{r}-\langle\hat{\mathbf{r}}\rangle_{p}\right)\cdot\left(\langle\hat{\mathbf{r}}\rangle_{r}-\langle\hat{\mathbf{r}}\rangle_{p}\right)^\intercal\rrangle
\end{gather*}
where $\llangle\cdots\rrangle$ is the statistical average over all the time trace slices, and $\mathbf{\hat r} = \begin{bmatrix} \cdots\hat X_i, \hat Y_i\cdots
\end{bmatrix}^\intercal$. The covariance matrix is then diagonalized to retrieve the symplectic transformation for the collective modes, and the thermal occupations of these modes. We find that in the new quadrature basis based on the diagonalized covariance matrix, the modified defect mode is only weakly modified, and the thermal occupancy can reach that of the single-mode optimal limit $\overline{n}_{\mathrm{cond,single}}=0.88$. The transformation coefficicents for the defect mode $\hat{X}$ is shown in the following table. Note that the contributions from $\hat{Y}_i$ are attributed to statistical uncertainties.
\begin{center}
\begin{tabular}{|c | c| c|} 
 \hline
 $\Omega_i-\Omega_m$ (\SI{}{kHz}) & Coefficients $\hat{X}$ & Coefficients $\hat{Y}$\\ [0.5ex] 
 \hline
 0 & 0.999 & \SI{5.05e-7}{} \\ 
 \hline
 15.1 & \SI{1.53e-5}{} & \SI{-2.11e-6}{} \\ 
 \hline
 13.9 & \SI{-2.54e-7}{} & \SI{-2.33e-6}{} \\ 
 \hline
 -17.4 & \SI{2.16e-4}{} & \SI{5.06e-6}{} \\ 
 \hline
 -19.3 & \SI{1.23e-4}{} & \SI{2.11e-6}{} \\ 
 \hline
 -33.3 & \SI{3.04e-4}{} & \SI{3.14e-6}{} \\ 
 \hline
 -34.4 & \SI{2.96e-2}{} & \SI{5.27e-4}{} \\ 
 \hline
 -36.6 & \SI{1.15e-4}{} & \SI{-8.88e-6}{} \\ 
 \hline
 -42.1 & \SI{4.18e-2}{} & \SI{-2.13e-4}{} \\ 
 \hline
 -43.0 & \SI{1.50e-4}{} & \SI{3.14e-6}{} \\ 
 \hline
\end{tabular}
\end{center}

\section{Sideband cooling limit}\label{sec:SI_SC}
\begin{figure}[!t]
 	 	\includegraphics[width = 0.5\textwidth]{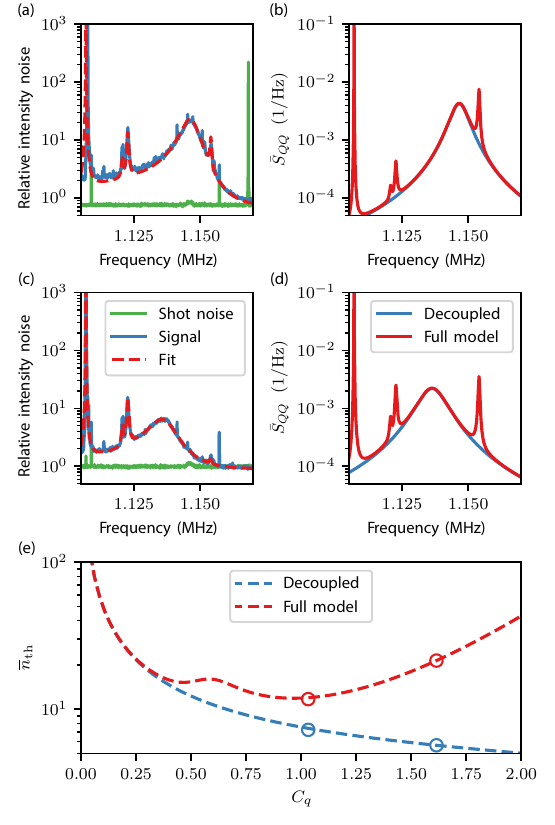}
  \caption{Sideband cooling limit study using the \SI{862.2}{nm} optical mode. \textbf{(a)} Measured mechanical spectrum at $\mathcal{C}_q=1.0$, showing spectral overlap between membrane modes. \textbf{(b)} From the fit, the position noise density of the defect mode is reconstructed, and shows correlations to the nearby auxiliary modes. \textbf{(c,d)} Same measurement and analysis but for $\mathcal{C}_q = 1.6$. \textbf{(e)} Calibrated sideband cooling phonon occupancy at different $\mathcal{C}_q$.}
	\label{fig:SE_SC}
\end{figure}

In our system, due to the reflectivity wavelength dependence of the dielectric coatings of the cavity, we can choose the optical linewidth by tuning the laser to a different optical wavelength. We test the sideband cooling phonon occupancy limit of our system with an optical mode at \SI{862.2}{nm} with optical linewidth $\kappa/2\pi = \SI{13.5}{MHz}$, the lowest among all the optical modes we characterized. The idealized theoretical cooling limit is: 
\[\overline{n}_\mathrm{ideal} = \frac{(\Omega_m+\bar{\Delta})^2 + (\kappa/2)^2}{-4\overline{\Delta}\Omega_m} = 2.9,\]
evaluated with the laser pump applied at the magic detuning. We pump the cavity with \SI{2}{mW} of input power and measure the output spectrum of the cavity (see Fig.~\ref{fig:SE_SC}(a,c)). Using Eq.~\ref{eq:S_xx} to fit the measured spectrum, we can reliably extract all the system parameters ($\mathcal{C}_q$, $\eta$, $\theta$), and also the noise from the nearby spurious mechanical modes $\bar{S}_{\Delta\Delta}(\omega)$. Using the extracted quantities, we can reconstruct the mechanical position noise density $\bar{S}_{QQ}(\omega)$ using Eq.~\ref{eq:S_qq} (see Fig.~\ref{fig:SE_SC}(b,d)). 

To estimate the lowest achievable phonon occupancy, we consider two mechanical mode bases. In the original mode basis, with the full model described by Eq.~\ref{eq:S_xx} and Eq.~\ref{eq:S_qq}, the sideband cooling mediates a coupling between mechanical modes. As a result, the correlation between the defect mode and the spurious modes is non-negligible, and leads to a phonon occupancy of the defect mode greater than the theoretical ideal. Under our experimental conditions, the best phonon occupancy with this optical mode was $\overline{n}_\mathrm{eff} = 11.7$ at relatively low optical powers  ($\mathcal{C}_q = 1.0$, see Fig.~\ref{fig:SE_SC}(e)). 

Next, we analyze the system in the framework of multimode optomechanics~\cite{massel2012multimode} where the cavity-induced coupling between mechanical modes results in mode hybridization and a new set of decoupled normal mode basis. In the decoupled basis, where the cooling-induced correlations between the mechanical modes are nulled, the decoupled defect mode achieves a phonon occupancy $\overline{n}_\mathrm{eff} = 5.7$ at $\mathcal{C}_q = 1.6$. This is a result of the small number of correlation quanta ($\sim10$) compared to the large phonon occupancy ($\gg10$) of the spurious modes in the original basis. Such analysis is similar to those presented in the previous cooling experiments~\cite{guo2019feedback,saarinen2023laser}, where spurious mechanical modes near the target mode couple to the optomechanical cavity. Essentially, by fitting a Lorentzian to the defect mode and ignoring nearby spurious modes, one attains the same phonon occupancy as the decoupled mode. The obtained phonon occupancy is also identical to that which would be retrieved by an independent sideband asymmetry measurement~\cite{magrini2021realtime}. We stress that the phonon occupancy evaluated this way is only associated with the decoupled mode, which is defined in the presence of the cooling pump. In applications that require removal of the cooling pump, the mechanical modes decouple and the effective phonon occupancy goes back to that of the full model. At the experimental condition of $\mathcal{C}_q=1.6$, the full model would yield an increased phonon occupancy of $\overline{n}_\mathrm{eff} = 21.4$ (see Fig.~\ref{fig:SE_SC}(e)), still among the lowest phonon occupancy achieved in nanomechanical devices. Generally, the cooling-induced multimode hybridization results in higher total phonon occupancy compared to the idealized case, but in some special cases the decoupled modes can have lower occupancies.

We further stress that the correlated quanta evaluated from the full model do not impose any limit on measurement-based feedback schemes. In the presence of sideband cooling, the cooling-induced correlations to the spurious modes can be completely cancelled in a measurement-based feedback scheme. When the defect mode experiences both a thermal force $\sqrt{2\Gamma_m}\hat{P}_\mathrm{in}$ and a feedback force $\hat{F}_\mathrm{fb}$, where the feedback force is determined by the measurement record $\hat{F}_\mathrm{fb}(\omega) = H(\omega)\hat X_\mathrm{det}^\theta(\omega)$, a linear filter $H_0(\omega)=-2g/\sqrt{\eta\kappa_a}$ completely cancels the cooling induced correlations. Therefore, for any measurement-based control scheme that requires a filter $H_1(\omega)$, the constructed filter $H(\omega) = H_0(\omega)+H_1(\omega)$ achieves the desired state with no correlation to spurious modes.

\section{Mechanical simulations of density-modulated PnC membranes}

The quality factor of the vibrational mode of a strained resonator subject to dissipation dilution is given by $Q = D_Q\times Q_\mathrm{int}$, where $Q_\mathrm{int}$ is the inverse of the resonator material's loss tangent, and $D_Q$ is the dilution factor \cite{fedorov2019generalized}. The dilution factor depends on the mode deformation pattern, $\vec{u} = \vec{u}(x,y)$,  and on the resonator geometry. It can be estimated analytically or numerically. For high aspect ratio resonators (large lateral extent, small thickness) and for flexural vibrations, the dilution factor can be extremely large (up to $10^5 - 10^6$ \cite{beccari2022strained,bereyhi2022perimeter}) and the quality factor is correspondingly enhanced beyond the material inverse loss tangent. To compute dilution factors, we carry out pre-stressed eigenfrequency analyses in COMSOL Multiphysics, using the ``Shell'' interface. This finite elements study is particularly suitable for the simulation of high aspect ratio nanomechanical objects.

The dilution factor depends strongly on the mode curvature close to the clamping points \cite{yu2012control}. For this reason, we refine the mesh for the finite element model close to the membrane clamped boundaries (outer edges), using a typical element size around $\lambda L/30$. Here, $L$ is the largest lateral dimension of the membrane, $\lambda = \sqrt{E h^2 /(12\sigma L^2)} \ll 1$ is the strain parameter, $E$ is the membrane Young's modulus, $\sigma$ is the material stress, and $h$ is the membrane thickness. Fixed boundary conditions are imposed at the membrane clamping points, i.e. $\vec{u} = \partial \vec{u}/\partial \vec{n} = 0$ (where $\partial \vec{u}/ \partial \vec{n}$ is the displacement field derivative normal to the boundary).

The dilution factors are calculated with the ratio of kinetic and linear elastic energies \cite{fedorov2019generalized}. For out-of-plane bending modes:

\begin{widetext}    
\begin{equation}
    D_Q = \frac{12\rho \Omega^2}{E h^2} \frac{\left(1-\nu^2\right)\int w^2\ dx\,dy}{\int \left(\partial^2 w/\partial x^2 + \partial^2 w/\partial y^2 \right)^2 + 2\left(1 - \nu \right)\left( \left(\partial^2 w / \partial x \partial y\right)^2 - \partial^2 w / \partial x^2 \cdot\partial^2 w / \partial y^2 \right)dxdy},
\label{eqn:dilfactor}
\end{equation}
\end{widetext}

\noindent where $w$ is the ${u}$ component in the out-of-plane direction, $\Omega$ is the mechanical frequency, $\rho$ is the mass density of the membrane material and $\nu$ is Poisson's ratio. A PnC nanomechanical structure can enhance the $D_Q$ of some bending resonances, by localizing the displacement to a defect in the spatial-periodic modulation of the speed of sound and suppressing its curvature close to the clamping boundaries. This idea, introduced by Tsaturyan et al. in 2016 \cite{tsaturyan2017ultracoherent}, was the first incarnation of the ``soft clamping'' effect. 

In our case, the PnC is realized with a modulation of the resonator's effective mass density, described by $\rho_\mathrm{eff}(x,y) = g(x,y)\rho$ \cite{hoj2022ultracoherent}. In practice, the density modulation is implemented by patterning nanopillars over the nanomechanical membrane, with a thickness much larger than that of the membrane. Pillars are confined to circular regions of diameter commensurate with the membrane acoustic wavelength (see Fig. \ref{fig:fig1}(f)), and are arranged in a triangular pattern. The nanopillars locally mass-load the membrane to realize a density modulation, but they also introduce additional mechanical dissipation. When a bending wave impinges on the pillar, it induces deformation carrying linear strain energy without a significant geometrically-nonlinear contribution \cite{fedorov2019generalized}. The pillars will thus reduce the dissipation dilution of the \SiN\ membrane, by an amount that depends sensitively on the pillars' geometrical dimensions and on the elastic frequency of the flexural mode. The lower the elastic frequency and the smaller the pillars, the lower the susceptibility of the pillar displacement, and the lower the added dissipation induced by the individual pillars will be. The pillar damping contribution can be dominant, or sufficiently low that it is negligible compared to the finite dissipation dilution of the soft clamped mode, depending on the oscillation frequency and on the pillars' geometry \cite{hoj2022ultracoherent}.

\begin{figure}
 	\includegraphics[width = 0.5\textwidth]{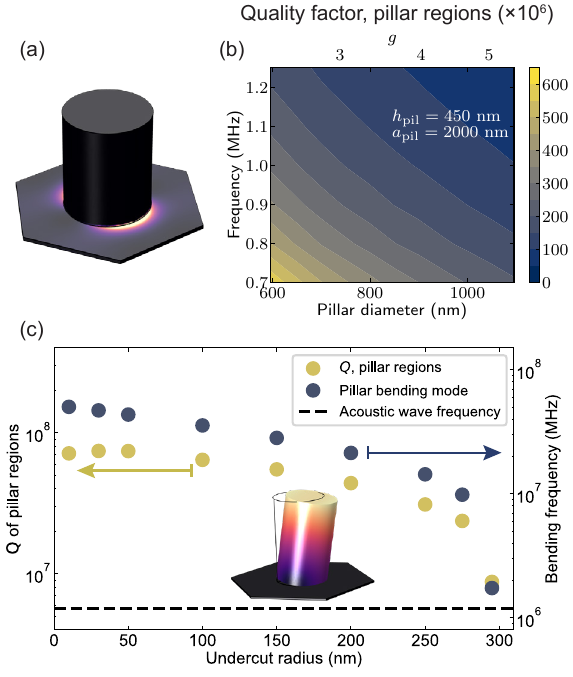}	\caption{Losses induced by bending of nanopillars used to modulate $\rho_\mathrm{eff}$. \textbf{(a)}, Visualization of the linear elastic energy density, which is associated with mechanical dissipation, close to the nanopillar. The largest contribution to dissipation is seen to occur close to the nanopillar base. \textbf{(b)}, $Q$ of a uniform pillar array with infinite extent as a function of the frequency of the acoustic flexural wave and of the pillar diameter. The pillar separation $a_\mathrm{pil}$ and the pillar thickness $h_\mathrm{pil}$ are kept constant to the specified values. The horizontal axis is expressed in terms of the pillar diameter and of the resulting density modulation $g = \rho_\mathrm{eff}/\rho$. \textbf{(c)}, Effect of an undercut at the pillar base on the pillar $Q$. For this simulation, the pillar has a diameter of \SI{600}{nm} and a thickness of \SI{1000}{nm}, and the separation between nearest-neighbours pillars is fixed to \SI{1.5}{\micro\meter}. The undercut layer is hardly visible in the illustration, as it is only \SI{6}{nm}-thick. The pillar array's $Q$ (ochre dots) is seen to drop sharply when the first flexural resonance frequency (blue dots, the inset illustrates the displacement field) of the nanopillar approaches the acoustic excitation frequency, marked by a horizontal dashed line.}
	\label{fig:pillar_bending}
\end{figure}

We investigated this damping contribution with 3D finite element simulations of a single nanopillar unit cell. From now on, we assume that the pillars are arranged in a triangular lattice with a lattice constant $a_\mathrm{pil}$ (separation between the nearest neighbours), and we identify the pillar diameter and height (thickness) with $d_\mathrm{pil},\  h_\mathrm{pil}$. The effective density in the circular regions patterned with the nanopillars is then:

\begin{equation}
    \rho_\mathrm{eff} = \rho\left[1 + \frac{\pi}{2\sqrt{3}}\frac{\rho_\mathrm{pil}h_\mathrm{pil}}{\rho h}\left(\frac{d_\mathrm{pil}}{a_\mathrm{pil}}\right)^2\right],
    \label{eqn:rho_eff_triangular_lattice}
\end{equation}

\noindent where $h$ and $\rho$ are the thickness and density of the membrane film (\SiN). The periodicity of the pillar array is embedded in the simulation by restricting the domain to an hexagon-shaped unit cell (see Fig. \ref{fig:pillar_bending}(a)) with Floquet boundary conditions [$\vec{u}(\vec{x} + \vec{R}) = \vec{u}(\vec{x})e^{-i\vec{k}\cdot\vec{R}}$]
on opposing sides of the hexagon. The magnitude of the elastic wavevector that defines the boundary conditions, $k$, is chosen in order to produce a flexural eigenmode at the mechanical frequency of interest ($ka_\mathrm{pil} \ll 1$):

\begin{equation}
    k \approx \Omega\sqrt{\frac{\rho_\mathrm{eff}}{\sigma}},
\end{equation}

\noindent where $\sigma$ is the membrane deposition stress and $\Omega$ is the mechanical frequency in angular units. The model is then solved for its first eigenmode, which represents the pillar displacement upon the arrival of the flexural wave. The dissipation dilution of an infinite pillar lattice is evaluated using Equation \ref{eqn:dilfactor}, with $w$ corresponding now to the $z$ component of the displacement field of the nanopillar. The results of the computation of the infinitely-extended pillar lattice's quality factor are shown in Fig. \ref{fig:pillar_bending}(b), for a fixed pillar thickness of $h_\mathrm{pil} = \SI{450}{nm}$ and separation $a_\mathrm{pil} = \SI{2}{\micro\meter}$, and variable diameters and mechanical frequencies. The method was also benchmarked by verifying that when the pillar thickness is set to 0, the extracted dilution factor is given by the ``clampless'' expression \cite{fedorov2019generalized}:
\begin{equation}
    D_Q \approx \frac{12(1-\nu^2)\sigma^2}{E \rho_\mathrm{min} h^2 \Omega^2},
    \label{eqn:ultimate_DQ}
\end{equation}
Note that the aSi loss angle is an unknown parameter in the FEM simulations, that we arbitrarily set by choosing $Q_\mathrm{int, aSi} = 10000$. The results in Fig. \ref{fig:pillar_bending}(b) do not depend sensitively on $Q_\mathrm{int, aSi}$; nevertheless one should treat them as a rough estimation of the pillar damping contribution.

Another pillar geometry that is experimentally relevant is the case of a cylindrical pillar with a thin undercut layer formed at the base. As described in the next section, this is a possible outcome of the nanopillar microfabrication, in case the pillar is grown on top of an etch-stop layer with a different chemical identity. In this scenario, the frequency of the first bending mode of the nanopillars is decreased, as shown by the blue dots in Fig. \ref{fig:pillar_bending}(c), and the pillar motion can start to hybridize with the soft clamped membrane mode. This induces a large reduction in the overall quality factor when the undercut covers a significant portion of the pillar base (see the yellow dots in Fig. \ref{fig:pillar_bending}(c)). Note that the undercut is not visible in the figure inset, as it is carved in a layer of only \SI{6}{nm} at the pillar base, which is similar to the fabricated pillar geometrical parameters.

In real density-modulated PnC devices, the nanopillar lattice does not cover the full membrane but is limited to the periodic circular regions of high density (see Fig. \ref{fig:fig1}(e)). To compute the actual damping contribution due to the pillars, it is then necessary to weigh the quality factor calculated as before with a \textit{participation ratio} of the linear elastic energy developed in the high-$\rho_\mathrm{eff}$ regions. The overall dissipation limit is then estimated using the following formula:

\begin{widetext}
\begin{equation}
\begin{split}
    Q_\mathrm{pillar\ bend} &= Q_\mathrm{single\  pillar}/ p_\mathrm{hd} \\
    p_\mathrm{hd} &= \frac{\int_\mathrm{hd} dS\ \left(\left(\frac{\partial^2 w}{\partial x^2} + \frac{\partial^2 w}{\partial y^2}\right)^2 + 2(1-\nu)\left(\left(\frac{\partial^2 w}{\partial x \partial y}\right)^2 - \frac{\partial^2 w}{\partial x^2}\frac{\partial^2 w}{\partial y^2}\right)\right)}{\int dS\ \left(\left(\frac{\partial^2 w}{\partial x^2} + \frac{\partial^2 w}{\partial y^2}\right)^2 + 2(1-\nu)\left(\left(\frac{\partial^2 w}{\partial x \partial y}\right)^2 - \frac{\partial^2 w}{\partial x^2}\frac{\partial^2 w}{\partial y^2}\right)\right)},
\end{split}
\end{equation}
\end{widetext}

\noindent where the integral at the numerator of the participation ratio $p_\mathrm{hd}$ is computed only in the high-density regions, while the one at the denominator is extended to the whole membrane surface. The energy participation ratio depends sensitively on the PnC defect geometry.

For the PnC membrane device employed in these experiments, the macroscopic estimate of the quality factor, neglecting pillar dissipation, is $D_Q\times Q_\mathrm{int} \approx 6\cdot10^8$, slightly lower than the design value due to a low density modulation, $g_\mathrm{max} \approx 2.9$, an accidental result of fabrication drifts. The dissipation contribution due to the nanopillars turned out to be comparable, for the measured pillar dimensions: $Q_\mathrm{pillar\ bend}\approx 6\cdot 10^8$. Overall, the two dissipation contributions combine to give an estimate for the overall intrinsic loss-dominated quality factor $Q \approx 3\cdot 10^8$, slightly higher than what was measured with the membrane device. We also cannot exclude a small contribution to the observed dissipation due to collisions with the residual gas molecules in the vacuum chamber where the MIM cavity is located \cite{bao2002energy}.

\section{Microfabrication of density-modulated PnC membranes}

\begin{figure}[!t]
 	\includegraphics[width = 0.5\textwidth]{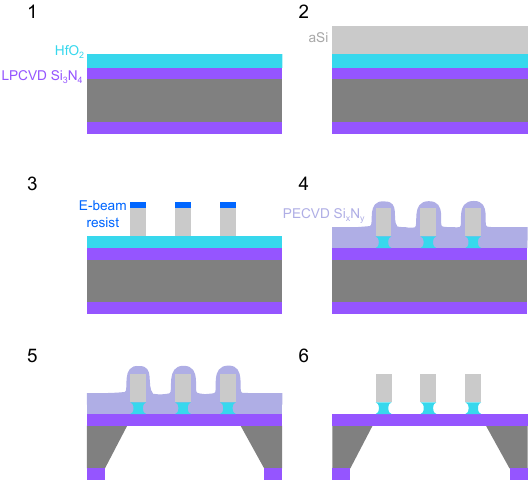}	\caption{Simplified fabrication process for density-modulated membrane resonators. Layer thicknesses and lateral dimensions are not to scale. 1 -- Etch stop layer growth. 2 -- aSi pillar layer growth. 3 -- Electron beam lithography and pillar pattern transfer with RIE. 4 -- Growth of PECVD nitride encapsulation layer. 5 -- Chip separation and \ch{KOH} membrane release. 6 -- Removal of the encapsulation layer in buffered \ch{HF}.}
	\label{fig:PnC_PF}
\end{figure}

We implement density-modulated PnC membranes by fabricating amorphous silicon (aSi) nanopillars on a high aspect ratio \SiN\ membrane. This method has some advantages over the original density-modulated membrane samples \cite{hoj2022ultracoherent}, where the pillars are made of plasma-enhanced chemical vapor deposition (PECVD) silicon nitride. For example, the dimensional accuracy of the fabricated nanopillars is improved. We also found that density-modulated membranes with PECVD nitride pillars caused a strong optical bistability of the MIM cavity \cite{an1997optical}, probably due to excess optical absorption induced in the \SiN\ layer during our particular PECVD process. This experimental obstacle seemingly disappeared, when we adopted new membrane samples with aSi pillars. However, the process becomes a little more laborious, due to the need of protecting the pillars during the membrane undercut. The aSi pillars and the substrate have a very similar chemical composition, and they will be dissolved during the undercut process, if exposed to the etchant (KOH). We devised a PECVD \ch{Si_xN_y} encapsulation layer for the protection of aSi pillars, that can be removed selectively with respect to the pillars and the membrane as the last step of microfabrication. 

In our PnC membranes, we fabricated pillars with diameters between $d_\mathrm{pil} = \SIrange[]{300}{800}{nm}$ and nearest-neighbour distances between $a_\mathrm{pil} = \SIrange[]{1.0}{2.0}{\micro\meter}$. We chose to use electron beam lithography to pattern the pillars, for the highest versatility of prototyping. Dry etching transfers the mask pattern to the underlying film, while maintaining smooth and vertical sidewalls and keeping the pattern dimensions faithful to the original design. However, there are few dry etching recipes with a good selectivity between \SiN\ and \ch{Si}; as soon as the $\sim \SI{20}{nm}$ \SiN\ membrane is uncovered, it would rapidly get consumed by the etching process. One solution is to stop the dry etching step just short of uncovering the \SiN\ layer, and finishing with a high-selectivity step such as wet etching \cite{hoj2022ultracoherent}. Unfortunately, wet etching is typically isotropic, and would shrink the pillar dimensions from the design values, bringing forth issues in the reproducibility and control of the effective density and pillar damping. We decided to instead employ an etch-stop layer, much more resistant to dry etching than \SiN,  on top of the membrane film, that allows for adequate overetching and process tolerances. Such an etch-stop material is not difficult to find: for \ch{SF6}-based processes, oxides such as \ch{SiO2}, \ch{Al2O3} or \ch{HfO2} (hafnium oxide) have been tested and found to provide a suitable selectivity. However, another important requirement is that no significant undercut in the etch-stop layer is created in subsequent steps of microfabrication. Undercut at the pillar base can reduce the frequency of the pillar bending resonances dramatically, thus increasing the pillar dissipation at MHz frequencies. \ch{HfO2} proved to satisfy this requirement, as a very thin layer with thickness $< \SI{5}{nm}$ can completely block the dry etching step for a sufficient amount of time, and it is dissolved in many acid and base solutions.

The process starts with $(100)$-oriented silicon wafers on which a $\sim$ 20-nm layer of stoichiometric, high stress \SiN\ has been grown via low-pressure chemical vapor deposition. After cleaning and dehydrating the wafer with an \ch{O2} plasma, we proceed to grow with atomic layer deposition (ALD) a $\sim$6-nm \ch{HfO2} etch-stop film (step 1 in Fig. \ref{fig:PnC_PF}). The growth takes place with a reactor temperature of \SI{200}{\celsius}, using \ch{H2O} and Tetrakis(ethylmethylamido)hafnium (TEMAHf) as precursors. TEMAHf is pre-heated to \SI{80}{\celsius} before starting the deposition process. We then proceed with the deposition of the aSi pillar layer. We employ a PECVD tool (Oxford PlasmaLabSystem 100), with silane (\ch{SiH4}) as the only precursor (step 2 of Fig. \ref{fig:PnC_PF}). The chamber temperature is set to \SI{300}{\celsius} during the deposition, and a 2\% \ch{SiH4}:\ch{N2} mixture is flowed in the chamber at \ch{1000}{sccm}; plasma is generated using \SI{30}{W} of RF power. The pressure in the chamber is kept around \SI{1500}{mtorr} during the process. The typical pillar thickness we target is about $\SI{600}{nm}$.

We then proceed to define the nanopillar pattern with electron beam lithography. We spin-coat flowable oxide FOx16 resist (a formulation of HSQ) at \SI{2000}{RPM}, resulting in a mask layer approximately 800-nm thick. FOx is exposed with a dose of about \SI{1400}{\micro\coulomb\per\centi\meter^2} and developed with TMAH25\% (2 minutes of immersion with agitation). Before the e-beam writing, the pattern is corrected for proximity effects in electron beam exposure that would lead to nonuniformity within the pillar lattice regions. The pattern is transferred to the aSi layer with reactive ion etching (step 3 in Fig. \ref{fig:PnC_PF}), using a recipe in which \ch{SF6} and \ch{C4F8} gases flow simultaneously in the plasma chamber, where the wafer is kept at \SI{20}{\celsius}. The etch is monitored \textit{in situ} using a 670-nm laser beam reflected from the thin film stack: as the aSi layer is gradually thinned down, fringes are observed, due to thin film interference. The endpoint, occurring when the \ch{HfO2} layer is exposed, is clearly visible from the sudden dip of the interference signal slope. We let the dry etching process run for about \SI{30}{s} after the endpoint to ensure that the pillars are fully defined on the whole wafer, then we stop the process. Finally, we remove the FOx mask and the residual etch-stop layer by dipping the wafer in \ch{HF} 1\% for about \SI{3.5}{min}.

After patterning the pillars, we encapsulate them in a dielectric layer to protect them during the silicon deep etching step (step 4 in Fig. \ref{fig:PnC_PF}). We require a layer that can conformally cover the pillar topography, without defects or pinholes and with tensile deposition strain, such that it does not destructively buckle when it is suspended. Owing to previous experience in other microfabrication projects \cite{beccari2022strained},  we employ PECVD \ch{Si_xN_y}. We first grow a thin ($\sim \SI{20}{nm}$), protective layer of \ch{Al2O3} with ALD, to shield the membrane layer from plasma bombardment during PECVD. Then, approximately \SI{125}{nm} of \ch{Si_xN_y} is grown in our Oxford PlasmalabSystem100 PECVD with 2\% \ch{SiH4}:\ch{N2} and \ch{NH3} as the precursors. The flow rates are set to \SI{975}{} and \SI{30}{sccm}. The chamber pressure is \ch{800}{mtorr} and the reactor temperature is kept to \SI{300}{\degreeCelsius} during the deposition. $\SI{40}{W}$ of RF power excites the plasma during deposition, and the deposited layer has been characterized to have a tensile stress of around $+\SI{300}{MPa}$ at room temperature. This PECVD layer has proved to perfectly seal the nanopillars and withstand several hours of immersion in hot \ch{KOH} without significant consumption.

At this point the process proceeds analogously to that of conventional stress-modulated PnC membranes (see the Supplementary Material of \cite{fedorov2020thermal}). A thick ($\sim \SI{3}{\micro\meter}$) layer of positive tone photoresist is spun on the frontside for protection during the backside lithography process that we perform with an MLA150 laser writer (Heidelberg Instruments), with alignment to 8 frontside markers. Membrane windows must be appropriately resized in order to account for the \ch{KOH} slow-etching $\langle 111 \rangle$ planes. UV lithography is followed by \SiN\ dry etching with a plasma of \ch{CHF_3} and \ch{SF_6}. After the resist mask and protection layer removal with N-Methyl-2-Pyrrolidone (NMP) and \ch{O2} plasma, we deep-etch with \ch{KOH} from the membrane windows while keeping the frontside protected, by installing the wafer in a watertight PEEK holder where only the backside is exposed (see \cite{tsaturyan2017ultracoherent,fedorov2020thermal}). \ch{KOH}40\% at \SI{70}{\celsius} is employed, and the etching is interrupted when about \SI{30}{}-\SI{40}{\micro\meter} of silicon remain. The wafer is then rinsed and cleaned with hot \ch{HCl} of the residues formed during \ch{KOH} etching. Then, the wafer is separated into individual dies before concluding the process: a protective layer of positive-tone resist is coated on the frontside before cutting the wafer with a dicing saw, and the process continues chipwise. Chips are again cleaned with NMP and \ch{O2} plasma, and the deep-etch is concluded with a second immersion in \ch{KOH} 40\% at a lower temperature of \SI{55}{\celsius} (step 5 in Fig. \ref{fig:PnC_PF}), followed by cleaning in \ch{HCl}. From the end of the \ch{KOH} etching step, the composite membranes are suspended, and great care must be adopted in displacing and immersing the samples in liquid; nevertheless, the presence of a relatively thick PECVD nitride layer ensures that the survival yield is quite high after this step ($> 90\%$). We dry the samples by moving them to an ultrapure isopropyl alcohol (IPA) bath after water rinsing. IPA has a low surface tension and a high vapour pressure, therefore it can be easily dried off the membranes after a few minutes of immersion, with the help of a (cautiously operated) \ch{N2} gun.

Finally, the PECVD nitride and \ch{Al2O3} layers can be removed selectively with wet etching in buffered HF (BHF; step 6 in Fig. \ref{fig:PnC_PF}). The etch rates of the encapsulation layers in BHF are orders of magnitude higher than the etch rate of stoichiometric \SiN\ and \ch{HfO2}, therefore even though the membrane backside is exposed, the membrane thinning during this step remains limited (few nanometers). Chips are loaded in a Teflon carrier where they are vertically mounted, and immersed for about \SI{3}{min} \SI{20}{s} in BHF 7:1. It is crucial not to overetch more than necessary to fully remove the encapsulation films: membranes become extremely fragile and the survival yield drops sharply when their thickness is reduced below $\sim \SI{15}{nm}$. The membranes are then carefully rinsed, transferred in an ethanol bath and dried in a critical point dryer, where the liquids can be evacuated gently and with little contamination.

\section{Fabrication and simulation of phononic crystal mirrors}

\begin{figure}[!t]
 	\includegraphics[width = 0.5\textwidth]{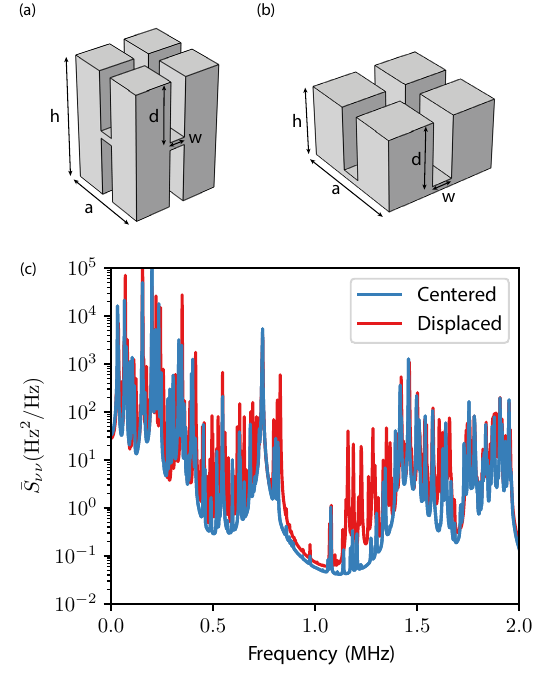}	\caption{\textbf{a} Square lattice unit cell of the top mirror. The periodicity is along the horizontal plane. a=1.4mm, h=4mm, d=1.88mm, w=250µm. \textbf{b} Square lattice unit cell of the bottom mirror. The periodicity is along the horizontal plane. a=1.6mm, h=1mm, d=900µm, w=300µm. \textbf{c} Simulated frequency noise spectra of an empty cavity with PnC top and bottom mirrors. We show both the case of an optical mode perfectly centered on the bottom mirror (blue), and displaced \SI{200}{\micro\meter} away from the center (red), accounting for a realistic assembly misalignment.}
	\label{fig:SI_MIR}
\end{figure}

Using PnC structures to reduce cavity frequency noise from thermomechanical motion of the mirrors was previously explored in Ref.~\cite{saarinen2023laser}. In that work, the bottom mirror is fabricated by first patterning a silicon wafer with a periodic lattice of holes, and then bonding it with a Pyrex substrate on which the mirror coating is sputtered. Because of the technical difficulty in applying the same procedure for the curved top mirror, a fiber mirror is used instead, which also serves as the high transmission port of the cavity. The use of a fiber mirror resulted in excess broadband cavity frequency fluctuations. 
Also, the mode mismatch between the fiber and the cavity mode led to a low cavity output efficiency of approximately \SI{4}{\%}. 

We circumvent these problems by directly patterning PnC structures on both the top and bottom mirrors. The top and bottom mirror substrates are respectively fused silica and borosilicate glass, with a high-reflection coating sputtered on one side and an anti-reflection layer coating the other side. We use a dicing saw for glass machining to dice a regular array of lines into the mirror substrates. The blade is continuously cooled by a pressurized water jet during the dicing process. 
The maximum cut depth allowed for our blade is \SI{2.5}{mm} and we constrain the designed PnC accordingly. We cut the flat bottom mirror only from one side (its thickness is only \SI{1}{mm}), and the top mirror is diced symmetrically with parallel cuts from both sides, since it is 4-mm thick. The relatively deep cuts in the top mirror need to be patterned over multiple passes, with gradually increasing depths. After dicing one mirror side, the piece is flipped and the other side is diced after aligning to the first cuts, visible through the glass substrate. Lines are arranged in a square lattice for simplicity (see Fig.~\ref{fig:fig2}(d,f)), although more complex patterns can be machined with the dicing saw. We simulate the band diagrams of the unit cells of both the top and the bottom mirrors in COMSOL Multiphysics with the Structural Mechanics module. 
The simulation result is shown in Fig.~\ref{fig:fig2} of the main text. We optimized the lattice constant and cut depths in order to maximize the bandgap width, while centering the bandgap around \SI{1}{MHz} and making sure that the remaining glass thickness is sufficient to maintain a reasonable level of structural stiffness.

Due to the finite size of the mirrors, we expect to observe edge modes within the mechanical bandgap frequency range. The thermal vibrations of these modes penetrate into the PnC structure with exponentially decaying amplitudes. To account for their noise contributions, we simulated the frequency noise spectrum of the MIM assembly, consisting of the bottom and the top mirrors in contact with a silicon spacer chip, shown in Fig.~\ref{fig:SI_MIR}. Displacement noise at the location of the cavity optical mode is estimated using the fluctuation-dissipation theorem and the eigenmode parameters obtained from the COMSOL eigenfrequency solver, then converted to frequency noise. From previous measurements of mirror modes, we assume a uniform quality factor of $10^3$ for all the modes. The eigenfrequency solution confirmed the existence of edge modes with frequencies within the mechanical bandgap, but did not predict any significant contribution to the cavity frequency noise: the PnC is sufficiently large to reduce their contribution at the cavity mode position. The bandgap noise is mostly contributed from the off-resonant tail of the thermal mechanical noise from the modes outside the bandgap. We also observe that the noise peaks at the upper edge of the bandgap are sensitive to the relative displacement of the optical mode from the mirror center (see Fig.~\ref{fig:SI_MIR}). This is attributed to the spatial symmetries of these modes.

After patterning the PnC structures on the mirrors, we assembled a cavity with a spacer chip in place of a membrane, and observed that the TE\textsubscript{00} linewidth with the diced mirrors is identical to that of the original cavity. This indicates that our fabrication process does not cause measurable excess roughness or damage to the mirror surfaces. On the other hand, we noticed that when the assembly was clamped too tightly, excess cavity loss occurred due to significant deformation of the PnC mirrors, with a reduced stiffness. We mitigate this detrimental effect in the experiment by gently clamping the MIM cavity, with a spring tension sufficient to guarantee the structural stability of the assembly. We also ensure that the cavity mode is well-centered on the bottom mirror, to reduce the aforementioned upper band-edge modes thermal noise contribution. For the MIM experiment discussed in the main text, we did not observe any mirror modes within the mechanical bandgap of the membrane chip. We can distinguish membrane modes from mirror modes by exploiting the fact that the coupling rates of membrane modes vary between different cavity resonances, while this is not the case for mirror modes.

\section{TiSa laser phase noise characterization}
\begin{figure}[!t]
 	\includegraphics[width = 0.5\textwidth]{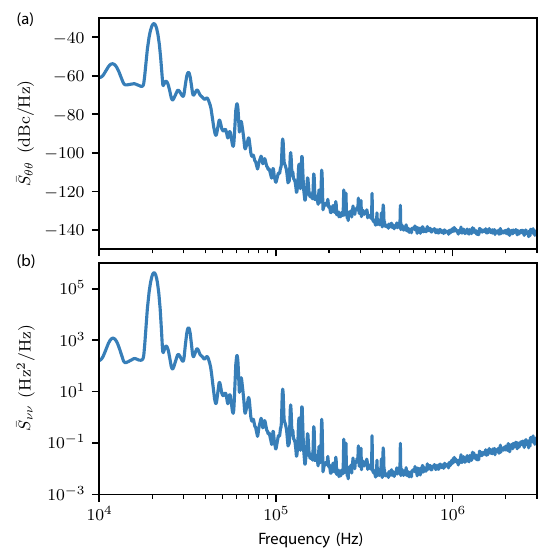}	\caption{Measured noise density of the laser relative \textbf{(a)} phase and \textbf{(b)} frequency fluctuations of the two TiSa lasers.}
	\label{fig:SI_tisa}
\end{figure}
We upper-bound the phase noise of the TiSa laser (Sirah Lasertechnik) used in the experiment by beating it with another TiSa laser (M Squared). The frequency difference of the two lasers is passively stable around \SI{9}{MHz}, and the beating signal is recorded digitally with a sampling rate of \SI{56}{MHz}. The signal is then demodulated at \SI{9}{MHz} to retrieve the I/Q data, from which the phase noise is retrieved. The noise densities of both the laser phase and frequency are computed and shown in Fig.~\ref{fig:SI_tisa}. At \SI{1}{MHz} where the experiment is conducted, the measured noise density is primarily limited by the detector noise and laser shot noise. Therefore, we upper-bound the laser frequency noise at $\bar{S}_{\nu\nu}(\SI{1}{MHz})<\SI{3e-2}{Hz^2/Hz}$.

\bibliographystyle{apsrev4-2}
\bibliography{RT_refs}

%apsrev4-2.bst 2019-01-14 (MD) hand-edited version of apsrev4-1.bst
%Control: key (0)
%Control: author (72) initials jnrlst
%Control: editor formatted (1) identically to author
%Control: production of article title (-1) disabled
%Control: page (0) single
%Control: year (1) truncated
%Control: production of eprint (0) enabled
\providecommand{\noopsort}[1]{}
\begin{thebibliography}{64}%
\makeatletter
\providecommand \@ifxundefined [1]{%
 \@ifx{#1\undefined}
}%
\providecommand \@ifnum [1]{%
 \ifnum #1\expandafter \@firstoftwo
 \else \expandafter \@secondoftwo
 \fi
}%
\providecommand \@ifx [1]{%
 \ifx #1\expandafter \@firstoftwo
 \else \expandafter \@secondoftwo
 \fi
}%
\providecommand \natexlab [1]{#1}%
\providecommand \enquote  [1]{``#1''}%
\providecommand \bibnamefont  [1]{#1}%
\providecommand \bibfnamefont [1]{#1}%
\providecommand \citenamefont [1]{#1}%
\providecommand \href@noop [0]{\@secondoftwo}%
\providecommand \href [0]{\begingroup \@sanitize@url \@href}%
\providecommand \@href[1]{\@@startlink{#1}\@@href}%
\providecommand \@@href[1]{\endgroup#1\@@endlink}%
\providecommand \@sanitize@url [0]{\catcode `\\12\catcode `\$12\catcode
  `\&12\catcode `\#12\catcode `\^12\catcode `\_12\catcode `\%12\relax}%
\providecommand \@@startlink[1]{}%
\providecommand \@@endlink[0]{}%
\providecommand \url  [0]{\begingroup\@sanitize@url \@url }%
\providecommand \@url [1]{\endgroup\@href {#1}{\urlprefix }}%
\providecommand \urlprefix  [0]{URL }%
\providecommand \Eprint [0]{\href }%
\providecommand \doibase [0]{https://doi.org/}%
\providecommand \selectlanguage [0]{\@gobble}%
\providecommand \bibinfo  [0]{\@secondoftwo}%
\providecommand \bibfield  [0]{\@secondoftwo}%
\providecommand \translation [1]{[#1]}%
\providecommand \BibitemOpen [0]{}%
\providecommand \bibitemStop [0]{}%
\providecommand \bibitemNoStop [0]{.\EOS\space}%
\providecommand \EOS [0]{\spacefactor3000\relax}%
\providecommand \BibitemShut  [1]{\csname bibitem#1\endcsname}%
\let\auto@bib@innerbib\@empty
%</preamble>
\bibitem [{\citenamefont {Purdy}\ \emph
  {et~al.}(2013{\natexlab{a}})\citenamefont {Purdy}, \citenamefont {Peterson},\
  and\ \citenamefont {Regal}}]{purdy2013observation}%
  \BibitemOpen
  \bibfield  {author} {\bibinfo {author} {\bibfnamefont {T.~P.}\ \bibnamefont
  {Purdy}}, \bibinfo {author} {\bibfnamefont {R.~W.}\ \bibnamefont
  {Peterson}},\ and\ \bibinfo {author} {\bibfnamefont {C.~A.}\ \bibnamefont
  {Regal}},\ }\href {https://doi.org/10.1126/science.1231282} {\bibfield
  {journal} {\bibinfo  {journal} {Science}\ }\textbf {\bibinfo {volume}
  {339}},\ \bibinfo {pages} {801} (\bibinfo {year}
  {2013}{\natexlab{a}})}\BibitemShut {NoStop}%
\bibitem [{\citenamefont {Magrini}\ \emph {et~al.}(2022)\citenamefont
  {Magrini}, \citenamefont {{Camarena-Ch{\'a}vez}}, \citenamefont {Bach},
  \citenamefont {Johnson},\ and\ \citenamefont
  {Aspelmeyer}}]{magrini2022squeezed}%
  \BibitemOpen
  \bibfield  {author} {\bibinfo {author} {\bibfnamefont {L.}~\bibnamefont
  {Magrini}}, \bibinfo {author} {\bibfnamefont {V.~A.}\ \bibnamefont
  {{Camarena-Ch{\'a}vez}}}, \bibinfo {author} {\bibfnamefont {C.}~\bibnamefont
  {Bach}}, \bibinfo {author} {\bibfnamefont {A.}~\bibnamefont {Johnson}},\ and\
  \bibinfo {author} {\bibfnamefont {M.}~\bibnamefont {Aspelmeyer}},\ }\href
  {https://doi.org/10.1103/PhysRevLett.129.053601} {\bibfield  {journal}
  {\bibinfo  {journal} {Physical Review Letters}\ }\textbf {\bibinfo {volume}
  {129}},\ \bibinfo {pages} {053601} (\bibinfo {year} {2022})}\BibitemShut
  {NoStop}%
\bibitem [{\citenamefont {Aggarwal}\ \emph {et~al.}(2020)\citenamefont
  {Aggarwal}, \citenamefont {Cullen}, \citenamefont {Cripe}, \citenamefont
  {Cole}, \citenamefont {Lanza}, \citenamefont {Libson}, \citenamefont
  {Follman}, \citenamefont {Heu}, \citenamefont {Corbitt},\ and\ \citenamefont
  {Mavalvala}}]{aggarwal2020roomtemperature}%
  \BibitemOpen
  \bibfield  {author} {\bibinfo {author} {\bibfnamefont {N.}~\bibnamefont
  {Aggarwal}}, \bibinfo {author} {\bibfnamefont {T.~J.}\ \bibnamefont
  {Cullen}}, \bibinfo {author} {\bibfnamefont {J.}~\bibnamefont {Cripe}},
  \bibinfo {author} {\bibfnamefont {G.~D.}\ \bibnamefont {Cole}}, \bibinfo
  {author} {\bibfnamefont {R.}~\bibnamefont {Lanza}}, \bibinfo {author}
  {\bibfnamefont {A.}~\bibnamefont {Libson}}, \bibinfo {author} {\bibfnamefont
  {D.}~\bibnamefont {Follman}}, \bibinfo {author} {\bibfnamefont
  {P.}~\bibnamefont {Heu}}, \bibinfo {author} {\bibfnamefont {T.}~\bibnamefont
  {Corbitt}},\ and\ \bibinfo {author} {\bibfnamefont {N.}~\bibnamefont
  {Mavalvala}},\ }\href {https://doi.org/10.1038/s41567-020-0877-x} {\bibfield
  {journal} {\bibinfo  {journal} {Nature Physics}\ }\textbf {\bibinfo {volume}
  {16}},\ \bibinfo {pages} {784} (\bibinfo {year} {2020})}\BibitemShut
  {NoStop}%
\bibitem [{\citenamefont {Tsaturyan}\ \emph {et~al.}(2017)\citenamefont
  {Tsaturyan}, \citenamefont {Barg}, \citenamefont {Polzik},\ and\
  \citenamefont {Schliesser}}]{tsaturyan2017ultracoherent}%
  \BibitemOpen
  \bibfield  {author} {\bibinfo {author} {\bibfnamefont {Y.}~\bibnamefont
  {Tsaturyan}}, \bibinfo {author} {\bibfnamefont {A.}~\bibnamefont {Barg}},
  \bibinfo {author} {\bibfnamefont {E.~S.}\ \bibnamefont {Polzik}},\ and\
  \bibinfo {author} {\bibfnamefont {A.}~\bibnamefont {Schliesser}},\ }\href
  {https://doi.org/10.1038/nnano.2017.101} {\bibfield  {journal} {\bibinfo
  {journal} {Nature Nanotechnology}\ }\textbf {\bibinfo {volume} {12}},\
  \bibinfo {pages} {776} (\bibinfo {year} {2017})}\BibitemShut {NoStop}%
\bibitem [{\citenamefont {Ghadimi}\ \emph {et~al.}(2018)\citenamefont
  {Ghadimi}, \citenamefont {Fedorov}, \citenamefont {Engelsen}, \citenamefont
  {Bereyhi}, \citenamefont {Schilling}, \citenamefont {Wilson},\ and\
  \citenamefont {Kippenberg}}]{ghadimi2018elastic}%
  \BibitemOpen
  \bibfield  {author} {\bibinfo {author} {\bibfnamefont {A.~H.}\ \bibnamefont
  {Ghadimi}}, \bibinfo {author} {\bibfnamefont {S.~A.}\ \bibnamefont
  {Fedorov}}, \bibinfo {author} {\bibfnamefont {N.~J.}\ \bibnamefont
  {Engelsen}}, \bibinfo {author} {\bibfnamefont {M.~J.}\ \bibnamefont
  {Bereyhi}}, \bibinfo {author} {\bibfnamefont {R.}~\bibnamefont {Schilling}},
  \bibinfo {author} {\bibfnamefont {D.~J.}\ \bibnamefont {Wilson}},\ and\
  \bibinfo {author} {\bibfnamefont {T.~J.}\ \bibnamefont {Kippenberg}},\ }\href
  {https://doi.org/10.1126/science.aar6939} {\bibfield  {journal} {\bibinfo
  {journal} {Science}\ }\textbf {\bibinfo {volume} {360}},\ \bibinfo {pages}
  {764} (\bibinfo {year} {2018})}\BibitemShut {NoStop}%
\bibitem [{\citenamefont {H{\o}j}\ \emph {et~al.}(2021)\citenamefont {H{\o}j},
  \citenamefont {Wang}, \citenamefont {Gao}, \citenamefont {Hoff},
  \citenamefont {Sigmund},\ and\ \citenamefont
  {Andersen}}]{hoj2021ultracoherent}%
  \BibitemOpen
  \bibfield  {author} {\bibinfo {author} {\bibfnamefont {D.}~\bibnamefont
  {H{\o}j}}, \bibinfo {author} {\bibfnamefont {F.}~\bibnamefont {Wang}},
  \bibinfo {author} {\bibfnamefont {W.}~\bibnamefont {Gao}}, \bibinfo {author}
  {\bibfnamefont {U.~B.}\ \bibnamefont {Hoff}}, \bibinfo {author}
  {\bibfnamefont {O.}~\bibnamefont {Sigmund}},\ and\ \bibinfo {author}
  {\bibfnamefont {U.~L.}\ \bibnamefont {Andersen}},\ }\href
  {https://doi.org/10.1038/s41467-021-26102-4} {\bibfield  {journal} {\bibinfo
  {journal} {Nature Communications}\ }\textbf {\bibinfo {volume} {12}},\
  \bibinfo {pages} {5766} (\bibinfo {year} {2021})}\BibitemShut {NoStop}%
\bibitem [{\citenamefont {Bereyhi}\ \emph
  {et~al.}(2022{\natexlab{a}})\citenamefont {Bereyhi}, \citenamefont {Beccari},
  \citenamefont {Groth}, \citenamefont {Fedorov}, \citenamefont {Arabmoheghi},
  \citenamefont {Kippenberg},\ and\ \citenamefont
  {Engelsen}}]{bereyhi2022hierarchical}%
  \BibitemOpen
  \bibfield  {author} {\bibinfo {author} {\bibfnamefont {M.~J.}\ \bibnamefont
  {Bereyhi}}, \bibinfo {author} {\bibfnamefont {A.}~\bibnamefont {Beccari}},
  \bibinfo {author} {\bibfnamefont {R.}~\bibnamefont {Groth}}, \bibinfo
  {author} {\bibfnamefont {S.~A.}\ \bibnamefont {Fedorov}}, \bibinfo {author}
  {\bibfnamefont {A.}~\bibnamefont {Arabmoheghi}}, \bibinfo {author}
  {\bibfnamefont {T.~J.}\ \bibnamefont {Kippenberg}},\ and\ \bibinfo {author}
  {\bibfnamefont {N.~J.}\ \bibnamefont {Engelsen}},\ }\href
  {https://doi.org/10.1038/s41467-022-30586-z} {\bibfield  {journal} {\bibinfo
  {journal} {Nature Communications}\ }\textbf {\bibinfo {volume} {13}},\
  \bibinfo {pages} {3097} (\bibinfo {year} {2022}{\natexlab{a}})}\BibitemShut
  {NoStop}%
\bibitem [{\citenamefont {Bereyhi}\ \emph
  {et~al.}(2022{\natexlab{b}})\citenamefont {Bereyhi}, \citenamefont
  {Arabmoheghi}, \citenamefont {Beccari}, \citenamefont {Fedorov},
  \citenamefont {Huang}, \citenamefont {Kippenberg},\ and\ \citenamefont
  {Engelsen}}]{bereyhi2022perimeter}%
  \BibitemOpen
  \bibfield  {author} {\bibinfo {author} {\bibfnamefont {M.~J.}\ \bibnamefont
  {Bereyhi}}, \bibinfo {author} {\bibfnamefont {A.}~\bibnamefont
  {Arabmoheghi}}, \bibinfo {author} {\bibfnamefont {A.}~\bibnamefont
  {Beccari}}, \bibinfo {author} {\bibfnamefont {S.~A.}\ \bibnamefont
  {Fedorov}}, \bibinfo {author} {\bibfnamefont {G.}~\bibnamefont {Huang}},
  \bibinfo {author} {\bibfnamefont {T.~J.}\ \bibnamefont {Kippenberg}},\ and\
  \bibinfo {author} {\bibfnamefont {N.~J.}\ \bibnamefont {Engelsen}},\ }\href
  {https://doi.org/10.1103/PhysRevX.12.021036} {\bibfield  {journal} {\bibinfo
  {journal} {Physical Review X}\ }\textbf {\bibinfo {volume} {12}},\ \bibinfo
  {pages} {021036} (\bibinfo {year} {2022}{\natexlab{b}})}\BibitemShut
  {NoStop}%
\bibitem [{\citenamefont {Shin}\ \emph {et~al.}(2022)\citenamefont {Shin},
  \citenamefont {Cupertino}, \citenamefont {{\noopsort{jong}}{de Jong}},
  \citenamefont {Steeneken}, \citenamefont {Bessa},\ and\ \citenamefont
  {Norte}}]{shin2022spiderweb}%
  \BibitemOpen
  \bibfield  {author} {\bibinfo {author} {\bibfnamefont {D.}~\bibnamefont
  {Shin}}, \bibinfo {author} {\bibfnamefont {A.}~\bibnamefont {Cupertino}},
  \bibinfo {author} {\bibfnamefont {M.~H.~J.}\ \bibnamefont
  {{\noopsort{jong}}{de Jong}}}, \bibinfo {author} {\bibfnamefont {P.~G.}\
  \bibnamefont {Steeneken}}, \bibinfo {author} {\bibfnamefont {M.~A.}\
  \bibnamefont {Bessa}},\ and\ \bibinfo {author} {\bibfnamefont {R.~A.}\
  \bibnamefont {Norte}},\ }\href {https://doi.org/10.1002/adma.202106248}
  {\bibfield  {journal} {\bibinfo  {journal} {Advanced Materials}\ }\textbf
  {\bibinfo {volume} {34}},\ \bibinfo {pages} {2106248} (\bibinfo {year}
  {2022})},\ \Eprint {https://arxiv.org/abs/2108.04809} {arxiv:2108.04809}
  \BibitemShut {NoStop}%
\bibitem [{\citenamefont {Saarinen}\ \emph {et~al.}(2023)\citenamefont
  {Saarinen}, \citenamefont {Kralj}, \citenamefont {Langman}, \citenamefont
  {Tsaturyan},\ and\ \citenamefont {Schliesser}}]{saarinen2023laser}%
  \BibitemOpen
  \bibfield  {author} {\bibinfo {author} {\bibfnamefont {S.~A.}\ \bibnamefont
  {Saarinen}}, \bibinfo {author} {\bibfnamefont {N.}~\bibnamefont {Kralj}},
  \bibinfo {author} {\bibfnamefont {E.~C.}\ \bibnamefont {Langman}}, \bibinfo
  {author} {\bibfnamefont {Y.}~\bibnamefont {Tsaturyan}},\ and\ \bibinfo
  {author} {\bibfnamefont {A.}~\bibnamefont {Schliesser}},\ }\href
  {https://doi.org/10.1364/OPTICA.468590} {\bibfield  {journal} {\bibinfo
  {journal} {Optica}\ }\textbf {\bibinfo {volume} {10}},\ \bibinfo {pages}
  {364} (\bibinfo {year} {2023})}\BibitemShut {NoStop}%
\bibitem [{\citenamefont {Fedorov}\ \emph {et~al.}(2020)\citenamefont
  {Fedorov}, \citenamefont {Beccari}, \citenamefont {Arabmoheghi},
  \citenamefont {Wilson}, \citenamefont {Engelsen},\ and\ \citenamefont
  {Kippenberg}}]{fedorov2020thermal}%
  \BibitemOpen
  \bibfield  {author} {\bibinfo {author} {\bibfnamefont {S.~A.}\ \bibnamefont
  {Fedorov}}, \bibinfo {author} {\bibfnamefont {A.}~\bibnamefont {Beccari}},
  \bibinfo {author} {\bibfnamefont {A.}~\bibnamefont {Arabmoheghi}}, \bibinfo
  {author} {\bibfnamefont {D.~J.}\ \bibnamefont {Wilson}}, \bibinfo {author}
  {\bibfnamefont {N.~J.}\ \bibnamefont {Engelsen}},\ and\ \bibinfo {author}
  {\bibfnamefont {T.~J.}\ \bibnamefont {Kippenberg}},\ }\href
  {https://doi.org/10.1364/OPTICA.402449} {\bibfield  {journal} {\bibinfo
  {journal} {Optica}\ }\textbf {\bibinfo {volume} {7}},\ \bibinfo {pages}
  {1609} (\bibinfo {year} {2020})}\BibitemShut {NoStop}%
\bibitem [{\citenamefont {Pluchar}\ \emph {et~al.}(2023)\citenamefont
  {Pluchar}, \citenamefont {Agrawal},\ and\ \citenamefont
  {Wilson}}]{pluchar2023thermal}%
  \BibitemOpen
  \bibfield  {author} {\bibinfo {author} {\bibfnamefont {C.~M.}\ \bibnamefont
  {Pluchar}}, \bibinfo {author} {\bibfnamefont {A.~R.}\ \bibnamefont
  {Agrawal}},\ and\ \bibinfo {author} {\bibfnamefont {D.~J.}\ \bibnamefont
  {Wilson}},\ }\href {https://doi.org/10.48550/arXiv.2307.03309} {\bibinfo
  {title} {Thermal intermodulation backaction in a high-cooperativity
  optomechanical system}} (\bibinfo {year} {2023}),\ \Eprint
  {https://arxiv.org/abs/2307.03309} {arxiv:2307.03309 [physics,
  physics:quant-ph]} \BibitemShut {NoStop}%
\bibitem [{\citenamefont {H{\o}j}\ \emph {et~al.}(2022)\citenamefont {H{\o}j},
  \citenamefont {Hoff},\ and\ \citenamefont {Andersen}}]{hoj2022ultracoherent}%
  \BibitemOpen
  \bibfield  {author} {\bibinfo {author} {\bibfnamefont {D.}~\bibnamefont
  {H{\o}j}}, \bibinfo {author} {\bibfnamefont {U.~B.}\ \bibnamefont {Hoff}},\
  and\ \bibinfo {author} {\bibfnamefont {U.~L.}\ \bibnamefont {Andersen}},\
  }\href {https://doi.org/10.48550/arXiv.2207.06703} {\bibinfo {title}
  {Ultra-coherent nanomechanical resonators based on density phononic crystal
  engineering}} (\bibinfo {year} {2022}),\ \Eprint
  {https://arxiv.org/abs/2207.06703} {arxiv:2207.06703 [cond-mat]} \BibitemShut
  {NoStop}%
\bibitem [{\citenamefont {Aspelmeyer}\ \emph {et~al.}(2014)\citenamefont
  {Aspelmeyer}, \citenamefont {Kippenberg},\ and\ \citenamefont
  {Marquardt}}]{aspelmeyer2014cavity}%
  \BibitemOpen
  \bibfield  {author} {\bibinfo {author} {\bibfnamefont {M.}~\bibnamefont
  {Aspelmeyer}}, \bibinfo {author} {\bibfnamefont {T.~J.}\ \bibnamefont
  {Kippenberg}},\ and\ \bibinfo {author} {\bibfnamefont {F.}~\bibnamefont
  {Marquardt}},\ }\href {https://doi.org/10.1103/RevModPhys.86.1391} {\bibfield
   {journal} {\bibinfo  {journal} {Reviews of Modern Physics}\ }\textbf
  {\bibinfo {volume} {86}},\ \bibinfo {pages} {1391} (\bibinfo {year}
  {2014})}\BibitemShut {NoStop}%
\bibitem [{\citenamefont {Haroche}\ and\ \citenamefont
  {Raimond}(2006)}]{haroche2006exploring}%
  \BibitemOpen
  \bibfield  {author} {\bibinfo {author} {\bibfnamefont {S.}~\bibnamefont
  {Haroche}}\ and\ \bibinfo {author} {\bibfnamefont {J.-M.}\ \bibnamefont
  {Raimond}},\ }\href@noop {} {\emph {\bibinfo {title} {Exploring the
  {{Quantum}}: {{Atoms}}, {{Cavities}}, and {{Photons}}}}}\ (\bibinfo
  {publisher} {{Oxford University Press}},\ \bibinfo {year} {2006})\BibitemShut
  {NoStop}%
\bibitem [{\citenamefont {Chu}\ \emph {et~al.}(2017)\citenamefont {Chu},
  \citenamefont {Kharel}, \citenamefont {Renninger}, \citenamefont {Burkhart},
  \citenamefont {Frunzio}, \citenamefont {Rakich},\ and\ \citenamefont
  {Schoelkopf}}]{chu2017quantum}%
  \BibitemOpen
  \bibfield  {author} {\bibinfo {author} {\bibfnamefont {Y.}~\bibnamefont
  {Chu}}, \bibinfo {author} {\bibfnamefont {P.}~\bibnamefont {Kharel}},
  \bibinfo {author} {\bibfnamefont {W.~H.}\ \bibnamefont {Renninger}}, \bibinfo
  {author} {\bibfnamefont {L.~D.}\ \bibnamefont {Burkhart}}, \bibinfo {author}
  {\bibfnamefont {L.}~\bibnamefont {Frunzio}}, \bibinfo {author} {\bibfnamefont
  {P.~T.}\ \bibnamefont {Rakich}},\ and\ \bibinfo {author} {\bibfnamefont
  {R.~J.}\ \bibnamefont {Schoelkopf}},\ }\href@noop {} {\bibfield  {journal}
  {\bibinfo  {journal} {Science}\ }\textbf {\bibinfo {volume} {358}},\ \bibinfo
  {pages} {199} (\bibinfo {year} {2017})}\BibitemShut {NoStop}%
\bibitem [{\citenamefont {Satzinger}\ \emph {et~al.}(2018)\citenamefont
  {Satzinger}, \citenamefont {Zhong}, \citenamefont {Chang}, \citenamefont
  {Peairs}, \citenamefont {Bienfait}, \citenamefont {Chou}, \citenamefont
  {Cleland}, \citenamefont {Conner}, \citenamefont {Dumur}, \citenamefont
  {Grebel}, \citenamefont {Gutierrez}, \citenamefont {November}, \citenamefont
  {Povey}, \citenamefont {Whiteley}, \citenamefont {Awschalom}, \citenamefont
  {Schuster},\ and\ \citenamefont {Cleland}}]{satzinger2018quantum}%
  \BibitemOpen
  \bibfield  {author} {\bibinfo {author} {\bibfnamefont {K.~J.}\ \bibnamefont
  {Satzinger}}, \bibinfo {author} {\bibfnamefont {Y.~P.}\ \bibnamefont
  {Zhong}}, \bibinfo {author} {\bibfnamefont {H.-S.}\ \bibnamefont {Chang}},
  \bibinfo {author} {\bibfnamefont {G.~A.}\ \bibnamefont {Peairs}}, \bibinfo
  {author} {\bibfnamefont {A.}~\bibnamefont {Bienfait}}, \bibinfo {author}
  {\bibfnamefont {M.-H.}\ \bibnamefont {Chou}}, \bibinfo {author}
  {\bibfnamefont {A.~Y.}\ \bibnamefont {Cleland}}, \bibinfo {author}
  {\bibfnamefont {C.~R.}\ \bibnamefont {Conner}}, \bibinfo {author}
  {\bibfnamefont {{\'E}.}~\bibnamefont {Dumur}}, \bibinfo {author}
  {\bibfnamefont {J.}~\bibnamefont {Grebel}}, \bibinfo {author} {\bibfnamefont
  {I.}~\bibnamefont {Gutierrez}}, \bibinfo {author} {\bibfnamefont {B.~H.}\
  \bibnamefont {November}}, \bibinfo {author} {\bibfnamefont {R.~G.}\
  \bibnamefont {Povey}}, \bibinfo {author} {\bibfnamefont {S.~J.}\ \bibnamefont
  {Whiteley}}, \bibinfo {author} {\bibfnamefont {D.~D.}\ \bibnamefont
  {Awschalom}}, \bibinfo {author} {\bibfnamefont {D.~I.}\ \bibnamefont
  {Schuster}},\ and\ \bibinfo {author} {\bibfnamefont {A.~N.}\ \bibnamefont
  {Cleland}},\ }\href {https://doi.org/10.1038/s41586-018-0719-5} {\bibfield
  {journal} {\bibinfo  {journal} {Nature}\ }\textbf {\bibinfo {volume} {563}},\
  \bibinfo {pages} {661} (\bibinfo {year} {2018})}\BibitemShut {NoStop}%
\bibitem [{\citenamefont {Chan}\ \emph {et~al.}(2011)\citenamefont {Chan},
  \citenamefont {Alegre}, \citenamefont {{Safavi-Naeini}}, \citenamefont
  {Hill}, \citenamefont {Krause}, \citenamefont {Gr{\"o}blacher}, \citenamefont
  {Aspelmeyer},\ and\ \citenamefont {Painter}}]{chan2011laser}%
  \BibitemOpen
  \bibfield  {author} {\bibinfo {author} {\bibfnamefont {J.}~\bibnamefont
  {Chan}}, \bibinfo {author} {\bibfnamefont {T.~P.~M.}\ \bibnamefont {Alegre}},
  \bibinfo {author} {\bibfnamefont {A.~H.}\ \bibnamefont {{Safavi-Naeini}}},
  \bibinfo {author} {\bibfnamefont {J.~T.}\ \bibnamefont {Hill}}, \bibinfo
  {author} {\bibfnamefont {A.}~\bibnamefont {Krause}}, \bibinfo {author}
  {\bibfnamefont {S.}~\bibnamefont {Gr{\"o}blacher}}, \bibinfo {author}
  {\bibfnamefont {M.}~\bibnamefont {Aspelmeyer}},\ and\ \bibinfo {author}
  {\bibfnamefont {O.}~\bibnamefont {Painter}},\ }\href
  {https://doi.org/10.1038/nature10461} {\bibfield  {journal} {\bibinfo
  {journal} {Nature}\ }\textbf {\bibinfo {volume} {478}},\ \bibinfo {pages}
  {89} (\bibinfo {year} {2011})}\BibitemShut {NoStop}%
\bibitem [{\citenamefont {Rossi}\ \emph {et~al.}(2018)\citenamefont {Rossi},
  \citenamefont {Mason}, \citenamefont {Chen}, \citenamefont {Tsaturyan},\ and\
  \citenamefont {Schliesser}}]{rossi2018measurementbased}%
  \BibitemOpen
  \bibfield  {author} {\bibinfo {author} {\bibfnamefont {M.}~\bibnamefont
  {Rossi}}, \bibinfo {author} {\bibfnamefont {D.}~\bibnamefont {Mason}},
  \bibinfo {author} {\bibfnamefont {J.}~\bibnamefont {Chen}}, \bibinfo {author}
  {\bibfnamefont {Y.}~\bibnamefont {Tsaturyan}},\ and\ \bibinfo {author}
  {\bibfnamefont {A.}~\bibnamefont {Schliesser}},\ }\href
  {https://doi.org/10.1038/s41586-018-0643-8} {\bibfield  {journal} {\bibinfo
  {journal} {Nature}\ }\textbf {\bibinfo {volume} {563}},\ \bibinfo {pages}
  {53} (\bibinfo {year} {2018})}\BibitemShut {NoStop}%
\bibitem [{\citenamefont {{Safavi-Naeini}}\ \emph {et~al.}(2013)\citenamefont
  {{Safavi-Naeini}}, \citenamefont {Gr{\"o}blacher}, \citenamefont {Hill},
  \citenamefont {Chan}, \citenamefont {Aspelmeyer},\ and\ \citenamefont
  {Painter}}]{safavi-naeini2013squeezed}%
  \BibitemOpen
  \bibfield  {author} {\bibinfo {author} {\bibfnamefont {A.~H.}\ \bibnamefont
  {{Safavi-Naeini}}}, \bibinfo {author} {\bibfnamefont {S.}~\bibnamefont
  {Gr{\"o}blacher}}, \bibinfo {author} {\bibfnamefont {J.~T.}\ \bibnamefont
  {Hill}}, \bibinfo {author} {\bibfnamefont {J.}~\bibnamefont {Chan}}, \bibinfo
  {author} {\bibfnamefont {M.}~\bibnamefont {Aspelmeyer}},\ and\ \bibinfo
  {author} {\bibfnamefont {O.}~\bibnamefont {Painter}},\ }\href
  {https://doi.org/10.1038/nature12307} {\bibfield  {journal} {\bibinfo
  {journal} {Nature}\ }\textbf {\bibinfo {volume} {500}},\ \bibinfo {pages}
  {185} (\bibinfo {year} {2013})}\BibitemShut {NoStop}%
\bibitem [{\citenamefont {Nielsen}\ \emph {et~al.}(2017)\citenamefont
  {Nielsen}, \citenamefont {Tsaturyan}, \citenamefont {M{\o}ller},
  \citenamefont {Polzik},\ and\ \citenamefont
  {Schliesser}}]{nielsen2017multimode}%
  \BibitemOpen
  \bibfield  {author} {\bibinfo {author} {\bibfnamefont {W.~H.~P.}\
  \bibnamefont {Nielsen}}, \bibinfo {author} {\bibfnamefont {Y.}~\bibnamefont
  {Tsaturyan}}, \bibinfo {author} {\bibfnamefont {C.~B.}\ \bibnamefont
  {M{\o}ller}}, \bibinfo {author} {\bibfnamefont {E.~S.}\ \bibnamefont
  {Polzik}},\ and\ \bibinfo {author} {\bibfnamefont {A.}~\bibnamefont
  {Schliesser}},\ }\href {https://doi.org/10.1073/pnas.1608412114} {\bibfield
  {journal} {\bibinfo  {journal} {Proceedings of the National Academy of
  Sciences}\ }\textbf {\bibinfo {volume} {114}},\ \bibinfo {pages} {62}
  (\bibinfo {year} {2017})}\BibitemShut {NoStop}%
\bibitem [{\citenamefont {Purdy}\ \emph
  {et~al.}(2013{\natexlab{b}})\citenamefont {Purdy}, \citenamefont {Yu},
  \citenamefont {Peterson}, \citenamefont {Kampel},\ and\ \citenamefont
  {Regal}}]{purdy2013strong}%
  \BibitemOpen
  \bibfield  {author} {\bibinfo {author} {\bibfnamefont {T.~P.}\ \bibnamefont
  {Purdy}}, \bibinfo {author} {\bibfnamefont {P.-L.}\ \bibnamefont {Yu}},
  \bibinfo {author} {\bibfnamefont {R.~W.}\ \bibnamefont {Peterson}}, \bibinfo
  {author} {\bibfnamefont {N.~S.}\ \bibnamefont {Kampel}},\ and\ \bibinfo
  {author} {\bibfnamefont {C.~A.}\ \bibnamefont {Regal}},\ }\href
  {https://doi.org/10.1103/PhysRevX.3.031012} {\bibfield  {journal} {\bibinfo
  {journal} {Physical Review X}\ }\textbf {\bibinfo {volume} {3}},\ \bibinfo
  {pages} {031012} (\bibinfo {year} {2013}{\natexlab{b}})}\BibitemShut
  {NoStop}%
\bibitem [{\citenamefont {Mason}\ \emph {et~al.}(2019)\citenamefont {Mason},
  \citenamefont {Chen}, \citenamefont {Rossi}, \citenamefont {Tsaturyan},\ and\
  \citenamefont {Schliesser}}]{mason2019continuous}%
  \BibitemOpen
  \bibfield  {author} {\bibinfo {author} {\bibfnamefont {D.}~\bibnamefont
  {Mason}}, \bibinfo {author} {\bibfnamefont {J.}~\bibnamefont {Chen}},
  \bibinfo {author} {\bibfnamefont {M.}~\bibnamefont {Rossi}}, \bibinfo
  {author} {\bibfnamefont {Y.}~\bibnamefont {Tsaturyan}},\ and\ \bibinfo
  {author} {\bibfnamefont {A.}~\bibnamefont {Schliesser}},\ }\href
  {https://doi.org/10.1038/s41567-019-0533-5} {\bibfield  {journal} {\bibinfo
  {journal} {Nature Physics}\ }\textbf {\bibinfo {volume} {15}},\ \bibinfo
  {pages} {745} (\bibinfo {year} {2019})}\BibitemShut {NoStop}%
\bibitem [{\citenamefont {Riedinger}\ \emph {et~al.}(2018)\citenamefont
  {Riedinger}, \citenamefont {Wallucks}, \citenamefont {Marinkovi{\'c}},
  \citenamefont {L{\"o}schnauer}, \citenamefont {Aspelmeyer}, \citenamefont
  {Hong},\ and\ \citenamefont {Gr{\"o}blacher}}]{riedinger2018remote}%
  \BibitemOpen
  \bibfield  {author} {\bibinfo {author} {\bibfnamefont {R.}~\bibnamefont
  {Riedinger}}, \bibinfo {author} {\bibfnamefont {A.}~\bibnamefont {Wallucks}},
  \bibinfo {author} {\bibfnamefont {I.}~\bibnamefont {Marinkovi{\'c}}},
  \bibinfo {author} {\bibfnamefont {C.}~\bibnamefont {L{\"o}schnauer}},
  \bibinfo {author} {\bibfnamefont {M.}~\bibnamefont {Aspelmeyer}}, \bibinfo
  {author} {\bibfnamefont {S.}~\bibnamefont {Hong}},\ and\ \bibinfo {author}
  {\bibfnamefont {S.}~\bibnamefont {Gr{\"o}blacher}},\ }\href
  {https://doi.org/10.1038/s41586-018-0036-z} {\bibfield  {journal} {\bibinfo
  {journal} {Nature}\ }\textbf {\bibinfo {volume} {556}},\ \bibinfo {pages}
  {473} (\bibinfo {year} {2018})}\BibitemShut {NoStop}%
\bibitem [{\citenamefont {{Mercier de L{\'e}pinay}}\ \emph
  {et~al.}(2021)\citenamefont {{Mercier de L{\'e}pinay}}, \citenamefont
  {{Ockeloen-Korppi}}, \citenamefont {Woolley},\ and\ \citenamefont
  {Sillanp{\"a}{\"a}}}]{mercierdelepinay2021quantum}%
  \BibitemOpen
  \bibfield  {author} {\bibinfo {author} {\bibfnamefont {L.}~\bibnamefont
  {{Mercier de L{\'e}pinay}}}, \bibinfo {author} {\bibfnamefont {C.~F.}\
  \bibnamefont {{Ockeloen-Korppi}}}, \bibinfo {author} {\bibfnamefont {M.~J.}\
  \bibnamefont {Woolley}},\ and\ \bibinfo {author} {\bibfnamefont {M.~A.}\
  \bibnamefont {Sillanp{\"a}{\"a}}},\ }\href
  {https://doi.org/10.1126/science.abf5389} {\bibfield  {journal} {\bibinfo
  {journal} {Science}\ }\textbf {\bibinfo {volume} {372}},\ \bibinfo {pages}
  {625} (\bibinfo {year} {2021})}\BibitemShut {NoStop}%
\bibitem [{\citenamefont {Kotler}\ \emph {et~al.}(2021)\citenamefont {Kotler},
  \citenamefont {Peterson}, \citenamefont {Shojaee}, \citenamefont {Lecocq},
  \citenamefont {Cicak}, \citenamefont {Kwiatkowski}, \citenamefont {Geller},
  \citenamefont {Glancy}, \citenamefont {Knill}, \citenamefont {Simmonds},
  \citenamefont {Aumentado},\ and\ \citenamefont {Teufel}}]{kotler2021direct}%
  \BibitemOpen
  \bibfield  {author} {\bibinfo {author} {\bibfnamefont {S.}~\bibnamefont
  {Kotler}}, \bibinfo {author} {\bibfnamefont {G.~A.}\ \bibnamefont
  {Peterson}}, \bibinfo {author} {\bibfnamefont {E.}~\bibnamefont {Shojaee}},
  \bibinfo {author} {\bibfnamefont {F.}~\bibnamefont {Lecocq}}, \bibinfo
  {author} {\bibfnamefont {K.}~\bibnamefont {Cicak}}, \bibinfo {author}
  {\bibfnamefont {A.}~\bibnamefont {Kwiatkowski}}, \bibinfo {author}
  {\bibfnamefont {S.}~\bibnamefont {Geller}}, \bibinfo {author} {\bibfnamefont
  {S.}~\bibnamefont {Glancy}}, \bibinfo {author} {\bibfnamefont
  {E.}~\bibnamefont {Knill}}, \bibinfo {author} {\bibfnamefont {R.~W.}\
  \bibnamefont {Simmonds}}, \bibinfo {author} {\bibfnamefont {J.}~\bibnamefont
  {Aumentado}},\ and\ \bibinfo {author} {\bibfnamefont {J.~D.}\ \bibnamefont
  {Teufel}},\ }\href {https://doi.org/10.1126/science.abf2998} {\bibfield
  {journal} {\bibinfo  {journal} {Science}\ }\textbf {\bibinfo {volume}
  {372}},\ \bibinfo {pages} {622} (\bibinfo {year} {2021})}\BibitemShut
  {NoStop}%
\bibitem [{\citenamefont {Delaney}\ \emph {et~al.}(2022)\citenamefont
  {Delaney}, \citenamefont {Urmey}, \citenamefont {Mittal}, \citenamefont
  {Brubaker}, \citenamefont {Kindem}, \citenamefont {Burns}, \citenamefont
  {Regal},\ and\ \citenamefont {Lehnert}}]{delaney2022superconductingqubit}%
  \BibitemOpen
  \bibfield  {author} {\bibinfo {author} {\bibfnamefont {R.~D.}\ \bibnamefont
  {Delaney}}, \bibinfo {author} {\bibfnamefont {M.~D.}\ \bibnamefont {Urmey}},
  \bibinfo {author} {\bibfnamefont {S.}~\bibnamefont {Mittal}}, \bibinfo
  {author} {\bibfnamefont {B.~M.}\ \bibnamefont {Brubaker}}, \bibinfo {author}
  {\bibfnamefont {J.~M.}\ \bibnamefont {Kindem}}, \bibinfo {author}
  {\bibfnamefont {P.~S.}\ \bibnamefont {Burns}}, \bibinfo {author}
  {\bibfnamefont {C.~A.}\ \bibnamefont {Regal}},\ and\ \bibinfo {author}
  {\bibfnamefont {K.~W.}\ \bibnamefont {Lehnert}},\ }\href
  {https://doi.org/10.1038/s41586-022-04720-2} {\bibfield  {journal} {\bibinfo
  {journal} {Nature}\ }\textbf {\bibinfo {volume} {606}},\ \bibinfo {pages}
  {489} (\bibinfo {year} {2022})}\BibitemShut {NoStop}%
\bibitem [{\citenamefont {Schrinski}\ \emph {et~al.}(2023)\citenamefont
  {Schrinski}, \citenamefont {Yang}, \citenamefont {{\noopsort{l{\"u}pke}}{von
  L{\"u}pke}}, \citenamefont {Bild}, \citenamefont {Chu}, \citenamefont
  {Hornberger}, \citenamefont {Nimmrichter},\ and\ \citenamefont
  {Fadel}}]{schrinski2023macroscopic}%
  \BibitemOpen
  \bibfield  {author} {\bibinfo {author} {\bibfnamefont {B.}~\bibnamefont
  {Schrinski}}, \bibinfo {author} {\bibfnamefont {Y.}~\bibnamefont {Yang}},
  \bibinfo {author} {\bibfnamefont {U.}~\bibnamefont
  {{\noopsort{l{\"u}pke}}{von L{\"u}pke}}}, \bibinfo {author} {\bibfnamefont
  {M.}~\bibnamefont {Bild}}, \bibinfo {author} {\bibfnamefont {Y.}~\bibnamefont
  {Chu}}, \bibinfo {author} {\bibfnamefont {K.}~\bibnamefont {Hornberger}},
  \bibinfo {author} {\bibfnamefont {S.}~\bibnamefont {Nimmrichter}},\ and\
  \bibinfo {author} {\bibfnamefont {M.}~\bibnamefont {Fadel}},\ }\href
  {https://doi.org/10.1103/PhysRevLett.130.133604} {\bibfield  {journal}
  {\bibinfo  {journal} {Physical Review Letters}\ }\textbf {\bibinfo {volume}
  {130}},\ \bibinfo {pages} {133604} (\bibinfo {year} {2023})}\BibitemShut
  {NoStop}%
\bibitem [{\citenamefont {Alferov}(2001)}]{alferov2001nobel}%
  \BibitemOpen
  \bibfield  {author} {\bibinfo {author} {\bibfnamefont {Z.~I.}\ \bibnamefont
  {Alferov}},\ }\href {https://doi.org/10.1103/RevModPhys.73.767} {\bibfield
  {journal} {\bibinfo  {journal} {Reviews of Modern Physics}\ }\textbf
  {\bibinfo {volume} {73}},\ \bibinfo {pages} {767} (\bibinfo {year}
  {2001})}\BibitemShut {NoStop}%
\bibitem [{\citenamefont {Bloch}\ \emph {et~al.}(2008)\citenamefont {Bloch},
  \citenamefont {Dalibard},\ and\ \citenamefont {Zwerger}}]{bloch2008manybody}%
  \BibitemOpen
  \bibfield  {author} {\bibinfo {author} {\bibfnamefont {I.}~\bibnamefont
  {Bloch}}, \bibinfo {author} {\bibfnamefont {J.}~\bibnamefont {Dalibard}},\
  and\ \bibinfo {author} {\bibfnamefont {W.}~\bibnamefont {Zwerger}},\ }\href
  {https://doi.org/10.1103/RevModPhys.80.885} {\bibfield  {journal} {\bibinfo
  {journal} {Reviews of Modern Physics}\ }\textbf {\bibinfo {volume} {80}},\
  \bibinfo {pages} {885} (\bibinfo {year} {2008})}\BibitemShut {NoStop}%
\bibitem [{\citenamefont {Bongs}\ \emph {et~al.}(2019)\citenamefont {Bongs},
  \citenamefont {Holynski}, \citenamefont {Vovrosh}, \citenamefont {Bouyer},
  \citenamefont {Condon}, \citenamefont {Rasel}, \citenamefont {Schubert},
  \citenamefont {Schleich},\ and\ \citenamefont {Roura}}]{bongs2019taking}%
  \BibitemOpen
  \bibfield  {author} {\bibinfo {author} {\bibfnamefont {K.}~\bibnamefont
  {Bongs}}, \bibinfo {author} {\bibfnamefont {M.}~\bibnamefont {Holynski}},
  \bibinfo {author} {\bibfnamefont {J.}~\bibnamefont {Vovrosh}}, \bibinfo
  {author} {\bibfnamefont {P.}~\bibnamefont {Bouyer}}, \bibinfo {author}
  {\bibfnamefont {G.}~\bibnamefont {Condon}}, \bibinfo {author} {\bibfnamefont
  {E.}~\bibnamefont {Rasel}}, \bibinfo {author} {\bibfnamefont
  {C.}~\bibnamefont {Schubert}}, \bibinfo {author} {\bibfnamefont {W.~P.}\
  \bibnamefont {Schleich}},\ and\ \bibinfo {author} {\bibfnamefont
  {A.}~\bibnamefont {Roura}},\ }\href
  {https://doi.org/10.1038/s42254-019-0117-4} {\bibfield  {journal} {\bibinfo
  {journal} {Nature Reviews Physics}\ }\textbf {\bibinfo {volume} {1}},\
  \bibinfo {pages} {731} (\bibinfo {year} {2019})}\BibitemShut {NoStop}%
\bibitem [{\citenamefont {M{\o}ller}\ \emph {et~al.}(2017)\citenamefont
  {M{\o}ller}, \citenamefont {Thomas}, \citenamefont {Vasilakis}, \citenamefont
  {Zeuthen}, \citenamefont {Tsaturyan}, \citenamefont {Balabas}, \citenamefont
  {Jensen}, \citenamefont {Schliesser}, \citenamefont {Hammerer},\ and\
  \citenamefont {Polzik}}]{moller2017quantum}%
  \BibitemOpen
  \bibfield  {author} {\bibinfo {author} {\bibfnamefont {C.~B.}\ \bibnamefont
  {M{\o}ller}}, \bibinfo {author} {\bibfnamefont {R.~A.}\ \bibnamefont
  {Thomas}}, \bibinfo {author} {\bibfnamefont {G.}~\bibnamefont {Vasilakis}},
  \bibinfo {author} {\bibfnamefont {E.}~\bibnamefont {Zeuthen}}, \bibinfo
  {author} {\bibfnamefont {Y.}~\bibnamefont {Tsaturyan}}, \bibinfo {author}
  {\bibfnamefont {M.}~\bibnamefont {Balabas}}, \bibinfo {author} {\bibfnamefont
  {K.}~\bibnamefont {Jensen}}, \bibinfo {author} {\bibfnamefont
  {A.}~\bibnamefont {Schliesser}}, \bibinfo {author} {\bibfnamefont
  {K.}~\bibnamefont {Hammerer}},\ and\ \bibinfo {author} {\bibfnamefont
  {E.~S.}\ \bibnamefont {Polzik}},\ }\href
  {https://doi.org/10.1038/nature22980} {\bibfield  {journal} {\bibinfo
  {journal} {Nature}\ }\textbf {\bibinfo {volume} {547}},\ \bibinfo {pages}
  {191} (\bibinfo {year} {2017})}\BibitemShut {NoStop}%
\bibitem [{\citenamefont {H{\"a}lg}\ \emph {et~al.}(2021)\citenamefont
  {H{\"a}lg}, \citenamefont {Gisler}, \citenamefont {Tsaturyan}, \citenamefont
  {Catalini}, \citenamefont {Grob}, \citenamefont {Krass}, \citenamefont
  {H{\'e}ritier}, \citenamefont {Mattiat}, \citenamefont {Thamm}, \citenamefont
  {Schirhagl}, \citenamefont {Langman}, \citenamefont {Schliesser},
  \citenamefont {Degen},\ and\ \citenamefont
  {Eichler}}]{halg2021membranebased}%
  \BibitemOpen
  \bibfield  {author} {\bibinfo {author} {\bibfnamefont {D.}~\bibnamefont
  {H{\"a}lg}}, \bibinfo {author} {\bibfnamefont {T.}~\bibnamefont {Gisler}},
  \bibinfo {author} {\bibfnamefont {Y.}~\bibnamefont {Tsaturyan}}, \bibinfo
  {author} {\bibfnamefont {L.}~\bibnamefont {Catalini}}, \bibinfo {author}
  {\bibfnamefont {U.}~\bibnamefont {Grob}}, \bibinfo {author} {\bibfnamefont
  {M.-D.}\ \bibnamefont {Krass}}, \bibinfo {author} {\bibfnamefont
  {M.}~\bibnamefont {H{\'e}ritier}}, \bibinfo {author} {\bibfnamefont
  {H.}~\bibnamefont {Mattiat}}, \bibinfo {author} {\bibfnamefont {A.-K.}\
  \bibnamefont {Thamm}}, \bibinfo {author} {\bibfnamefont {R.}~\bibnamefont
  {Schirhagl}}, \bibinfo {author} {\bibfnamefont {E.~C.}\ \bibnamefont
  {Langman}}, \bibinfo {author} {\bibfnamefont {A.}~\bibnamefont {Schliesser}},
  \bibinfo {author} {\bibfnamefont {C.~L.}\ \bibnamefont {Degen}},\ and\
  \bibinfo {author} {\bibfnamefont {A.}~\bibnamefont {Eichler}},\ }\href
  {https://doi.org/10.1103/PhysRevApplied.15.L021001} {\bibfield  {journal}
  {\bibinfo  {journal} {Physical Review Applied}\ }\textbf {\bibinfo {volume}
  {15}},\ \bibinfo {pages} {L021001} (\bibinfo {year} {2021})}\BibitemShut
  {NoStop}%
\bibitem [{\citenamefont {Chang}\ \emph {et~al.}(2010)\citenamefont {Chang},
  \citenamefont {Regal}, \citenamefont {Papp}, \citenamefont {Wilson},
  \citenamefont {Ye}, \citenamefont {Painter}, \citenamefont {Kimble},\ and\
  \citenamefont {Zoller}}]{chang2010cavity}%
  \BibitemOpen
  \bibfield  {author} {\bibinfo {author} {\bibfnamefont {D.~E.}\ \bibnamefont
  {Chang}}, \bibinfo {author} {\bibfnamefont {C.~A.}\ \bibnamefont {Regal}},
  \bibinfo {author} {\bibfnamefont {S.~B.}\ \bibnamefont {Papp}}, \bibinfo
  {author} {\bibfnamefont {D.~J.}\ \bibnamefont {Wilson}}, \bibinfo {author}
  {\bibfnamefont {J.}~\bibnamefont {Ye}}, \bibinfo {author} {\bibfnamefont
  {O.}~\bibnamefont {Painter}}, \bibinfo {author} {\bibfnamefont {H.~J.}\
  \bibnamefont {Kimble}},\ and\ \bibinfo {author} {\bibfnamefont
  {P.}~\bibnamefont {Zoller}},\ }\href
  {https://doi.org/10.1073/pnas.0912969107} {\bibfield  {journal} {\bibinfo
  {journal} {Proceedings of the National Academy of Sciences}\ }\textbf
  {\bibinfo {volume} {107}},\ \bibinfo {pages} {1005} (\bibinfo {year}
  {2010})}\BibitemShut {NoStop}%
\bibitem [{\citenamefont {Corbitt}\ \emph {et~al.}(2006)\citenamefont
  {Corbitt}, \citenamefont {Chen}, \citenamefont {Khalili}, \citenamefont
  {Ottaway}, \citenamefont {Vyatchanin}, \citenamefont {Whitcomb},\ and\
  \citenamefont {Mavalvala}}]{corbitt2006squeezedstate}%
  \BibitemOpen
  \bibfield  {author} {\bibinfo {author} {\bibfnamefont {T.}~\bibnamefont
  {Corbitt}}, \bibinfo {author} {\bibfnamefont {Y.}~\bibnamefont {Chen}},
  \bibinfo {author} {\bibfnamefont {F.}~\bibnamefont {Khalili}}, \bibinfo
  {author} {\bibfnamefont {D.}~\bibnamefont {Ottaway}}, \bibinfo {author}
  {\bibfnamefont {S.}~\bibnamefont {Vyatchanin}}, \bibinfo {author}
  {\bibfnamefont {S.}~\bibnamefont {Whitcomb}},\ and\ \bibinfo {author}
  {\bibfnamefont {N.}~\bibnamefont {Mavalvala}},\ }\href
  {https://doi.org/10.1103/PhysRevA.73.023801} {\bibfield  {journal} {\bibinfo
  {journal} {Physical Review A}\ }\textbf {\bibinfo {volume} {73}},\ \bibinfo
  {pages} {023801} (\bibinfo {year} {2006})}\BibitemShut {NoStop}%
\bibitem [{\citenamefont {Cripe}\ \emph {et~al.}(2019)\citenamefont {Cripe},
  \citenamefont {Aggarwal}, \citenamefont {Lanza}, \citenamefont {Libson},
  \citenamefont {Singh}, \citenamefont {Heu}, \citenamefont {Follman},
  \citenamefont {Cole}, \citenamefont {Mavalvala},\ and\ \citenamefont
  {Corbitt}}]{cripe2019measurement}%
  \BibitemOpen
  \bibfield  {author} {\bibinfo {author} {\bibfnamefont {J.}~\bibnamefont
  {Cripe}}, \bibinfo {author} {\bibfnamefont {N.}~\bibnamefont {Aggarwal}},
  \bibinfo {author} {\bibfnamefont {R.}~\bibnamefont {Lanza}}, \bibinfo
  {author} {\bibfnamefont {A.}~\bibnamefont {Libson}}, \bibinfo {author}
  {\bibfnamefont {R.}~\bibnamefont {Singh}}, \bibinfo {author} {\bibfnamefont
  {P.}~\bibnamefont {Heu}}, \bibinfo {author} {\bibfnamefont {D.}~\bibnamefont
  {Follman}}, \bibinfo {author} {\bibfnamefont {G.~D.}\ \bibnamefont {Cole}},
  \bibinfo {author} {\bibfnamefont {N.}~\bibnamefont {Mavalvala}},\ and\
  \bibinfo {author} {\bibfnamefont {T.}~\bibnamefont {Corbitt}},\ }\href
  {https://doi.org/10.1038/s41586-019-1051-4} {\bibfield  {journal} {\bibinfo
  {journal} {Nature}\ }\textbf {\bibinfo {volume} {568}},\ \bibinfo {pages}
  {364} (\bibinfo {year} {2019})}\BibitemShut {NoStop}%
\bibitem [{\citenamefont {Magrini}\ \emph {et~al.}(2021)\citenamefont
  {Magrini}, \citenamefont {Rosenzweig}, \citenamefont {Bach}, \citenamefont
  {{Deutschmann-Olek}}, \citenamefont {Hofer}, \citenamefont {Hong},
  \citenamefont {Kiesel}, \citenamefont {Kugi},\ and\ \citenamefont
  {Aspelmeyer}}]{magrini2021realtime}%
  \BibitemOpen
  \bibfield  {author} {\bibinfo {author} {\bibfnamefont {L.}~\bibnamefont
  {Magrini}}, \bibinfo {author} {\bibfnamefont {P.}~\bibnamefont {Rosenzweig}},
  \bibinfo {author} {\bibfnamefont {C.}~\bibnamefont {Bach}}, \bibinfo {author}
  {\bibfnamefont {A.}~\bibnamefont {{Deutschmann-Olek}}}, \bibinfo {author}
  {\bibfnamefont {S.~G.}\ \bibnamefont {Hofer}}, \bibinfo {author}
  {\bibfnamefont {S.}~\bibnamefont {Hong}}, \bibinfo {author} {\bibfnamefont
  {N.}~\bibnamefont {Kiesel}}, \bibinfo {author} {\bibfnamefont
  {A.}~\bibnamefont {Kugi}},\ and\ \bibinfo {author} {\bibfnamefont
  {M.}~\bibnamefont {Aspelmeyer}},\ }\href
  {https://doi.org/10.1038/s41586-021-03602-3} {\bibfield  {journal} {\bibinfo
  {journal} {Nature}\ }\textbf {\bibinfo {volume} {595}},\ \bibinfo {pages}
  {373} (\bibinfo {year} {2021})}\BibitemShut {NoStop}%
\bibitem [{\citenamefont {Metzger}\ \emph {et~al.}(2008)\citenamefont
  {Metzger}, \citenamefont {Ludwig}, \citenamefont {Neuenhahn}, \citenamefont
  {Ortlieb}, \citenamefont {Favero}, \citenamefont {Karrai},\ and\
  \citenamefont {Marquardt}}]{metzger2008selfinduced}%
  \BibitemOpen
  \bibfield  {author} {\bibinfo {author} {\bibfnamefont {C.}~\bibnamefont
  {Metzger}}, \bibinfo {author} {\bibfnamefont {M.}~\bibnamefont {Ludwig}},
  \bibinfo {author} {\bibfnamefont {C.}~\bibnamefont {Neuenhahn}}, \bibinfo
  {author} {\bibfnamefont {A.}~\bibnamefont {Ortlieb}}, \bibinfo {author}
  {\bibfnamefont {I.}~\bibnamefont {Favero}}, \bibinfo {author} {\bibfnamefont
  {K.}~\bibnamefont {Karrai}},\ and\ \bibinfo {author} {\bibfnamefont
  {F.}~\bibnamefont {Marquardt}},\ }\href
  {https://doi.org/10.1103/PhysRevLett.101.133903} {\bibfield  {journal}
  {\bibinfo  {journal} {Physical Review Letters}\ }\textbf {\bibinfo {volume}
  {101}},\ \bibinfo {pages} {133903} (\bibinfo {year} {2008})}\BibitemShut
  {NoStop}%
\bibitem [{\citenamefont {Bowen}\ and\ \citenamefont
  {Milburn}(2015)}]{bowen2015quantum}%
  \BibitemOpen
  \bibfield  {author} {\bibinfo {author} {\bibfnamefont {W.~P.}\ \bibnamefont
  {Bowen}}\ and\ \bibinfo {author} {\bibfnamefont {G.~J.}\ \bibnamefont
  {Milburn}},\ }\href {https://doi.org/10.1201/b19379} {\emph {\bibinfo {title}
  {Quantum {{Optomechanics}}}}}\ (\bibinfo  {publisher} {{CRC Press}},\
  \bibinfo {address} {{Boca Raton}},\ \bibinfo {year} {2015})\BibitemShut
  {NoStop}%
\bibitem [{\citenamefont {Thompson}\ \emph {et~al.}(2008)\citenamefont
  {Thompson}, \citenamefont {Zwickl}, \citenamefont {Jayich}, \citenamefont
  {Marquardt}, \citenamefont {Girvin},\ and\ \citenamefont
  {Harris}}]{thompson2008strong}%
  \BibitemOpen
  \bibfield  {author} {\bibinfo {author} {\bibfnamefont {J.~D.}\ \bibnamefont
  {Thompson}}, \bibinfo {author} {\bibfnamefont {B.~M.}\ \bibnamefont
  {Zwickl}}, \bibinfo {author} {\bibfnamefont {A.~M.}\ \bibnamefont {Jayich}},
  \bibinfo {author} {\bibfnamefont {F.}~\bibnamefont {Marquardt}}, \bibinfo
  {author} {\bibfnamefont {S.~M.}\ \bibnamefont {Girvin}},\ and\ \bibinfo
  {author} {\bibfnamefont {J.~G.~E.}\ \bibnamefont {Harris}},\ }\href
  {https://doi.org/10.1038/nature06715} {\bibfield  {journal} {\bibinfo
  {journal} {Nature}\ }\textbf {\bibinfo {volume} {452}},\ \bibinfo {pages}
  {72} (\bibinfo {year} {2008})}\BibitemShut {NoStop}%
\bibitem [{\citenamefont {Rabl}\ \emph {et~al.}(2009)\citenamefont {Rabl},
  \citenamefont {Genes}, \citenamefont {Hammerer},\ and\ \citenamefont
  {Aspelmeyer}}]{rabl2009phasenoise}%
  \BibitemOpen
  \bibfield  {author} {\bibinfo {author} {\bibfnamefont {P.}~\bibnamefont
  {Rabl}}, \bibinfo {author} {\bibfnamefont {C.}~\bibnamefont {Genes}},
  \bibinfo {author} {\bibfnamefont {K.}~\bibnamefont {Hammerer}},\ and\
  \bibinfo {author} {\bibfnamefont {M.}~\bibnamefont {Aspelmeyer}},\ }\href
  {https://doi.org/10.1103/PhysRevA.80.063819} {\bibfield  {journal} {\bibinfo
  {journal} {Physical Review A}\ }\textbf {\bibinfo {volume} {80}},\ \bibinfo
  {pages} {063819} (\bibinfo {year} {2009})}\BibitemShut {NoStop}%
\bibitem [{\citenamefont {An}\ \emph {et~al.}(1997)\citenamefont {An},
  \citenamefont {Sones}, \citenamefont {{Fang-Yen}}, \citenamefont {Dasari},\
  and\ \citenamefont {Feld}}]{an1997optical}%
  \BibitemOpen
  \bibfield  {author} {\bibinfo {author} {\bibfnamefont {K.}~\bibnamefont
  {An}}, \bibinfo {author} {\bibfnamefont {B.~A.}\ \bibnamefont {Sones}},
  \bibinfo {author} {\bibfnamefont {C.}~\bibnamefont {{Fang-Yen}}}, \bibinfo
  {author} {\bibfnamefont {R.~R.}\ \bibnamefont {Dasari}},\ and\ \bibinfo
  {author} {\bibfnamefont {M.~S.}\ \bibnamefont {Feld}},\ }\href
  {https://doi.org/10.1364/OL.22.001433} {\bibfield  {journal} {\bibinfo
  {journal} {Optics Letters}\ }\textbf {\bibinfo {volume} {22}},\ \bibinfo
  {pages} {1433} (\bibinfo {year} {1997})}\BibitemShut {NoStop}%
\bibitem [{\citenamefont {Clerk}\ \emph {et~al.}(2010)\citenamefont {Clerk},
  \citenamefont {Devoret}, \citenamefont {Girvin}, \citenamefont {Marquardt},\
  and\ \citenamefont {Schoelkopf}}]{clerk2010introduction}%
  \BibitemOpen
  \bibfield  {author} {\bibinfo {author} {\bibfnamefont {A.~A.}\ \bibnamefont
  {Clerk}}, \bibinfo {author} {\bibfnamefont {M.~H.}\ \bibnamefont {Devoret}},
  \bibinfo {author} {\bibfnamefont {S.~M.}\ \bibnamefont {Girvin}}, \bibinfo
  {author} {\bibfnamefont {F.}~\bibnamefont {Marquardt}},\ and\ \bibinfo
  {author} {\bibfnamefont {R.~J.}\ \bibnamefont {Schoelkopf}},\ }\href
  {https://doi.org/10.1103/RevModPhys.82.1155} {\bibfield  {journal} {\bibinfo
  {journal} {Reviews of Modern Physics}\ }\textbf {\bibinfo {volume} {82}},\
  \bibinfo {pages} {1155} (\bibinfo {year} {2010})}\BibitemShut {NoStop}%
\bibitem [{\citenamefont {Wieczorek}\ \emph {et~al.}(2015)\citenamefont
  {Wieczorek}, \citenamefont {Hofer}, \citenamefont {{Hoelscher-Obermaier}},
  \citenamefont {Riedinger}, \citenamefont {Hammerer},\ and\ \citenamefont
  {Aspelmeyer}}]{wieczorek2015optimal}%
  \BibitemOpen
  \bibfield  {author} {\bibinfo {author} {\bibfnamefont {W.}~\bibnamefont
  {Wieczorek}}, \bibinfo {author} {\bibfnamefont {S.~G.}\ \bibnamefont
  {Hofer}}, \bibinfo {author} {\bibfnamefont {J.}~\bibnamefont
  {{Hoelscher-Obermaier}}}, \bibinfo {author} {\bibfnamefont {R.}~\bibnamefont
  {Riedinger}}, \bibinfo {author} {\bibfnamefont {K.}~\bibnamefont
  {Hammerer}},\ and\ \bibinfo {author} {\bibfnamefont {M.}~\bibnamefont
  {Aspelmeyer}},\ }\href {https://doi.org/10.1103/PhysRevLett.114.223601}
  {\bibfield  {journal} {\bibinfo  {journal} {Physical Review Letters}\
  }\textbf {\bibinfo {volume} {114}},\ \bibinfo {pages} {223601} (\bibinfo
  {year} {2015})}\BibitemShut {NoStop}%
\bibitem [{\citenamefont {Zhang}\ and\ \citenamefont
  {M{\o}lmer}(2017)}]{zhang2017prediction}%
  \BibitemOpen
  \bibfield  {author} {\bibinfo {author} {\bibfnamefont {J.}~\bibnamefont
  {Zhang}}\ and\ \bibinfo {author} {\bibfnamefont {K.}~\bibnamefont
  {M{\o}lmer}},\ }\href {https://doi.org/10.1103/PhysRevA.96.062131} {\bibfield
   {journal} {\bibinfo  {journal} {Physical Review A}\ }\textbf {\bibinfo
  {volume} {96}},\ \bibinfo {pages} {062131} (\bibinfo {year}
  {2017})}\BibitemShut {NoStop}%
\bibitem [{\citenamefont {Rossi}\ \emph {et~al.}(2019)\citenamefont {Rossi},
  \citenamefont {Mason}, \citenamefont {Chen},\ and\ \citenamefont
  {Schliesser}}]{rossi2019observing}%
  \BibitemOpen
  \bibfield  {author} {\bibinfo {author} {\bibfnamefont {M.}~\bibnamefont
  {Rossi}}, \bibinfo {author} {\bibfnamefont {D.}~\bibnamefont {Mason}},
  \bibinfo {author} {\bibfnamefont {J.}~\bibnamefont {Chen}},\ and\ \bibinfo
  {author} {\bibfnamefont {A.}~\bibnamefont {Schliesser}},\ }\href
  {https://doi.org/10.1103/PhysRevLett.123.163601} {\bibfield  {journal}
  {\bibinfo  {journal} {Physical Review Letters}\ }\textbf {\bibinfo {volume}
  {123}},\ \bibinfo {pages} {163601} (\bibinfo {year} {2019})}\BibitemShut
  {NoStop}%
\bibitem [{\citenamefont {Meng}\ \emph {et~al.}(2022)\citenamefont {Meng},
  \citenamefont {Brawley}, \citenamefont {Khademi}, \citenamefont {Bridge},
  \citenamefont {Bennett},\ and\ \citenamefont
  {Bowen}}]{meng2022measurementbased}%
  \BibitemOpen
  \bibfield  {author} {\bibinfo {author} {\bibfnamefont {C.}~\bibnamefont
  {Meng}}, \bibinfo {author} {\bibfnamefont {G.~A.}\ \bibnamefont {Brawley}},
  \bibinfo {author} {\bibfnamefont {S.}~\bibnamefont {Khademi}}, \bibinfo
  {author} {\bibfnamefont {E.~M.}\ \bibnamefont {Bridge}}, \bibinfo {author}
  {\bibfnamefont {J.~S.}\ \bibnamefont {Bennett}},\ and\ \bibinfo {author}
  {\bibfnamefont {W.~P.}\ \bibnamefont {Bowen}},\ }\href
  {https://doi.org/10.1126/sciadv.abm7585} {\bibfield  {journal} {\bibinfo
  {journal} {Science Advances}\ }\textbf {\bibinfo {volume} {8}},\ \bibinfo
  {pages} {eabm7585} (\bibinfo {year} {2022})}\BibitemShut {NoStop}%
\bibitem [{\citenamefont {Szorkovszky}\ \emph {et~al.}(2011)\citenamefont
  {Szorkovszky}, \citenamefont {Doherty}, \citenamefont {Harris},\ and\
  \citenamefont {Bowen}}]{szorkovszky2011mechanical}%
  \BibitemOpen
  \bibfield  {author} {\bibinfo {author} {\bibfnamefont {A.}~\bibnamefont
  {Szorkovszky}}, \bibinfo {author} {\bibfnamefont {A.~C.}\ \bibnamefont
  {Doherty}}, \bibinfo {author} {\bibfnamefont {G.~I.}\ \bibnamefont
  {Harris}},\ and\ \bibinfo {author} {\bibfnamefont {W.~P.}\ \bibnamefont
  {Bowen}},\ }\href {https://doi.org/10.1103/PhysRevLett.107.213603} {\bibfield
   {journal} {\bibinfo  {journal} {Physical Review Letters}\ }\textbf {\bibinfo
  {volume} {107}},\ \bibinfo {pages} {213603} (\bibinfo {year}
  {2011})}\BibitemShut {NoStop}%
\bibitem [{\citenamefont {Galinskiy}\ \emph {et~al.}(2020)\citenamefont
  {Galinskiy}, \citenamefont {Tsaturyan}, \citenamefont {Parniak},\ and\
  \citenamefont {Polzik}}]{Galinskiy2020Phonon}%
  \BibitemOpen
  \bibfield  {author} {\bibinfo {author} {\bibfnamefont {I.}~\bibnamefont
  {Galinskiy}}, \bibinfo {author} {\bibfnamefont {Y.}~\bibnamefont
  {Tsaturyan}}, \bibinfo {author} {\bibfnamefont {M.}~\bibnamefont {Parniak}},\
  and\ \bibinfo {author} {\bibfnamefont {E.~S.}\ \bibnamefont {Polzik}},\
  }\href {https://doi.org/10.1364/OPTICA.390939} {\bibfield  {journal}
  {\bibinfo  {journal} {Optica}\ }\textbf {\bibinfo {volume} {7}},\ \bibinfo
  {pages} {718} (\bibinfo {year} {2020})}\BibitemShut {NoStop}%
\bibitem [{\citenamefont {Treutlein}\ \emph {et~al.}(2014)\citenamefont
  {Treutlein}, \citenamefont {Genes}, \citenamefont {Hammerer}, \citenamefont
  {Poggio},\ and\ \citenamefont {Rabl}}]{treutlein2014hybrid}%
  \BibitemOpen
  \bibfield  {author} {\bibinfo {author} {\bibfnamefont {P.}~\bibnamefont
  {Treutlein}}, \bibinfo {author} {\bibfnamefont {C.}~\bibnamefont {Genes}},
  \bibinfo {author} {\bibfnamefont {K.}~\bibnamefont {Hammerer}}, \bibinfo
  {author} {\bibfnamefont {M.}~\bibnamefont {Poggio}},\ and\ \bibinfo {author}
  {\bibfnamefont {P.}~\bibnamefont {Rabl}},\ }in\ \href
  {https://doi.org/10.1007/978-3-642-55312-7\_14} {\emph {\bibinfo {booktitle}
  {Cavity {{Optomechanics}}: {{Nano-}} and {{Micromechanical Resonators
  Interacting}} with {{Light}}}}},\ \bibinfo {series and number} {Quantum
  {{Science}} and {{Technology}}},\ \bibinfo {editor} {edited by\ \bibinfo
  {editor} {\bibfnamefont {M.}~\bibnamefont {Aspelmeyer}}, \bibinfo {editor}
  {\bibfnamefont {T.~J.}\ \bibnamefont {Kippenberg}},\ and\ \bibinfo {editor}
  {\bibfnamefont {F.}~\bibnamefont {Marquardt}}}\ (\bibinfo  {publisher}
  {{Springer}},\ \bibinfo {address} {{Berlin, Heidelberg}},\ \bibinfo {year}
  {2014})\ pp.\ \bibinfo {pages} {327--351}\BibitemShut {NoStop}%
\bibitem [{\citenamefont {Karg}\ \emph {et~al.}(2020)\citenamefont {Karg},
  \citenamefont {Gouraud}, \citenamefont {Ngai}, \citenamefont {Schmid},
  \citenamefont {Hammerer},\ and\ \citenamefont
  {Treutlein}}]{karg2020lightmediated}%
  \BibitemOpen
  \bibfield  {author} {\bibinfo {author} {\bibfnamefont {T.~M.}\ \bibnamefont
  {Karg}}, \bibinfo {author} {\bibfnamefont {B.}~\bibnamefont {Gouraud}},
  \bibinfo {author} {\bibfnamefont {C.~T.}\ \bibnamefont {Ngai}}, \bibinfo
  {author} {\bibfnamefont {G.-L.}\ \bibnamefont {Schmid}}, \bibinfo {author}
  {\bibfnamefont {K.}~\bibnamefont {Hammerer}},\ and\ \bibinfo {author}
  {\bibfnamefont {P.}~\bibnamefont {Treutlein}},\ }\href
  {https://doi.org/10.1126/science.abb0328} {\bibfield  {journal} {\bibinfo
  {journal} {Science}\ }\textbf {\bibinfo {volume} {369}},\ \bibinfo {pages}
  {174} (\bibinfo {year} {2020})}\BibitemShut {NoStop}%
\bibitem [{\citenamefont {Fischer}\ \emph {et~al.}(2019)\citenamefont
  {Fischer}, \citenamefont {McNally}, \citenamefont {Reetz}, \citenamefont
  {Assump{\c c}{\~a}o}, \citenamefont {Knief}, \citenamefont {Lin},\ and\
  \citenamefont {Regal}}]{fischer2019spin}%
  \BibitemOpen
  \bibfield  {author} {\bibinfo {author} {\bibfnamefont {R.}~\bibnamefont
  {Fischer}}, \bibinfo {author} {\bibfnamefont {D.~P.}\ \bibnamefont
  {McNally}}, \bibinfo {author} {\bibfnamefont {C.}~\bibnamefont {Reetz}},
  \bibinfo {author} {\bibfnamefont {G.~G.~T.}\ \bibnamefont {Assump{\c
  c}{\~a}o}}, \bibinfo {author} {\bibfnamefont {T.}~\bibnamefont {Knief}},
  \bibinfo {author} {\bibfnamefont {Y.}~\bibnamefont {Lin}},\ and\ \bibinfo
  {author} {\bibfnamefont {C.~A.}\ \bibnamefont {Regal}},\ }\href
  {https://doi.org/10.1088/1367-2630/ab117a} {\bibfield  {journal} {\bibinfo
  {journal} {New Journal of Physics}\ }\textbf {\bibinfo {volume} {21}},\
  \bibinfo {pages} {043049} (\bibinfo {year} {2019})}\BibitemShut {NoStop}%
\bibitem [{\citenamefont {Ko{\v s}ata}\ \emph {et~al.}(2020)\citenamefont
  {Ko{\v s}ata}, \citenamefont {Zilberberg}, \citenamefont {Degen},
  \citenamefont {Chitra},\ and\ \citenamefont {Eichler}}]{kosata2020spin}%
  \BibitemOpen
  \bibfield  {author} {\bibinfo {author} {\bibfnamefont {J.}~\bibnamefont
  {Ko{\v s}ata}}, \bibinfo {author} {\bibfnamefont {O.}~\bibnamefont
  {Zilberberg}}, \bibinfo {author} {\bibfnamefont {C.~L.}\ \bibnamefont
  {Degen}}, \bibinfo {author} {\bibfnamefont {R.}~\bibnamefont {Chitra}},\ and\
  \bibinfo {author} {\bibfnamefont {A.}~\bibnamefont {Eichler}},\ }\href
  {https://doi.org/10.1103/PhysRevApplied.14.014042} {\bibfield  {journal}
  {\bibinfo  {journal} {Physical Review Applied}\ }\textbf {\bibinfo {volume}
  {14}},\ \bibinfo {pages} {014042} (\bibinfo {year} {2020})}\BibitemShut
  {NoStop}%
\bibitem [{\citenamefont {Simonsen}\ \emph {et~al.}(2019)\citenamefont
  {Simonsen}, \citenamefont {Saarinen}, \citenamefont {Sanchez}, \citenamefont
  {{Ardenkj{\ae}r-Larsen}}, \citenamefont {Schliesser},\ and\ \citenamefont
  {Polzik}}]{simonsen2019sensitive}%
  \BibitemOpen
  \bibfield  {author} {\bibinfo {author} {\bibfnamefont {A.}~\bibnamefont
  {Simonsen}}, \bibinfo {author} {\bibfnamefont {S.~A.}\ \bibnamefont
  {Saarinen}}, \bibinfo {author} {\bibfnamefont {J.~D.}\ \bibnamefont
  {Sanchez}}, \bibinfo {author} {\bibfnamefont {J.~H.}\ \bibnamefont
  {{Ardenkj{\ae}r-Larsen}}}, \bibinfo {author} {\bibfnamefont {A.}~\bibnamefont
  {Schliesser}},\ and\ \bibinfo {author} {\bibfnamefont {E.~S.}\ \bibnamefont
  {Polzik}},\ }\href {https://doi.org/10.1364/OE.27.018561} {\bibfield
  {journal} {\bibinfo  {journal} {Optics Express}\ }\textbf {\bibinfo {volume}
  {27}},\ \bibinfo {pages} {18561} (\bibinfo {year} {2019})}\BibitemShut
  {NoStop}%
\bibitem [{\citenamefont {Zhou}\ \emph {et~al.}(2021)\citenamefont {Zhou},
  \citenamefont {Bao}, \citenamefont {Madugani}, \citenamefont {Long},
  \citenamefont {Gorman},\ and\ \citenamefont {LeBrun}}]{zhou2021broadband}%
  \BibitemOpen
  \bibfield  {author} {\bibinfo {author} {\bibfnamefont {F.}~\bibnamefont
  {Zhou}}, \bibinfo {author} {\bibfnamefont {Y.}~\bibnamefont {Bao}}, \bibinfo
  {author} {\bibfnamefont {R.}~\bibnamefont {Madugani}}, \bibinfo {author}
  {\bibfnamefont {D.~A.}\ \bibnamefont {Long}}, \bibinfo {author}
  {\bibfnamefont {J.~J.}\ \bibnamefont {Gorman}},\ and\ \bibinfo {author}
  {\bibfnamefont {T.~W.}\ \bibnamefont {LeBrun}},\ }\href
  {https://doi.org/10.1364/OPTICA.413117} {\bibfield  {journal} {\bibinfo
  {journal} {Optica}\ }\textbf {\bibinfo {volume} {8}},\ \bibinfo {pages} {350}
  (\bibinfo {year} {2021})}\BibitemShut {NoStop}%
\bibitem [{\citenamefont {Gorodetksy}\ \emph {et~al.}(2010)\citenamefont
  {Gorodetksy}, \citenamefont {Schliesser}, \citenamefont {Anetsberger},
  \citenamefont {Deleglise},\ and\ \citenamefont
  {Kippenberg}}]{gorodetksy2010determination}%
  \BibitemOpen
  \bibfield  {author} {\bibinfo {author} {\bibfnamefont {M.~L.}\ \bibnamefont
  {Gorodetksy}}, \bibinfo {author} {\bibfnamefont {A.}~\bibnamefont
  {Schliesser}}, \bibinfo {author} {\bibfnamefont {G.}~\bibnamefont
  {Anetsberger}}, \bibinfo {author} {\bibfnamefont {S.}~\bibnamefont
  {Deleglise}},\ and\ \bibinfo {author} {\bibfnamefont {T.~J.}\ \bibnamefont
  {Kippenberg}},\ }\href {https://doi.org/10.1364/OE.18.023236} {\bibfield
  {journal} {\bibinfo  {journal} {Optics Express}\ }\textbf {\bibinfo {volume}
  {18}},\ \bibinfo {pages} {23236} (\bibinfo {year} {2010})}\BibitemShut
  {NoStop}%
\bibitem [{\citenamefont {Weis}\ \emph {et~al.}(2010)\citenamefont {Weis},
  \citenamefont {Rivi{\`e}re}, \citenamefont {Del{\'e}glise}, \citenamefont
  {Gavartin}, \citenamefont {Arcizet}, \citenamefont {Schliesser},\ and\
  \citenamefont {Kippenberg}}]{weis2010optomechanically}%
  \BibitemOpen
  \bibfield  {author} {\bibinfo {author} {\bibfnamefont {S.}~\bibnamefont
  {Weis}}, \bibinfo {author} {\bibfnamefont {R.}~\bibnamefont {Rivi{\`e}re}},
  \bibinfo {author} {\bibfnamefont {S.}~\bibnamefont {Del{\'e}glise}}, \bibinfo
  {author} {\bibfnamefont {E.}~\bibnamefont {Gavartin}}, \bibinfo {author}
  {\bibfnamefont {O.}~\bibnamefont {Arcizet}}, \bibinfo {author} {\bibfnamefont
  {A.}~\bibnamefont {Schliesser}},\ and\ \bibinfo {author} {\bibfnamefont
  {T.~J.}\ \bibnamefont {Kippenberg}},\ }\href
  {https://doi.org/10.1126/science.1195596} {\bibfield  {journal} {\bibinfo
  {journal} {Science}\ }\textbf {\bibinfo {volume} {330}},\ \bibinfo {pages}
  {1520} (\bibinfo {year} {2010})}\BibitemShut {NoStop}%
\bibitem [{\citenamefont {Andersen}\ \emph {et~al.}(2010)\citenamefont
  {Andersen}, \citenamefont {Leuchs},\ and\ \citenamefont
  {Silberhorn}}]{andersen2010continuousvariable}%
  \BibitemOpen
  \bibfield  {author} {\bibinfo {author} {\bibfnamefont {U.}~\bibnamefont
  {Andersen}}, \bibinfo {author} {\bibfnamefont {G.}~\bibnamefont {Leuchs}},\
  and\ \bibinfo {author} {\bibfnamefont {C.}~\bibnamefont {Silberhorn}},\
  }\href {https://doi.org/10.1002/lpor.200910010} {\bibfield  {journal}
  {\bibinfo  {journal} {Laser \& Photonics Reviews}\ }\textbf {\bibinfo
  {volume} {4}},\ \bibinfo {pages} {337} (\bibinfo {year} {2010})}\BibitemShut
  {NoStop}%
\bibitem [{\citenamefont {Massel}\ \emph {et~al.}(2012)\citenamefont {Massel},
  \citenamefont {Cho}, \citenamefont {Pirkkalainen}, \citenamefont {Hakonen},
  \citenamefont {Heikkil{\"a}},\ and\ \citenamefont
  {Sillanp{\"a}{\"a}}}]{massel2012multimode}%
  \BibitemOpen
  \bibfield  {author} {\bibinfo {author} {\bibfnamefont {F.}~\bibnamefont
  {Massel}}, \bibinfo {author} {\bibfnamefont {S.~U.}\ \bibnamefont {Cho}},
  \bibinfo {author} {\bibfnamefont {J.-M.}\ \bibnamefont {Pirkkalainen}},
  \bibinfo {author} {\bibfnamefont {P.~J.}\ \bibnamefont {Hakonen}}, \bibinfo
  {author} {\bibfnamefont {T.~T.}\ \bibnamefont {Heikkil{\"a}}},\ and\ \bibinfo
  {author} {\bibfnamefont {M.~A.}\ \bibnamefont {Sillanp{\"a}{\"a}}},\ }\href
  {https://doi.org/10.1038/ncomms1993} {\bibfield  {journal} {\bibinfo
  {journal} {Nature Communications}\ }\textbf {\bibinfo {volume} {3}},\
  \bibinfo {pages} {987} (\bibinfo {year} {2012})}\BibitemShut {NoStop}%
\bibitem [{\citenamefont {Guo}\ \emph {et~al.}(2019)\citenamefont {Guo},
  \citenamefont {Norte},\ and\ \citenamefont
  {Gr{\"o}blacher}}]{guo2019feedback}%
  \BibitemOpen
  \bibfield  {author} {\bibinfo {author} {\bibfnamefont {J.}~\bibnamefont
  {Guo}}, \bibinfo {author} {\bibfnamefont {R.}~\bibnamefont {Norte}},\ and\
  \bibinfo {author} {\bibfnamefont {S.}~\bibnamefont {Gr{\"o}blacher}},\ }\href
  {https://doi.org/10.1103/PhysRevLett.123.223602} {\bibfield  {journal}
  {\bibinfo  {journal} {Physical Review Letters}\ }\textbf {\bibinfo {volume}
  {123}},\ \bibinfo {pages} {223602} (\bibinfo {year} {2019})}\BibitemShut
  {NoStop}%
\bibitem [{\citenamefont {Fedorov}\ \emph {et~al.}(2019)\citenamefont
  {Fedorov}, \citenamefont {Engelsen}, \citenamefont {Ghadimi}, \citenamefont
  {Bereyhi}, \citenamefont {Schilling}, \citenamefont {Wilson},\ and\
  \citenamefont {Kippenberg}}]{fedorov2019generalized}%
  \BibitemOpen
  \bibfield  {author} {\bibinfo {author} {\bibfnamefont {S.~A.}\ \bibnamefont
  {Fedorov}}, \bibinfo {author} {\bibfnamefont {N.~J.}\ \bibnamefont
  {Engelsen}}, \bibinfo {author} {\bibfnamefont {A.~H.}\ \bibnamefont
  {Ghadimi}}, \bibinfo {author} {\bibfnamefont {M.~J.}\ \bibnamefont
  {Bereyhi}}, \bibinfo {author} {\bibfnamefont {R.}~\bibnamefont {Schilling}},
  \bibinfo {author} {\bibfnamefont {D.~J.}\ \bibnamefont {Wilson}},\ and\
  \bibinfo {author} {\bibfnamefont {T.~J.}\ \bibnamefont {Kippenberg}},\ }\href
  {https://doi.org/10.1103/PhysRevB.99.054107} {\bibfield  {journal} {\bibinfo
  {journal} {Physical Review B}\ }\textbf {\bibinfo {volume} {99}},\ \bibinfo
  {pages} {054107} (\bibinfo {year} {2019})}\BibitemShut {NoStop}%
\bibitem [{\citenamefont {Beccari}\ \emph {et~al.}(2022)\citenamefont
  {Beccari}, \citenamefont {Visani}, \citenamefont {Fedorov}, \citenamefont
  {Bereyhi}, \citenamefont {Boureau}, \citenamefont {Engelsen},\ and\
  \citenamefont {Kippenberg}}]{beccari2022strained}%
  \BibitemOpen
  \bibfield  {author} {\bibinfo {author} {\bibfnamefont {A.}~\bibnamefont
  {Beccari}}, \bibinfo {author} {\bibfnamefont {D.~A.}\ \bibnamefont {Visani}},
  \bibinfo {author} {\bibfnamefont {S.~A.}\ \bibnamefont {Fedorov}}, \bibinfo
  {author} {\bibfnamefont {M.~J.}\ \bibnamefont {Bereyhi}}, \bibinfo {author}
  {\bibfnamefont {V.}~\bibnamefont {Boureau}}, \bibinfo {author} {\bibfnamefont
  {N.~J.}\ \bibnamefont {Engelsen}},\ and\ \bibinfo {author} {\bibfnamefont
  {T.~J.}\ \bibnamefont {Kippenberg}},\ }\href
  {https://doi.org/10.1038/s41567-021-01498-4} {\bibfield  {journal} {\bibinfo
  {journal} {Nature Physics}\ }\textbf {\bibinfo {volume} {18}},\ \bibinfo
  {pages} {436} (\bibinfo {year} {2022})}\BibitemShut {NoStop}%
\bibitem [{\citenamefont {Yu}\ \emph {et~al.}(2012)\citenamefont {Yu},
  \citenamefont {Purdy},\ and\ \citenamefont {Regal}}]{yu2012control}%
  \BibitemOpen
  \bibfield  {author} {\bibinfo {author} {\bibfnamefont {P.-L.}\ \bibnamefont
  {Yu}}, \bibinfo {author} {\bibfnamefont {{\relax TP}.}~\bibnamefont
  {Purdy}},\ and\ \bibinfo {author} {\bibfnamefont {{\relax CA}.}~\bibnamefont
  {Regal}},\ }\href@noop {} {\bibfield  {journal} {\bibinfo  {journal} {Phys.
  Rev. Lett.}\ }\textbf {\bibinfo {volume} {108}},\ \bibinfo {pages} {083603}
  (\bibinfo {year} {2012})}\BibitemShut {NoStop}%
\bibitem [{\citenamefont {Bao}\ \emph {et~al.}(2002)\citenamefont {Bao},
  \citenamefont {Yang}, \citenamefont {Yin},\ and\ \citenamefont
  {Sun}}]{bao2002energy}%
  \BibitemOpen
  \bibfield  {author} {\bibinfo {author} {\bibfnamefont {M.}~\bibnamefont
  {Bao}}, \bibinfo {author} {\bibfnamefont {H.}~\bibnamefont {Yang}}, \bibinfo
  {author} {\bibfnamefont {H.}~\bibnamefont {Yin}},\ and\ \bibinfo {author}
  {\bibfnamefont {Y.}~\bibnamefont {Sun}},\ }\href
  {https://doi.org/10.1088/0960-1317/12/3/322} {\bibfield  {journal} {\bibinfo
  {journal} {Journal of Micromechanics and Microengineering}\ }\textbf
  {\bibinfo {volume} {12}},\ \bibinfo {pages} {341} (\bibinfo {year}
  {2002})}\BibitemShut {NoStop}%
\end{thebibliography}%

\end{document}